\documentclass{JHEP3}

\usepackage{amsfonts,amsmath}               
\usepackage{longtable}                      
\usepackage{enumerate}                      
\usepackage{pstricks,pst-node}             
\usepackage[english]{babel}                 
\usepackage[vcentermath]{youngtab}          
\usepackage{enumitem}                       
\usepackage{multirow}                       
\usepackage{float}                          
\usepackage{lscape}                         
\usepackage[caption=false]{subfig}          
\usepackage{bm}                             
\usepackage{cite}   

\psset{unit=.6cm,linewidth=.5pt,radius=.2}  

\def\a{\alpha}
\def\b{\beta}

\def\d{\delta}
\def\e{\epsilon}

\newcommand{\pf}[1]{\boldsymbol{\mathit{#1}}}

\newcommand{\be}{\begin{equation}}
\newcommand{\ee}{\end{equation}}

\newcommand{\R}{\mathbb{R}}                 
\newcommand{\norm}[2]{(#1 \, | \, #2)}      
\newcommand{\bi}[1]{\textbf{\emph{#1}}}     
\newcommand\dash{\nobreakdash-\hspace{0pt}} 

\newcommand{\fund}{{\tiny \yng(1)}}
\newcommand{\twoform}{{\tiny \yng(1,1)}}
\newcommand{\threeform}{{\tiny \yng(1,1,1)}}
\newcommand{\fourform}{{\tiny \yng(1,1,1,1)}}

\newcommand{\symm}{{\tiny \yng(2)}}
\newcommand{\twohook}{{\tiny \yng(2,1)}}
\newcommand{\threehook}{{\tiny \yng(2,1,1)}}

\newcommand{\fivehook}{{\tiny \yng(2,1,1,1,1)}}

\newcommand{\normalNode}[2]{\Cnode(#1){#2}}

\newcommand{\dualityNode}[2]{\Cnode[fillstyle=solid,fillcolor=lightgray](#1){#2}}

\newcommand{\disabledNode}[2]{\Cnode[fillstyle=solid,fillcolor=black](#1){#2}}

\newcommand{\singleConnection}[2]{\ncline{-}{#1}{#2}}

\newcommand{\nodeLabel}[2]{\nput[labelsep=0.2]{-40}{#1}{\tiny{#2}}}

\title{Kac--Moody Spectrum of (Half--)Maximal Supergravities}

\author{Eric A. Bergshoeff$^1$, Joaquim Gomis$^{2,3}$, Teake A. Nutma$^1$ and Diederik Roest$^2$ \\
    ~ \\
    $^1$ Centre for Theoretical Physics, University of Groningen, \\
    \hspace*{0.15cm} Nijenborgh 4, 9747 AG Groningen, The Netherlands \\
    \hspace*{0.25cm} \email{E.A.Bergshoeff@rug.nl, T.A.Nutma@rug.nl} \\
    ~ \\
    $^2$ Departament Estructura i Constituents de la Materia \\
    \hspace*{0.15cm} Facultat de F\'{i}sica, Universitat de Barcelona \\
    \hspace*{0.15cm} Diagonal 647, 08028 Barcelona, Spain \\
    \hspace*{0.25cm} \email{gomis@ecm.ub.es, droest@ecm.ub.es }\\
    ~\\
    $^3$ PH-TH Division, CERN \\
     \hspace*{0.15cm} 1211 Geneva 23, Switzerland
}

\abstract{We establish the correspondence between, on one side, the
possible gaugings and massive deformations of half--maximal
supergravity coupled to vector multiplets and, on the other side,
certain generators of the associated very extended Kac--Moody
algebras. The difference between generators associated to gaugings
and to massive deformations is pointed out. Furthermore, we argue
that another set of generators are related to the so-called
quadratic constraints of the embedding tensor. Special emphasis is
placed on a truncation of the Kac--Moody algebra that is related to
the bosonic gauge transformations of supergravity. We give a separate
discussion of this truncation when non-zero deformations are present.
The new insights are also illustrated in the context of maximal
supergravity. }

\preprint{
    UG-07-06\\
    UB-ECM-PF-07/24\\
    CERN-PH-TH/2007-165
}

\begin{document}

\section{Introduction}

In describing the field theoretic representation of a supersymmetry
algebra, one usually specifies those fields that represent physical
states only. It is known that other fields can be added to the
supermultiplet that do not describe physical states but on which
nevertheless the full supersymmetry algebra can be realized (for an
early discussion of such potentials, see \cite{Gates:1980ay}). In this paper we
will focus on the following two classes of such fields.

The first class consists of $(D-1)$--form potentials in $D$
dimensions, which we will call ``deformation potentials''
for the following reason. The
equations of motion of these deformation potentials can be solved in
terms of integration constants that describe deformations of the
supersymmetric theory. The foremost example of a deformation
potential is the nine--form potential of type IIA string theory that
couples to the D8--brane \cite{Polchinski:1995mt,Polchinski:1995df,
Bergshoeff:1996ui}. The integration constant corresponding to this
nine--form potential is the masslike parameter $m$ of massive IIA
supergravity \cite{Romans:1985tz}. The relation between the two is given by
\be
d \,{}^\star F_{(10)}(A_{(9)}) = 0\hskip .5truecm
\Rightarrow\hskip .5truecm  {}^\star F_{(10)}(A_{(9)})\ \propto\ m\,.
\label{dualm}
\ee

The second class of fields that do not describe physical states
consists of $D$--form potentials in $D$ dimensions, which we will
call ``top--forms potentials'', or top-forms for short. The prime example of a top--form is the
Ramond--Ramond ten--form that couples to the D9--brane of type IIB
string theory \cite{Polchinski:1995mt}.
It turns out that this ten--form is part of a
quadruplet of ten-forms transforming according to the ${\bf 4}$
representation of the $SL(2,\R)$ duality group, while also a
doublet ${\bf 2}$ of ten-forms can be added in IIB supergravity
\cite{Bergshoeff:2005ac}.

It has been known for a number of years that one can reproduce the physical degrees of freedom
of maximal supergravity from the very extended Kac--Moody algebra $E_{11}$
\cite{West:2001as,Schnakenburg:2001ya,Kleinschmidt:2003mf}.
Furthermore, this Kac--Moody algebra contains generators corresponding
to the deformation potential of IIA \cite{Kleinschmidt:2003mf, West:2004st}
and the top--form potentials of IIB \cite{Kleinschmidt:2003mf, Kleinschmidt:2004rg, West:2005gu}.
Recently, the representations under the duality group of the
deformation and top--form potentials of all maximal supergravities
have been calculated \cite{Riccioni:2007au, Bergshoeff:2007qi}.
Remarkably, the $E_{11}$ results on deformation potentials are in agreement with those of
\cite{Nicolai:2001sv, deWit:2002vt, deWit:2003ja, deWit:2003hr, deWit:2005hv, firenze,
deWit:2004nw, Samtleben:2005bp, Bergshoeff:2002nv, Bergshoeff:2003ri}
where maximal gauged supergravities are classified within a supergravity approach\footnote{
An exception to this correspondence are the gauging of the `trombone' or scale
symmetry of the field equations and Bianchi identities \cite{Cremmer:1997xj},
as discussed in e.g.~\cite{HLW,Bergshoeff:2002nv, Bergshoeff:2003ri},
for which no corresponding deformation potentials have been identified in $E_{11}$.}.
In particular, this agreement shows that the components of the
embedding tensor
\cite{Nicolai:2001sv, deWit:2002vt,deWit:2003hr} can be identified with
the masslike deformation parameters of the supergravity theory. Therefore,
the field strength $F_{(D)}$ of the deformation
potential $A_{(D-1)}$ is proportional to the embedding tensor $\Theta$:
\be
{}^\star\, F_{(D)} \left(A_{(D-1)}\right)\ \propto\
 \Theta \, . \label{emb-ten}
\ee
This relation can be viewed as a duality relation, like the ones
between potentials and dual potentials.

It is natural to extend the analysis of
\cite{Riccioni:2007au, Bergshoeff:2007qi} to other cases. In this
paper we will do this for the class of half--maximal supergravity
theories. The Kac--Moody analysis for this case shows a number of new features. First of all, one
can add matter vector multiplets and consider matter--coupled
supergravity \cite{Kleinschmidt:2003mf, Schnakenburg:2004vd}.
Our results on the deformation and top--form
potentials will depend on the number of vector multiplets. Another
new feature is that one encounters duality groups that are
{\it not} maximal non-compact. Only a limited number of vector multiplets lead to a
maximal non--compact duality group. Finally, the duality groups are not
necessarily simply laced, and hence we will have to address the issue
of non-symmetric Cartan matrices and roots of different lengths. For
more details on the latter, see appendix \ref{app:group_theory}.

An additional motivation to study the case of half--maximal
supergravities is that for $D<10$, e.g. $D=4$ or $D=6$, the
corresponding matter-coupled supergravities  are related to
compactifications of string theory and M-theory with background
fluxes. The nonzero fluxes lead to the additional mass parameters.
Especially the $D=6$ case is interesting due to the existence of a
chiral and a non-chiral theory. These two theories
are related via S- and T-dualities between Type I string theory on $T^4$
and Type II string theory on K3. The
mass parameters of these theories have been investigated
\cite{Kaloper:1999yr,Haack:2001iz} and the massive dualities
between them have been studied \cite{Janssen:2001hy, Behrndt:2001ab}.

In this paper we will pay particular attention to the bosonic
algebra that the different $p$--form Kac--Moody
generators with $p>0$ satisfy amongst each
other. We will call this algebra the ``$p$--form algebra''. This
algebra, without the deformation and  top-form generators, also
occurs in
\cite{Cremmer:1998px,Lavrinenko:1999xi} as the bosonic gauge algebra
of supergravity. The $p$--form potentials
corresponding to these generators, together
with gravity and the scalar fields, constitute the part of the very
extended Kac--Moody spectrum that does not
require the introduction of the dual graviton.
We will show how the possible deformation and top--form potentials,
with which the $p$--form algebra can be extended, follow
from the Kac--Moody algebra. In particular, we will show that for the case of half--maximal supergravity
the deformation potentials of the $p$--form algebra, and hence also
the embedding tensor in generic dimensions, can be written in terms of the
fundamental and three-form representation of the duality group.

One encounters the following subtlety in establishing the connection
between the $p$--form algebra and supergravity:
whereas for each physical state the Kac--Moody algebra gives rise to both the
potential and the dual potential this is not the case for the
deformation potentials. The Kac--Moody algebra does give rise to the
deformation potentials but not to the
dual embedding tensor. Indeed the duality relation \eqref{emb-ten} does
not follow from the Kac--Moody approach. We know from supergravity
that the inclusion of a mass parameter or an embedding tensor leads to deformations of the transformation rules.
We will show that in specific cases these deformations cannot be captured by the $p$--form
algebra alone but that, instead, one is forced to introduce further mixed symmetry
generators whose interpretation has yet to be clarified.

This paper is organized as follows. In section \ref{sec:KM} we briefly
summarize the Kac--Moody approach to supergravity. In section \ref{sec:pfa} we introduce
the $p$--form algebra and uncover interesting properties of the
deformation and top--form potentials in the context of this algebra.
We will use the case of maximal supergravity to elucidate a few of these general properties.
In the next section we apply the Kac--Moody approach to the case of
half--maximal supergravity. In section \ref{sec:def} we will show that the
addition  of the embedding tensor leads to the introduction
of additional symmetry generators to obtain closure.
Finally, in the conclusions we
comment on our results. We have included four appendices. Appendix \ref{app:defs}
shortly summarizes the terminology we introduce in this paper.
Appendix \ref{app:half--max} contains
a brief summary of the physical degrees of freedom and duality groups of
matter-coupled half--maximal supergravity.
Appendix \ref{app:group_theory} covers some group--theoretical details
concerning the Kac--Moody algebras that are non--simply laced. Finally, appendix
\ref{app:results} contains lists of tables with the relevant low
level results of the spectrum of the relevant Kac--Moody algebra.

\section{The Kac--Moody approach to supergravity}\label{sec:KM}

The spectrum of physical states of the different maximal supergravity
theories can be obtained from the very extended Kac-Moody algebra $E_{11}$
\cite{West:2001as, Schnakenburg:2001ya, Kleinschmidt:2003mf}.
This has been extended to the set of all possible deformation and
top--form potentials in \cite{Riccioni:2007au, Bergshoeff:2007qi}.
A similar analysis could be done for $E_{10}$ \cite{Nicolai:2003fw, Damour:2002cu, Damour:2004zy}
except for the top--form potentials. In addition, non--maximal supergravity
and the associated very extended Kac--Moody algebras have been
discussed in \cite{Kleinschmidt:2003mf, Schnakenburg:2004vd, Riccioni:2007hm}.
In the present paper we will apply the ``Kac--Moody approach'' to extract
the deformation and top--form potentials of half--maximal supergravity.
In general this  approach breaks down into four steps:

\begin{enumerate}
    \item Reduce to $D=3$ over a torus and determine the  $G/K(G)$ scalar
            coset sigma model.
    \item Take the very extension $G^{+++}/K(G^{+++})$.
    \item Oxidize back to $3 \leq D \leq D_{\rm{max}}$.
    \item Read off the spectrum by means of a level decomposition.
\end{enumerate}

\noindent
As steps 2, 3, and 4 can be automatically carried out on the computer
\cite{SimpLie}, this approach is very simple to carry out in practice.
We will now take a close look at each of these steps.

The first step is to determine the $G/K(G)$ scalar coset sigma model in
three dimensions for the toroidally reduced supergravity in
question, where $K(G)$ is the maximal compact subgroup of $G$. If
there is no such a  sigma model, which often is the case for theories with less than 16 supercharges,
the Kac--Moody approach comes to a standstill. But when the coset does exist,
as is the case for maximal and half--maximal supergravity, we can go on and take
the very extension $G^{+++}/K(G^{+++})$. The first extension
corresponds to the (untwisted) affine version of $G$, which has been
shown to be the symmetry group of various supergravities in $D=2$
\cite{Julia:1982gx}. Also the second (over) extension and the third
(very) extension are conjectured to be symmetry groups of maximal
supergravity: the former has been employed for a $D=1$ coset
\cite{Nicolai:2003fw, Damour:2002cu, Damour:2004zy} while the latter
has been used for non-linear realisations of the higher-dimensional theory
\cite{West:2001as,Schnakenburg:2001ya,Kleinschmidt:2003mf}.

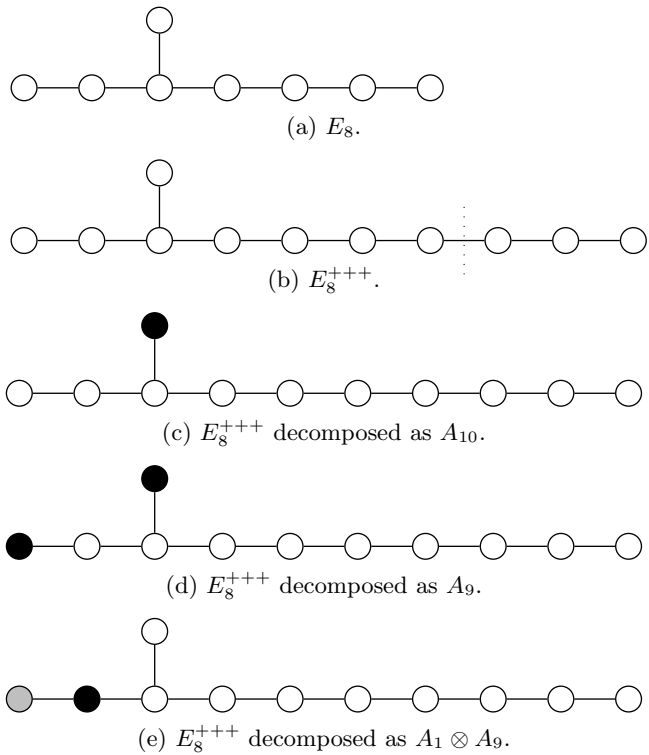
\begin{figure}[t]%
    \centering
    \psset{unit=.9cm,radius=.2}%
    \subfloat[$E_8$.]{\label{E8} \begin{pspicture}(0,-0.2)(9,1)
\normalNode{2,1}{N11213619404}
\normalNode{0,0}{N21213619404}
\normalNode{1,0}{N31213619404}
\normalNode{2,0}{N41213619404}
\normalNode{3,0}{N51213619404}
\normalNode{4,0}{N61213619404}
\normalNode{5,0}{N71213619404}
\normalNode{6,0}{N81213619404}
\ncline{-}{N21213619404}{N31213619404}
\ncline{-}{N31213619404}{N41213619404}
\ncline{-}{N41213619404}{N51213619404}
\ncline{-}{N51213619404}{N61213619404}
\ncline{-}{N61213619404}{N71213619404}
\ncline{-}{N71213619404}{N81213619404}
\ncline{-}{N11213619404}{N41213619404}
\end{pspicture}} \\
    \subfloat[$E_8^{+++}$.]{\label{veE8} \begin{pspicture}(0,-0.2)(9,1)
\normalNode{2,1}{N11213619404}
\normalNode{0,0}{N21213619404}
\normalNode{1,0}{N31213619404}
\normalNode{2,0}{N41213619404}
\normalNode{3,0}{N51213619404}
\normalNode{4,0}{N61213619404}
\normalNode{5,0}{N71213619404}
\normalNode{6,0}{N81213619404}
\normalNode{7,0}{N91213619404}
\normalNode{8,0}{N101213619404}
\normalNode{9,0}{N111213619404}
\ncline{-}{N21213619404}{N31213619404}
\ncline{-}{N31213619404}{N41213619404}
\ncline{-}{N41213619404}{N51213619404}
\ncline{-}{N51213619404}{N61213619404}
\ncline{-}{N61213619404}{N71213619404}
\ncline{-}{N71213619404}{N81213619404}
\ncline{-}{N81213619404}{N91213619404}
\ncline{-}{N91213619404}{N101213619404}
\ncline{-}{N101213619404}{N111213619404}
\ncline{-}{N11213619404}{N41213619404}
\psline[linestyle=dotted]{-}(6.5,0.5)(6.5,-0.5)
\end{pspicture}} \\
    \subfloat[$E_8^{+++}$ decomposed as $A_{10}$.]{\label{E11-11D}%
\begin{pspicture}(0,-0.2)(9,1)
\disabledNode{2,1}{N11723083412}
\normalNode{0,0}{N21723083412}
\normalNode{1,0}{N31723083412}
\normalNode{2,0}{N41723083412}
\normalNode{3,0}{N51723083412}
\normalNode{4,0}{N61723083412}
\normalNode{5,0}{N71723083412}
\normalNode{6,0}{N81723083412}
\normalNode{7,0}{N91723083412}
\normalNode{8,0}{N101723083412}
\normalNode{9,0}{N111723083412}
\singleConnection{N21723083412}{N31723083412}
\singleConnection{N31723083412}{N41723083412}
\singleConnection{N41723083412}{N51723083412}
\singleConnection{N51723083412}{N61723083412}
\singleConnection{N61723083412}{N71723083412}
\singleConnection{N71723083412}{N81723083412}
\singleConnection{N81723083412}{N91723083412}
\singleConnection{N91723083412}{N101723083412}
\singleConnection{N101723083412}{N111723083412}
\singleConnection{N11723083412}{N41723083412}
\end{pspicture}} \\
    \subfloat[$E_8^{+++}$ decomposed as $A_9$.]{\label{E11-IIA}%
\begin{pspicture}(0,-0.2)(9,1)
\disabledNode{2,1}{N11213619404}
\disabledNode{0,0}{N21213619404}
\normalNode{1,0}{N31213619404}
\normalNode{2,0}{N41213619404}
\normalNode{3,0}{N51213619404}
\normalNode{4,0}{N61213619404}
\normalNode{5,0}{N71213619404}
\normalNode{6,0}{N81213619404}
\normalNode{7,0}{N91213619404}
\normalNode{8,0}{N101213619404}
\normalNode{9,0}{N111213619404}
\singleConnection{N21213619404}{N31213619404}
\singleConnection{N31213619404}{N41213619404}
\singleConnection{N41213619404}{N51213619404}
\singleConnection{N51213619404}{N61213619404}
\singleConnection{N61213619404}{N71213619404}
\singleConnection{N71213619404}{N81213619404}
\singleConnection{N81213619404}{N91213619404}
\singleConnection{N91213619404}{N101213619404}
\singleConnection{N101213619404}{N111213619404}
\singleConnection{N11213619404}{N41213619404}
\end{pspicture}} \\
    \subfloat[$E_8^{+++}$ decomposed as $A_1 \otimes A_9$.]{\label{E11-IIB}%
\begin{pspicture}(0,-0.2)(9,1)
\normalNode{2,1}{N11213619404}
\dualityNode{0,0}{N21213619404}
\disabledNode{1,0}{N31213619404}
\normalNode{2,0}{N41213619404}
\normalNode{3,0}{N51213619404}
\normalNode{4,0}{N61213619404}
\normalNode{5,0}{N71213619404}
\normalNode{6,0}{N81213619404}
\normalNode{7,0}{N91213619404}
\normalNode{8,0}{N101213619404}
\normalNode{9,0}{N111213619404}
\ncline{-}{N21213619404}{N31213619404}
\ncline{-}{N31213619404}{N41213619404}
\ncline{-}{N41213619404}{N51213619404}
\ncline{-}{N51213619404}{N61213619404}
\ncline{-}{N61213619404}{N71213619404}
\ncline{-}{N71213619404}{N81213619404}
\ncline{-}{N81213619404}{N91213619404}
\ncline{-}{N91213619404}{N101213619404}
\ncline{-}{N101213619404}{N111213619404}
\ncline{-}{N11213619404}{N41213619404}
\end{pspicture}}%
    \caption[$E_8^{+++}$ decompositions.]{The Dynkin diagrams of $E_8$ \subref{E8}, the very extended $E_8^{+++}$
		\subref{veE8}, and its decompositions corresponding to 11D
		\subref{E11-11D}, IIA \subref{E11-IIA} and IIB \subref{E11-IIB} supergravity.
		In these decompositions the black nodes are disabled, the white
		nodes correspond to the gravity line $SL(D,\R)$ and the gray node in the last diagram
		corresponds to the duality group $A_1$.}%
    \label{fig:E8_decompositions}%
\end{figure}

Once $G^{+++}/K(G^{+++})$ has been constructed, we are in the
position to oxidize back to $3 \leq D \leq D_{\rm{max}}$ dimensions
using group disintegrations. The valid disintegrations for $G^{+++}$
are always of the type $G_D \otimes SL(D,\R)$, where $G_D$ is the
duality group in $D$ dimensions and $SL(D,\R)$ refers to the space-time
symmetries. Extended Dynkin diagrams are a useful tool to visualize these
group disintegrations: the disintegrations then correspond to
`disabling' certain nodes of the diagram in order to obtain two
disjoint parts, of which one is the $SL(D,\R)$ gravity line and the
other is the $G_D$ duality group. As an example we give the cases of maximal
supergravity in $D=11,10$ in figure \ref{fig:E8_decompositions}.
Note that the duality group $G_D$ contains an extra $\R^+$ factor whenever
there is a second disabled node. This explains why the duality group of
IIA supergravity is $\R^+$ and why those of IIB and $D=11$
supergravity do not have such a factor.

The maximum oxidization dimension is determined by the largest
$SL(D_{\rm{max}},\R)$ chain possible starting from the very extended
node in the (extended) Dynkin diagram of $G^{+++}$
\cite{Cremmer:1999du, Keurentjes:2002xc}. In our conventions these will
always start at the right hand side of the extended Dynkin diagram.
The lower limit on the oxidization dimension stems from the fact that
below $D=3$ the duality group $G_D$ becomes infinite-dimensional, and
there are currently no computer--based tools available to analyze these cases.

After the group disintegration has been fixed, the generators of
$G^{+++}/K(G^{+++})$ can be analyzed by means of a level
decomposition \cite{Damour:2002cu,Nicolai:2003fw}. A level
decomposition comes down to a branching of $G^{+++}$ with respect to the $G_D \otimes SL(D,\R)$ disintegration.
The disabled nodes then induce a grading on $G^{+++}$ which will be
indicated by the so-called levels. When $G^{+++}$ is of real split
form, i.e.~maximally non-compact, modding out by the subgroup
$K(G^{+++})$ implies truncating all the negative levels in the
representation and generically also modding out the scalars at level $0$
by the compact part of the duality group $G_D$.
For clarity we will restrict our discussion to the split forms,
although with some slight modifications everything also holds for the
non-split cases, as follows from \cite{Keurentjes:2002rc}.
Indeed, we have verified for various non-split cases that the
computer calculations give rise to the general results discussed in this
paper.

The spectrum is obtained by associating to each generator a supergravity
field in the same representation. This leads to the
following fields at each level:
At the lowest levels the physical states of the supergravity we
started out with appear together with their duals\footnote{More
precisely: corresponding to any $p$--form generator we also find a
$(D-p-2)$--form. In addition there is a $(D-3,1)$--form with mixed
symmetries and possibly $(D-2)$--form generators, which are
interpreted as the dual graviton and dual scalars, respectively.}.
The duality relations themselves are not
reproduced by the level decomposition: in the absence of dynamics
these relations have to be imposed by hand (for a discussion of dynamics
in the context of $E_{10}$ and $E_{11}$, see e.g.~\cite{Damour:2002cu, Damour:2004zy, Kleinschmidt:2004rg}
and \cite{West:2001as, Schnakenburg:2001ya} respectively). At higher levels
there are the so-called ``dual'' generators, which can be interpreted as
infinitely many exotic dual copies of the previously mentioned
fields \cite{Riccioni:2006az, Riccioni:2007hm}. The remaining
``non-dual'' generators do not correspond to any physical degrees of
freedom. Amongst these are the $(D-1)$-- and $D$--form potentials we
are interested in.

In short, once the relevant $G/K(G)$ coset in three dimensions is
known, all we have to do is consider the different decompositions of
its very extended Dynkin diagram.
The deformation and top--form potentials can then be read off from
the spectrum the computer has calculated.

\section{The $p$--form algebra}\label{sec:pfa}

In this section we will consider the bosonic algebra that the different
$p$--form Kac--Moody generators with $p>0$ satisfy amongst each other.
Subsequently it will be shown how the same algebra
arises in supergravity. In the following two subsections we
will discuss two classes of special generators. Frequently, we
will clarify general features of the algebra by the example
of maximal supergravities. In the next section we will discuss
the case of matter--coupled half--maximal supergravities.

\subsection{Truncation to $p$--forms}

It is convenient to introduce a special algebra, which we call the
``$p$--form algebra''. It can be obtained as a truncation from the very extended
Kac--Moody algebra in a particular $G_D \otimes SL(D,\R)$ decomposition
by deleting all generators except those at positive levels in a purely antisymmetric $SL(D,\R)$
tensor representation of rank\,\footnote{One could also
include the $p=0$ or scalar generators, which are the generators of the
duality group $G_D$.} $1 \leq p \leq D$.
Embedded within the Kac--Moody algebra this is generically not a proper
subalgebra (it is not closed), but on its own it nonetheless is
a Lie algebra. What one ends up with after the truncation is an algebra of generators
represented by components of $p$--forms that satisfy commutation relations of the form
\begin{equation}
[ A_{\mu_1 \cdots \mu_p} , B_{\nu_1 \cdots \nu_q} ] = C_{\mu_1 \cdots \mu_p \nu_1 \cdots \nu_q} \, .
\end{equation}
Suppressing the $SL(D,\R)$ indices, we will write this more concisely as
\begin{equation}
[\pf p , \pf q] = \pf r .
\end{equation}
Here we have introduced the shorthand notation ${\pf p}$,
which will be used throughout this paper\,\footnote{Note that ${\pf p}$
indicates the components of $p$--forms and not $p$--forms themselves. In this
way we avoid the anti--commutators  which are used in
\cite{Lavrinenko:1999xi}.}.
In the above commutator the ranks of the $p$--forms add up: $r = p + q$.
In other words, the rank of the third form is
equal to the sum of the ranks of the first and second forms.

An important property of the $p$--form algebra is the existence of
``fundamental'' $p$-forms whose multiple commutators give rise to the
whole algebra by using the Jacobi identity. These fundamental $p$--forms correspond to the
positive simple roots of the disabled nodes in the Kac--Moody algebra.
From the decomposed Dynkin diagram one can thus deduce
the number and type of these fundamental $p$-forms:
any disabled node connected to the $n$\textsuperscript{th} node of the
gravity line (counting from the very extended node) gives rise to a
$(D-n)$--form. Furthermore, if the disabled node in question is
also connected to a node of the duality group the $(D-n)$--form
carries a non--trivial representation of the duality group.

In the simplest case when there is only one disabled node,
and thus only one fundamental $p$--form,
we can schematically write for each $p$--form
\begin{equation}
\underset{\ell\ \text{times}}{\underbrace{[\pf q,\ldots,[\pf q[\pf q,\pf q]]\ldots]}} = \pf p\,,
\end{equation}
where $q$ is the rank of the fundamental form and $p = \ell q$. The number of times the
commutators are applied corresponds to the level $\ell$ at which the $p$-form
occurs in the level decomposition of the Kac--Moody algebra. By definition
the fundamental generators occur at level one.
This is the structure of e.g.~11D supergravity, which only has a
fundamental three--form $\pf 3$, see figure \ref{E11-11D}. In addition there is a
six-form $\pf 6$, which can be obtained from the $\pf 3$ by the commutation relation
\begin{equation}
[\pf 3,\pf 3] = \pf 6\,.
\end{equation}
According to the definition above the 6--form generator occurs at level $\ell = 2$.
Note that this $p$--form algebra is defined in any dimension $D$. It is
only after we impose dynamics, i.e.~duality relations, that we should restrict
to $D=11$ in order to make contact with $D=11$ supergravity.

Another example might further clarify the above. Consider again the Dynkin
diagram of $E_{11}$ and the embedding of an $SL(10,\R)$ gravity line
that corresponds to IIB supergravity, see figure \ref{E11-IIB}.
We now associate to each generator a supergravity field.
There are two nodes
outside of the white gravity line. One is the grey node not
connected to the gravity line. This node corresponds to the
$SL(2,\R)$ duality symmetry. The other is the black node
attached to the gravity line at the 8th position counted from the
right, and hence corresponds to
a fundamental two-form. Since this node is also
connected to the internal symmetry node the two-form is
in a non--trivial representation of $SL(2,\R)$:
the IIB theory contains a doublet of NS-NS and R-R two-forms.
We denote these two--forms by $\pf 2^{\alpha}$.
Using the same notation for the higher-rank forms we have
the following  $p$--form algebra
\begin{equation}\label{p-form}
\begin{split}
    [ \pf 2^\alpha, \pf 2^\beta ] & = \pf 4 \epsilon^{\alpha \beta} \,, \\
    [ \pf 2^\alpha, \pf 4] & = \pf 6^\alpha \,, \\
    [ \pf 2^\alpha, \pf 6^\beta ] & = \pf 8^{\alpha \beta} \,, \\
    [ \pf 2^\alpha, \pf 8^{\beta \gamma} ] & = \pf{10}^{\alpha \beta \gamma} +
\epsilon^{\alpha ( \beta} \pf{10}^{\gamma)}  \,,
\end{split}
\end{equation}
where all $SL(2,\R)$ representations are symmetric and
$\epsilon^{\alpha \beta}$ is the Levi-Civita tensor. There are other
non-zero commutators but, due to the Jacobi identity,
 they follow from these basic ones involving
the fundamental $\pf 2^\alpha$--form generators. The commutators \eqref{p-form}
specify the level $\ell$ at which each generator occurs. This level
can be read off from these commutators by counting the number of times
the fundamental generators $\pf 2^\alpha$ occur in the multiple commutator
that expresses the generator in terms of the fundamental ones. In this
way we obtain that the generators $\pf 4,\pf 6^\alpha, \pf 8^{\alpha\beta},
\pf{10}^{\alpha\beta\gamma}$ and $ \pf{10}^\alpha$
occur at level $\ell = 2,3,4,5$ and $5$, respectively.

The $p$--form algebra contains generators corresponding
to the following IIB supergravity
fields: a doublet of two--forms, a  singlet
four-form potential, the doublet of six-form potentials that are
dual to the two-forms, a triplet of eight-form potentials that are
dual to the scalars\,\footnote{Supersymmetry will imply a single constraint
on the nine--form field-strengths in order to produce the correct counting
 of physical degrees of freedom dual to the scalars \cite{Dall'Agata:1998va}.}
\cite{Dall'Agata:1998va} and, finally, a doublet and quadruplet of ten-form potentials
\cite{Bergshoeff:2005ac}.

It was shown in \cite{Bergshoeff:2006qw}
that the algebra of bosonic gauge transformations of IIB supergravity can
be brought to precisely the form \eqref{p-form}.  This was achieved after
making a number of redefinitions of the fields and gauge parameters,
as was also done in the ``doubled'' formalism of \cite{Cremmer:1998px}. The
correspondence goes as follows. The $p$--form gauge transformations
of IIB supergravity can be written as \cite{Bergshoeff:2006qw}
\begin{equation}
\begin{split}
    \d A^\a_{(2)}
        & = \Lambda^\a_{(2)} \,, \\
    \d A_{(4)}
        & = \Lambda_{(4)}+\e_{\gamma\d}\Lambda^\gamma_{(2)}A^\d_{(2)} \,,  \\
    \d A^\a_{(6)}
        & = \Lambda^\a_{(6)}
        +\Lambda_{(4)}A^\a_{(2)} -2 \Lambda^\a_{(2)} A_{(4)} \,,  \\
    \d A^{\a\b}_{(8)}
        & = \Lambda^{\a\b}_{(8)}
        +\Lambda^{(\a}_{(6)}\,A^{\b)}_{(2)}  -3 \Lambda^{(\a}_{(2)}A^{\b)}_{(6)} \,,  \\
    \d A^{\a\b\gamma}_{(10)}
        & = \Lambda^{\a\b\gamma}_{(10)} +\Lambda^{(\a \b}_{(8)}\,A^{\gamma)}_{(2)}
        -4  \Lambda^{(\a}_{(2)}A^{\b\gamma)}_{(8)} \,, \\
    \d A^{\a}_{(10)}
        & = \Lambda ^{\alpha}_{(10)}
        + \tfrac 5 {27} \epsilon_{\beta\gamma}\Lambda^{\alpha \beta}_{(8)}\, A^{\gamma}_{(2)}
        - \tfrac {20} {27}\epsilon_{\beta\gamma} \Lambda^{\beta}_{(2)}\, A^{\gamma\alpha}_{(8)}
        + \Lambda_{(4)}\,A_{(6)}^\alpha - \tfrac 2 3 \Lambda_{(6)}^\alpha A_{(4)} \,.
\label{bosgauge2}
\end{split}
\end{equation}
Here we use the notation $\Lambda_{(2n)}\equiv \partial\lambda_{(2n-1)}$,
following \cite{Cremmer:1998px}.
By definition each parameter $\Lambda$ is closed.
In contrast to \cite{Bergshoeff:2006qw} we have redefined
the gauge parameter of the doublet ten--form potential such that
the gauge transformations of this potential precisely agree with the
Kac--Moody algebra or its truncation to $p$--forms. This can always be done for top--form transformations.
Note that the same structure
 also follows from a superspace calculation \cite{Bergshoeff:2007ma}.

In order to compare with the $p$--form algebra we now truncate the bosonic
gauge algebra to a finite--dimensional subalgebra as follows:
\be
\Lambda_{(2n)}\ \text{is\ constant\hskip 2truecm or\hskip 2truecm}
\lambda_{(2n-1)} = x\cdot \Lambda_{(2n)}\,,\label{Lietruncation}
\ee
where it is understood that the spacetime coordinate $x^\mu$ is contracted
with the first index of $\Lambda_{(2n)}$. Note that this indeed is a
consistent truncation due to the fact that the local gauge parameters
$\lambda_{(2n-1)} = \lambda_{(2n-1)}(x)$ always occurs in the transformation rules \eqref{bosgauge2}
with a derivative acting on it. Furthermore, we could have included a constant part in
$\lambda_{(2n-1)}$, but this drops out of the gauge transformations for the same reason.

The commutator algebra corresponding to \eqref{bosgauge2}, for constant
$\Lambda$, is
now precisely of the form \eqref{p-form} provided we associate to
each $p$--form in \eqref{p-form} the gauge transformation,
with parameter $\Lambda_{(p)}$, of a $p$--form potential in \eqref{bosgauge2}.
The $p$--form algebra
arises as a Lie algebra truncation, defined by \eqref{Lietruncation},
of the bosonic gauge algebra\,\footnote{
The relation between the bosonic gauge algebra and the
$p$--form algebra, in the sense that the latter is a Lie algebra
truncation of the former, can also be introduced for
diffeomorphisms. Restricting the general coordinate
transformations to $x \rightarrow \Lambda x$, where $\Lambda$ is a
constant $GL(D)$ matrix, these span a Lie algebra.
}\,, see figure 2.
The
$p$--form gauge field transformations \eqref{bosgauge2} can now be viewed as
a nonlinear realisation
 of the $p$--form algebra \eqref{p-form}\,\footnote{
Note that the gauge transformations \eqref{bosgauge2} contain no terms
that are higher order in the potentials, as would occur in a generic non--linear
representation. It turns out that all higher order terms (but not the $0^{\rm th}$
order one) can be eliminated by making appropriate field redefinitions.}.
Note that the $p$--form gauge fields not only transform under
their own gauge transformations but also under those of the other gauge
fields. Consequently, the curvatures of these gauge fields will contain
Chern-Simons like terms.

\FIGURE[t]{
    \psset{unit=2cm}
    \begin{pspicture}(0,0)(6,2)
        \rput[l](0,2){\rnode{km}{\psshadowbox{Kac--Moody algebra}}}
        \rput(3,0){\rnode{pf}{\psshadowbox{$p$--form algebra}}}
        \rput[r](6,2){\rnode{bs}{\psshadowbox{Bosonic gauge algebra}}}
        \ncline[nodesep=6pt]{->}{km}{pf}
        \ncput*{$\begin{array}{c} (1 \leq p \leq D)\textrm{--form} \\ \textrm{truncation} \end{array}$}
        \ncline[nodesep=6pt]{->}{bs}{pf}
        \ncput*{$\begin{array}{c} \textrm{Constant gauge} \\
\textrm{parameters $\Lambda=\partial\lambda$} \end{array}$}
    \end{pspicture}
    \vspace{12pt}
    \caption{
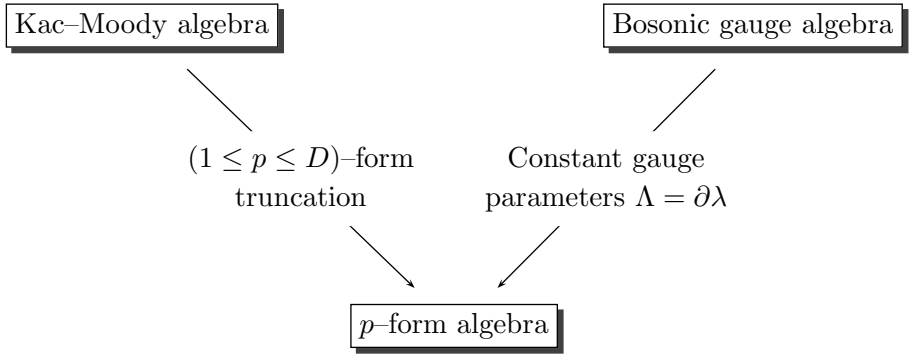
        The $p$-form algebra as the respective limits of the Kac--Moody
        and the bosonic gauge algebra.
    }
}

We would like to stress that the truncation of the bosonic gauge algebra
to a $p$--form Lie algebra
is only possible in a particular basis of the supergravity theory.
In particular, the gauge transformations need to be
expressed only in terms of the gauge potentials and not their
derivatives.
It will not always be possible to bring the gauge transformations
of any supergravity theory to such a form.
An example of this will be discussed in section \ref{sec:def} in
the context of gauged and massive supergravities.

The above reasoning can be applied to any very extended Kac--Moody
algebra. It provides a useful truncation to the part of the spectrum
that contains all $p$-form potentials, including the deformation and
top-form potentials, which will be discussed next.

\subsection{Deformation potentials}

We now wish to discuss  some properties of the deformation and top-form
potentials in the light of the $p$--form algebra introduced above.
We first consider the deformation potentials.
It has been argued in, e.g.,~\cite{Polchinski:1995mt, Bergshoeff:1996ui,
Riccioni:2007au, Bergshoeff:2007qi} that deformation potentials are in
one-to-one correspondence with deformations of the supergravity
theory, such as gaugings or massive deformations. Indeed, the
$D$--form curvatures of the
$(D-1)$--form potentials can be seen as the duals of the deformation
parameters: in the presence of a deformation, one can only realize
the supersymmetry algebra on a $(D-1)$--form potential provided its
field strength is the Hodge dual of the deformation parameter.
As far as the deformation parameters are concerned one can
distinguish between
gauged and massive deformations, as we will discuss below.

The most familiar class of deformed supergravities are the so-called
``gauged'' supergravities. They are special in the sense that the
deformations can be seen as the result of gauging a subgroup $H$ of
the duality group $G$. Not all deformed supergravities can be viewed
as gauged supergravities. In the case of maximal supergravity there is
one exception: massive IIA supergravity cannot be obtained by
gauging the $\R^+$ duality group \cite{Romans:1985tz}.
The gauged supergravities can be seen as the first in a series of ``type
$p$ deformations''. There is a a simple criterion
that defines
 to which type of deformation parameter  each deformation
potential gives rise to.
The central observation is that to each
$(D-1)$--form one can associate a unique commutator
\begin{equation}
    [ \pf p , \pf{(D-p-1)} ] = \pf{(D-1)} \,,
    \label{deform-commutator}
\end{equation}
where $\pf p$ corresponds to a fundamental $p$-form
and where we have suppressed the representation of the duality group.
The deformation potential corresponding to such a deformation
generator gives rise to a type $p$ deformation parameter.

We observe that each type $p$ deformation is characterized by the
fact that a fundamental $p$--form gauge field becomes massive.
For $p=1$ this leads to gauged supergravities, in which a vector can
become massive by absorbing a scalar degree of freedom\footnote{In the $p=0$ case there is only a massive
vector when an isometry of the scalar manifold is being gauged.}.
Note that other non-fundamental gauge fields may become massive as well.
The case $p=2$ entails a fundamental two--form that becomes massive by ``eating'' a vector.
The prime example of this is massive IIA supergravity in ten
dimensions \cite{Romans:1985tz}. Another example is the non-chiral
half--maximal supergravity in six dimensions \cite{Romans:1985tw}.
An example of a $p=3$ deformation is the half--maximal
supergravity theory of \cite{Townsend:1983kk} where a fundamental
three-form potential acquires a topological mass term. Due to the
restricted number of dimensions it can be seen that there are no $p
\geq 4$ deformations of supergravity theories.

\TABLE[h]{
\begin{tabular}{|c|c|}
    \hline
    \multirow{2}{*}{$D$}    & Fundamental  \\
                        & $p$--forms \\
    \hline
    \hline
    11  & $\pf 3$ \\
    IIA & $\pf 1,\pf 2$ \\
    IIB & $\pf 2$\\
    3--9& $\pf 1$ \\
    \hline
\end{tabular}
\caption{
    Fundamental $p$--forms for maximal supergravity
    \cite{West:2001as,Schnakenburg:2001ya,Riccioni:2007au,Bergshoeff:2007qi},
    where we have suppressed the duality group indices and multiplicities.}
\label{table:maximal_fund}
}

It is interesting to apply these general observations to the case
of maximal supergravity. In that case all deformations
are gauge deformations except massive IIA supergravity.
This can be easily understood from the
Kac--Moody approach. In $D\le 9$ all fundamental $p$--forms are
vectors (see table \ref{table:maximal_fund}) and thus one can never realize the commutation relation
\eqref{deform-commutator} for $p\ne 1$. Only in $D=10$ we have
a fundamental 2--form making massive supergravity
possible. In $D=10$ a deformation potential is a 9--form and
eq.~\eqref{deform-commutator} becomes $[\pf 7,\pf 2]=\pf 9$. Instead we have
that $[\pf 8,\pf 1]$ vanishes. Note that IIA supergravity allows a massive deformation
but IIB supergravity does not. The reason for this,
from the Kac--Moody point of view, is that in writing the commutator
$[\pf 7,\pf 2]=\pf 9$ it is understood that there is either a fundamental 7--form,
which is not the case, or this representation can be written as
a multiple commutator of fundamental $p$--forms. The latter is only
possible if there is at least one fundamental $p$--form with an odd
number of indices. This condition is only satisfied in the case of
IIA supergravity.

Note that in $D=11$ we have a single fundamental 3--form,
but in $D=11$ a deformation potential is a 10--form which cannot be
written as a multiple commutator of 3--forms. Therefore, there
is no massive deformation in $D=11$.

\subsection{Top--form potentials}

Finally, we consider  the top--form potentials and point out
an intriguing relation with the deformation potentials and parameters.
Given a supergravity theory with deformation parameters in different
representations of the duality group, it is not obvious that these
deformation parameters can be turned on all at the same time. In
fact, in the case of gauged supergravities it is known that the
deformation parameters satisfy certain quadratic constraints
\cite{deWit:2002vt, deWit:2003hr}.

To illustrate the need for
quadratic constraints on the deformation parameters and observe the relation
with the top--form potentials
it is instructive to consider $D=9$ maximal
gauged supergravity with duality group $\mathbb{R}^+ \times SL(2,\R)$
\cite{Bergshoeff:2002nv}. There
is a triplet $m_{\alpha \beta}$ of deformations, corresponding to
the gauging of a one-dimensional subgroup of
$SL(2,\R)$, and a doublet $m_\alpha$ of deformations,
corresponding to an $\mathbb{R}^+$ gauging:
\begin{subequations}
\begin{align}
 {\bf 3} & \;\;\; \Leftrightarrow \;\;\; m_{\alpha \beta}\,: \;\;
 &  \left. \begin{matrix} SO(2) \\ SO(1,1) \\ \mathbb{R}^+ \end{matrix} \right\} & \in SL(2,\R) &
 &\text{- IIB (and mIIA) origin.} \\
&&&&& \nonumber\\
 {\bf 2} & \;\;\; \Leftrightarrow \;\;\; m_\alpha\,:
 & &\mathbb{R}^+ &
 & \text{- IIA origin.}
\end{align}
\end{subequations}
All components of the triplet and of the doublet can be obtained via
generalized Scherk--Schwarz reductions of IIB and of massless IIA
supergravity, respectively.
In addition, one component of the triplet, corresponding to the
$\mathbb{R}^+ \in SL(2,\R)$ gauging, can be obtained via a Kaluza-Klein
reduction of the massive IIA theory. Note that it
is impossible to perform a generalized Scherk--Schwarz reduction in the
massive case, since the mass parameter breaks the relevant scale symmetry.

Due to the different origins of the triplet and the doublet it is impossible to obtain them
simultaneously from ten dimensions. However, one might wonder whether they
can be turned on at the same time, independent
of any higher-dimensional origin. This question has been answered in
the negative  \cite{Bergshoeff:2002nv}, which can be summarized by
imposing the following quadratic constraint:
\begin{align}
&\text{quadratic constraint:} & m_{\alpha\beta}\, m_\gamma & = 0\;\;\; \Leftrightarrow \;\;\;
{\bf 3}\  \times  \ {\bf 2} \ =\ {\bf 4}\  +\ {\bf 2}\,. &
 \label{9D-qc}
\end{align}
These constraints occur in the $\bf 4$ and $\bf 2$
representation which are in one-to-one correspondence with two of the
three representations of top--forms, as can be seen in \cite{Riccioni:2007au, Bergshoeff:2007qi}.
So for each constraint there is a corresponding top--form potential.

Also in lower dimensions, both for maximal supergravity in $3 \leq D \leq 7$
\cite{Nicolai:2001sv, deWit:2002vt, deWit:2003hr} and half--maximal
supergravity in $D=3,4,5$ \cite{deWit:2003ja, Schon:2006kz}, the quadratic constraints
have been calculated with the embedding
tensor approach. In each case we observe that there is an exact one-to-one
correspondence between the quadratic constraints and the top--form
potentials in terms of representations of the duality group.

In the embedding tensor
approach this correspondence can be explained as follows.
Starting with a gauged supergravity Lagrangian ${\cal L}_{\text{sugra}}$
one can always replace the constant
embedding tensor by a scalar field that
is constant only on--shell due to the field equation of a
deformation potential $A_{(D-1)}$, see also
e.g.~\cite{Bergshoeff:1996ui, Lavrinenko:1999xi}.
Furthermore, for each
quadratic constraint one introduces a top--form  Lagrange multiplier $A_{(D)}$ in the
same representation to enforce the constraint\,\footnote{
We thank Henning Samtleben for pointing this out to us. See also the recent paper \cite{deWit:2008ta}.}.
The total Lagrangian thus becomes (suppressing duality group indices)
\be\label{lm}
{\cal L}
    = {\cal L}_{\text{sugra}}
    + A_{(D-1)}\partial\Theta
    + A_{(D)}\Theta\Theta\, ,
\ee
where $\Theta$ is the embedding tensor.
One might wonder how a gauge field can act as a Lagrange multiplier.
It turns out that the gauge transformation of the top--form
in \eqref{lm} is cancelled
by adding an extra term to the gauge transformation of the deformation
potential in the following way
\be
\begin{split}
\delta A_{(D-1)} &= \partial \Lambda_{(D-2)} + \Lambda_{(D-1)}\Theta\,, \\
\delta A_{(D)} & = \partial\Lambda_{(D-1)}\,.
\end{split}
\ee
The field equation for the embedding tensor field leads to a duality
relation of the form \eqref{emb-ten}.
This provides a concrete way of explaining why the deformation potentials
and the embedding tensor must be in the same representation of the duality
group, and similarly for the top--form potentials and the quadratic
constraints.

We conclude that one can divide the top--form potentials into two
classes: the first class consists of all top--forms that are
Lagrange multipliers enforcing quadratic constraints on the
deformation parameters, while the second class contains all the other
independent top--forms whose role is unclear from the present point of view.
Examples of supergravity theories with independent top--form
potentials are the half--maximal chiral supergravity theory in six
dimensions and IIB and $D=9$ maximal supergravity. The first theory does not contain
deformation potentials and hence no quadratic constraints. The same
applies to IIB which contains an independent
quadruplet of potentials that is related to the D9--brane and an
independent doublet of top--forms that sofar has no brane
interpretation. Finally, in $D=9$ maximal supergravity there is another
top--form representation, in addition to \eqref{9D-qc}, that does not
correspond to a quadratic constraint.

\section{Matter coupled half--maximal supergravity} \label{sec:KM-hm}

We now proceed with the case of matter coupled half--maximal supergravity.
In subsection 4.1 we first investigate the structure of the Kac-Moody and $p$--form algebras.
In the next two subsections we discuss the deformation and
top--form potentials that are contained in it.

\subsection{Kac--Moody and $p$--form algebras}

Half--maximal supergravities, coupled to $D-10+n$
vector multiplets, reduce to the scalar coset $SO(8,8+n) / SO(8) \times SO(8+n)$ when
reduced to three dimensions.
In other words, the relevant groups for
supergravity theories with 16 supercharges are the $B$ and $D$
series in the above real form. Of these, only three are of split
real form, i.e.~maximally non-compact, which are given by
$n=-1,0,+1$. These correspond to the split forms of $B_7$, $D_8$ and
$B_8$, respectively.

We are interested in the decomposition of the very extensions
of these algebras with respect to the possible gravity lines.
An exhaustive list of the possibilities for the algebras of real split form is given in table
\ref{B7dec}. As can be seen from this table, these correspond to the unique $D$-dimensional
supergravity theory with 16 supercharges coupled to $m+n$ vector multiplets with $m=10-D$.
The corresponding duality groups $G_D$ in $D$ dimensions are also given in table \ref{B7dec}.
Note that there is no second disabled node and therefore no $\R^+$ factor
in the duality group for the 6b case and in $D=3,4$.

In appendix \ref{app:results} the result of the decomposition of the $D_8^{+++}$ algebras
with respect to the different $SL(D,\R)$ subalgebras is given. In addition,
the decompositions of the other two split forms $B_{7,8}^{+++}$ can be
found on the website of SimpLie \cite{SimpLie}. It can be seen that these
decompositions give rise to exactly the physical degrees of freedom
\cite{Kleinschmidt:2003mf}. In addition there are the deformation and
top-form potentials in the Kac--Moody spectrum. In particular, table
\ref{halfmax} summarises our results for the deformation
and top--form potentials for half--maximal supergravity in $D$ dimensions.

To discuss the $p$--form algebra it is easiest to start with
$8 \leq D \leq 10$ dimensions where there is a unified result valid
in any dimension $D=8,9,10$. We will refer to this as the
``generic'' situation. In lower dimensions this generic pattern remains
but there are extra generators specific for each dimension $D<8$, see
 table \ref{halfmax}. We will not discuss all the details of the lower
 dimensions but its should be clear that they follow the same pattern
as the higher dimensions except that the expressions involved are a bit messier.

In $8 \leq D \leq 10$ dimensions
the $p$--form algebra is
given by the following generators and commutation relations.
As can be seen from table \ref{B7dec}, except in a few special
(lower--dimensional) cases to be discussed below, in each dimension
the fundamental $p$--forms
of the algebra are a $\pf 1^M$ and a $\pf{(D-4)}$. The former is in the
fundamental representation of the duality group $G_D$ while the latter
is a singlet. The other generators follow from the following basic
commutators describing the $p$--form algebra:
\begin{equation}
\begin{split}
    [\pf 1^M, \pf 1^N ]         & = \eta^{MN} \pf 2 \,, \\
    [\pf 1^M, \pf{(D-4)}]        & = \pf{(D-3)}^M \,, \\
    [\pf 1^M, \pf{(D-3)}^N]      & = \pf{(D-2)}^{[MN]} + \eta^{MN} \pf{(D-2)}  \,, \\
    [\pf 1^M, \pf{(D-2)}^{[NP]}] & = \eta^{M [N} \pf{(D-1)}^{P]} + \pf{(D-1)}^{[MNP]} \,, \\
    [\pf 1^M, \pf{(D-2)}]        & = \pf{(D-1)}^M\,,\\
    [\pf 1^M, \pf{(D-1)}^N]      & = \eta^{MN} \pf D + \pf D^{[MN]} \,, \\
    [\pf 1^M, \pf{(D-1)}^{[NPQ]}]& = \eta^{M [N} \pf D^{PQ]} + \pf D^{[MNPQ]} \,,
\end{split}
 \label{gen-alg}
\end{equation}
where $\eta^{MN}$ is the $SO(m,m+n)$ invariant metric
and the straight brackets
indicate anti--symmetrization. In addition $[\pf 1^M, \pf 2]$ vanishes. From these commutators we read off the
levels $(\ell_1, \ell_2)$ of the different generators. Here $\ell_1, \ell_2$
is the number of times the fundamental $\pf{(D-4)}, \pf 1^M$
generators occur in the
multiple commutators expressing  the generator in terms

\begin{landscape}
\TABLE[p]{
    \begin{tabular}{|c|c|c|c|c|c|}
        \hline
        $D$ & $G_D$ & Multiplets & $B_7^{+++} \; \; (n=-1)$ & $D_8^{+++} \; \; (n=0)$ & $B_8^{+++} \; \; (n=1)$ \\
        \hline
        \hline
        &&&&& \\

        10  & $\R^+ \times SO(n)$ & $G V^{n}$
            & $-$
            & \begin{pspicture}(0,0)(8,1)
\Cnode[fillstyle=solid,fillcolor=black](1,1){N11784499420}
\Cnode[fillstyle=solid,fillcolor=black](5,1){N21784499420}
\Cnode(0,0){N31784499420}
\Cnode(1,0){N41784499420}
\Cnode(2,0){N51784499420}
\Cnode(3,0){N61784499420}
\Cnode(4,0){N71784499420}
\Cnode(5,0){N81784499420}
\Cnode(6,0){N91784499420}
\Cnode(7,0){N101784499420}
\Cnode(8,0){N111784499420}
\ncline{-}{N31784499420}{N41784499420}
\ncline{-}{N41784499420}{N51784499420}
\ncline{-}{N51784499420}{N61784499420}
\ncline{-}{N61784499420}{N71784499420}
\ncline{-}{N71784499420}{N81784499420}
\ncline{-}{N81784499420}{N91784499420}
\ncline{-}{N91784499420}{N101784499420}
\ncline{-}{N101784499420}{N111784499420}
\ncline{-}{N81784499420}{N21784499420}
\ncline{-}{N41784499420}{N11784499420}
\end{pspicture} ~
            & \begin{pspicture}(0,0)(8,1)
\Cnode[fillstyle=solid,fillcolor=black](0,1){N1330860211}
\Cnode[fillstyle=solid,fillcolor=black](5,1){N2330860211}
\Cnode(0,0){N3330860211}
\Cnode(1,0){N4330860211}
\Cnode(2,0){N5330860211}
\Cnode(3,0){N6330860211}
\Cnode(4,0){N7330860211}
\Cnode(5,0){N8330860211}
\Cnode(6,0){N9330860211}
\Cnode(7,0){N10330860211}
\Cnode(8,0){N11330860211}
\ncline{-}{N3330860211}{N4330860211}
\ncline{-}{N4330860211}{N5330860211}
\ncline{-}{N5330860211}{N6330860211}
\ncline{-}{N6330860211}{N7330860211}
\ncline{-}{N7330860211}{N8330860211}
\ncline{-}{N8330860211}{N9330860211}
\ncline{-}{N9330860211}{N10330860211}
\ncline{-}{N10330860211}{N11330860211}
\ncline{-}{N8330860211}{N2330860211}
\ncline[doubleline=true,doublesep=0.2,arrowsize=0.6,arrowlength=0.25,arrowinset=0.6]{->}{N3330860211}{N1330860211}
\end{pspicture} ~ \\ [20pt]

        9   &  $\R^+ \times SO(1,1+n)$ & $G V^{n+1}$
            & \begin{pspicture}(0,0)(7,1)
\Cnode[fillstyle=solid,fillcolor=black](0,1){N1950701429}
\Cnode[fillstyle=solid,fillcolor=black](4,1){N2950701429}
\Cnode(0,0){N3950701429}
\Cnode(1,0){N4950701429}
\Cnode(2,0){N5950701429}
\Cnode(3,0){N6950701429}
\Cnode(4,0){N7950701429}
\Cnode(5,0){N8950701429}
\Cnode(6,0){N9950701429}
\Cnode(7,0){N10950701429}
\ncline{-}{N3950701429}{N4950701429}
\ncline{-}{N4950701429}{N5950701429}
\ncline{-}{N5950701429}{N6950701429}
\ncline{-}{N6950701429}{N7950701429}
\ncline{-}{N7950701429}{N8950701429}
\ncline{-}{N8950701429}{N9950701429}
\ncline{-}{N9950701429}{N10950701429}
\ncline[doubleline=true,doublesep=0.2,arrowsize=0.6,arrowlength=0.25,arrowinset=0.6]{->}{N3950701429}{N1950701429}
\ncline{-}{N7950701429}{N2950701429}
\end{pspicture} ~
            & \begin{pspicture}(0,0)(8,1)
\Cnode[fillstyle=solid,fillcolor=black](1,1){N12116776400}
\Cnode[fillstyle=solid,fillcolor=black](5,1){N22116776400}
\Cnode[fillstyle=solid,fillcolor=black](0,0){N32116776400}
\Cnode(1,0){N42116776400}
\Cnode(2,0){N52116776400}
\Cnode(3,0){N62116776400}
\Cnode(4,0){N72116776400}
\Cnode(5,0){N82116776400}
\Cnode(6,0){N92116776400}
\Cnode(7,0){N102116776400}
\Cnode(8,0){N112116776400}
\ncline{-}{N32116776400}{N42116776400}
\ncline{-}{N42116776400}{N52116776400}
\ncline{-}{N52116776400}{N62116776400}
\ncline{-}{N62116776400}{N72116776400}
\ncline{-}{N72116776400}{N82116776400}
\ncline{-}{N82116776400}{N92116776400}
\ncline{-}{N92116776400}{N102116776400}
\ncline{-}{N102116776400}{N112116776400}
\ncline{-}{N82116776400}{N22116776400}
\ncline{-}{N42116776400}{N12116776400}
\end{pspicture} ~
            & \begin{pspicture}(0,0)(8,1)
\Cnode[fillstyle=solid,fillcolor=lightgray](0,1){N1237832825}
\Cnode[fillstyle=solid,fillcolor=black](5,1){N2237832825}
\Cnode[fillstyle=solid,fillcolor=black](0,0){N3237832825}
\Cnode(1,0){N4237832825}
\Cnode(2,0){N5237832825}
\Cnode(3,0){N6237832825}
\Cnode(4,0){N7237832825}
\Cnode(5,0){N8237832825}
\Cnode(6,0){N9237832825}
\Cnode(7,0){N10237832825}
\Cnode(8,0){N11237832825}
\ncline{-}{N3237832825}{N4237832825}
\ncline{-}{N4237832825}{N5237832825}
\ncline{-}{N5237832825}{N6237832825}
\ncline{-}{N6237832825}{N7237832825}
\ncline{-}{N7237832825}{N8237832825}
\ncline{-}{N8237832825}{N9237832825}
\ncline{-}{N9237832825}{N10237832825}
\ncline{-}{N10237832825}{N11237832825}
\ncline{-}{N8237832825}{N2237832825}
\ncline[doubleline=true,doublesep=0.2,arrowsize=0.6,arrowlength=0.25,arrowinset=0.6]{->}{N3237832825}{N1237832825}
\end{pspicture} ~ \\ [20pt]

        8   &  $\R^+ \times SO(2,2+n)$ & $G V^{n+2}$
            & \begin{pspicture}(0,0)(7,1)
\Cnode[fillstyle=solid,fillcolor=lightgray](0,1){N1299352491}
\Cnode[fillstyle=solid,fillcolor=black](4,1){N2299352491}
\Cnode[fillstyle=solid,fillcolor=black](0,0){N3299352491}
\Cnode(1,0){N4299352491}
\Cnode(2,0){N5299352491}
\Cnode(3,0){N6299352491}
\Cnode(4,0){N7299352491}
\Cnode(5,0){N8299352491}
\Cnode(6,0){N9299352491}
\Cnode(7,0){N10299352491}
\ncline{-}{N3299352491}{N4299352491}
\ncline{-}{N4299352491}{N5299352491}
\ncline{-}{N5299352491}{N6299352491}
\ncline{-}{N6299352491}{N7299352491}
\ncline{-}{N7299352491}{N8299352491}
\ncline{-}{N8299352491}{N9299352491}
\ncline{-}{N9299352491}{N10299352491}
\ncline[doubleline=true,doublesep=0.2,arrowsize=0.6,arrowlength=0.25,arrowinset=0.6]{->}{N3299352491}{N1299352491}
\ncline{-}{N7299352491}{N2299352491}
\end{pspicture} ~
            & \begin{pspicture}(0,0)(8,1)
\Cnode[fillstyle=solid,fillcolor=lightgray](1,1){N1817052144}
\Cnode[fillstyle=solid,fillcolor=black](5,1){N2817052144}
\Cnode[fillstyle=solid,fillcolor=lightgray](0,0){N3817052144}
\Cnode[fillstyle=solid,fillcolor=black](1,0){N4817052144}
\Cnode(2,0){N5817052144}
\Cnode(3,0){N6817052144}
\Cnode(4,0){N7817052144}
\Cnode(5,0){N8817052144}
\Cnode(6,0){N9817052144}
\Cnode(7,0){N10817052144}
\Cnode(8,0){N11817052144}
\ncline{-}{N3817052144}{N4817052144}
\ncline{-}{N4817052144}{N5817052144}
\ncline{-}{N5817052144}{N6817052144}
\ncline{-}{N6817052144}{N7817052144}
\ncline{-}{N7817052144}{N8817052144}
\ncline{-}{N8817052144}{N9817052144}
\ncline{-}{N9817052144}{N10817052144}
\ncline{-}{N10817052144}{N11817052144}
\ncline{-}{N8817052144}{N2817052144}
\ncline{-}{N4817052144}{N1817052144}
\end{pspicture} ~
            & \begin{pspicture}(0,0)(8,1)
\Cnode[fillstyle=solid,fillcolor=lightgray](0,1){N1296821337}
\Cnode[fillstyle=solid,fillcolor=black](5,1){N2296821337}
\Cnode[fillstyle=solid,fillcolor=lightgray](0,0){N3296821337}
\Cnode[fillstyle=solid,fillcolor=black](1,0){N4296821337}
\Cnode(2,0){N5296821337}
\Cnode(3,0){N6296821337}
\Cnode(4,0){N7296821337}
\Cnode(5,0){N8296821337}
\Cnode(6,0){N9296821337}
\Cnode(7,0){N10296821337}
\Cnode(8,0){N11296821337}
\ncline{-}{N3296821337}{N4296821337}
\ncline{-}{N4296821337}{N5296821337}
\ncline{-}{N5296821337}{N6296821337}
\ncline{-}{N6296821337}{N7296821337}
\ncline{-}{N7296821337}{N8296821337}
\ncline{-}{N8296821337}{N9296821337}
\ncline{-}{N9296821337}{N10296821337}
\ncline{-}{N10296821337}{N11296821337}
\ncline{-}{N8296821337}{N2296821337}
\ncline[doubleline=true,doublesep=0.2,arrowsize=0.6,arrowlength=0.25,arrowinset=0.6]{->}{N3296821337}{N1296821337}
\end{pspicture} ~ \\[20pt]

        7   &  $\R^+ \times SO(3,3+n)$ & $G V^{n+3}$
            & \begin{pspicture}(0,0)(7,1)
\Cnode[fillstyle=solid,fillcolor=lightgray](0,1){N11658816833}
\Cnode[fillstyle=solid,fillcolor=black](4,1){N21658816833}
\Cnode[fillstyle=solid,fillcolor=lightgray](0,0){N31658816833}
\Cnode[fillstyle=solid,fillcolor=black](1,0){N41658816833}
\Cnode(2,0){N51658816833}
\Cnode(3,0){N61658816833}
\Cnode(4,0){N71658816833}
\Cnode(5,0){N81658816833}
\Cnode(6,0){N91658816833}
\Cnode(7,0){N101658816833}
\ncline{-}{N31658816833}{N41658816833}
\ncline{-}{N41658816833}{N51658816833}
\ncline{-}{N51658816833}{N61658816833}
\ncline{-}{N61658816833}{N71658816833}
\ncline{-}{N71658816833}{N81658816833}
\ncline{-}{N81658816833}{N91658816833}
\ncline{-}{N91658816833}{N101658816833}
\ncline[doubleline=true,doublesep=0.2,arrowsize=0.6,arrowlength=0.25,arrowinset=0.6]{->}{N31658816833}{N11658816833}
\ncline{-}{N71658816833}{N21658816833}
\end{pspicture} ~
            & \begin{pspicture}(0,0)(8,1)
\Cnode[fillstyle=solid,fillcolor=lightgray](1,1){N1909476900}
\Cnode[fillstyle=solid,fillcolor=black](5,1){N2909476900}
\Cnode[fillstyle=solid,fillcolor=lightgray](0,0){N3909476900}
\Cnode[fillstyle=solid,fillcolor=lightgray](1,0){N4909476900}
\Cnode[fillstyle=solid,fillcolor=black](2,0){N5909476900}
\Cnode(3,0){N6909476900}
\Cnode(4,0){N7909476900}
\Cnode(5,0){N8909476900}
\Cnode(6,0){N9909476900}
\Cnode(7,0){N10909476900}
\Cnode(8,0){N11909476900}
\ncline{-}{N3909476900}{N4909476900}
\ncline{-}{N4909476900}{N5909476900}
\ncline{-}{N5909476900}{N6909476900}
\ncline{-}{N6909476900}{N7909476900}
\ncline{-}{N7909476900}{N8909476900}
\ncline{-}{N8909476900}{N9909476900}
\ncline{-}{N9909476900}{N10909476900}
\ncline{-}{N10909476900}{N11909476900}
\ncline{-}{N8909476900}{N2909476900}
\ncline{-}{N4909476900}{N1909476900}
\end{pspicture} ~
            & \begin{pspicture}(0,0)(8,1)
\Cnode[fillstyle=solid,fillcolor=lightgray](0,1){N11786757445}
\Cnode[fillstyle=solid,fillcolor=black](5,1){N21786757445}
\Cnode[fillstyle=solid,fillcolor=lightgray](0,0){N31786757445}
\Cnode[fillstyle=solid,fillcolor=lightgray](1,0){N41786757445}
\Cnode[fillstyle=solid,fillcolor=black](2,0){N51786757445}
\Cnode(3,0){N61786757445}
\Cnode(4,0){N71786757445}
\Cnode(5,0){N81786757445}
\Cnode(6,0){N91786757445}
\Cnode(7,0){N101786757445}
\Cnode(8,0){N111786757445}
\ncline{-}{N31786757445}{N41786757445}
\ncline{-}{N41786757445}{N51786757445}
\ncline{-}{N51786757445}{N61786757445}
\ncline{-}{N61786757445}{N71786757445}
\ncline{-}{N71786757445}{N81786757445}
\ncline{-}{N81786757445}{N91786757445}
\ncline{-}{N91786757445}{N101786757445}
\ncline{-}{N101786757445}{N111786757445}
\ncline{-}{N81786757445}{N21786757445}
\ncline[doubleline=true,doublesep=0.2,arrowsize=0.6,arrowlength=0.25,arrowinset=0.6]{->}{N31786757445}{N11786757445}
\end{pspicture} ~ \\ [20pt]

         6a  &  $\R^+ \times SO(4,4+n)$ & $G V^{n+4}$
            & \begin{pspicture}(0,0)(7,1)
\Cnode[fillstyle=solid,fillcolor=lightgray](0,1){N11531843743}
\Cnode[fillstyle=solid,fillcolor=black](4,1){N21531843743}
\Cnode[fillstyle=solid,fillcolor=lightgray](0,0){N31531843743}
\Cnode[fillstyle=solid,fillcolor=lightgray](1,0){N41531843743}
\Cnode[fillstyle=solid,fillcolor=black](2,0){N51531843743}
\Cnode(3,0){N61531843743}
\Cnode(4,0){N71531843743}
\Cnode(5,0){N81531843743}
\Cnode(6,0){N91531843743}
\Cnode(7,0){N101531843743}
\ncline{-}{N31531843743}{N41531843743}
\ncline{-}{N41531843743}{N51531843743}
\ncline{-}{N51531843743}{N61531843743}
\ncline{-}{N61531843743}{N71531843743}
\ncline{-}{N71531843743}{N81531843743}
\ncline{-}{N81531843743}{N91531843743}
\ncline{-}{N91531843743}{N101531843743}
\ncline[doubleline=true,doublesep=0.2,arrowsize=0.6,arrowlength=0.25,arrowinset=0.6]{->}{N31531843743}{N11531843743}
\ncline{-}{N71531843743}{N21531843743}
\end{pspicture} ~
            & \begin{pspicture}(0,0)(8,1)
\Cnode[fillstyle=solid,fillcolor=lightgray](1,1){N11980953788}
\Cnode[fillstyle=solid,fillcolor=black](5,1){N21980953788}
\Cnode[fillstyle=solid,fillcolor=lightgray](0,0){N31980953788}
\Cnode[fillstyle=solid,fillcolor=lightgray](1,0){N41980953788}
\Cnode[fillstyle=solid,fillcolor=lightgray](2,0){N51980953788}
\Cnode[fillstyle=solid,fillcolor=black](3,0){N61980953788}
\Cnode(4,0){N71980953788}
\Cnode(5,0){N81980953788}
\Cnode(6,0){N91980953788}
\Cnode(7,0){N101980953788}
\Cnode(8,0){N111980953788}
\ncline{-}{N31980953788}{N41980953788}
\ncline{-}{N41980953788}{N51980953788}
\ncline{-}{N51980953788}{N61980953788}
\ncline{-}{N61980953788}{N71980953788}
\ncline{-}{N71980953788}{N81980953788}
\ncline{-}{N81980953788}{N91980953788}
\ncline{-}{N91980953788}{N101980953788}
\ncline{-}{N101980953788}{N111980953788}
\ncline{-}{N81980953788}{N21980953788}
\ncline{-}{N41980953788}{N11980953788}
\end{pspicture} ~
            & \begin{pspicture}(0,0)(8,1)
\Cnode[fillstyle=solid,fillcolor=lightgray](0,1){N11963811941}
\Cnode[fillstyle=solid,fillcolor=black](5,1){N21963811941}
\Cnode[fillstyle=solid,fillcolor=lightgray](0,0){N31963811941}
\Cnode[fillstyle=solid,fillcolor=lightgray](1,0){N41963811941}
\Cnode[fillstyle=solid,fillcolor=lightgray](2,0){N51963811941}
\Cnode[fillstyle=solid,fillcolor=black](3,0){N61963811941}
\Cnode(4,0){N71963811941}
\Cnode(5,0){N81963811941}
\Cnode(6,0){N91963811941}
\Cnode(7,0){N101963811941}
\Cnode(8,0){N111963811941}
\ncline{-}{N31963811941}{N41963811941}
\ncline{-}{N41963811941}{N51963811941}
\ncline{-}{N51963811941}{N61963811941}
\ncline{-}{N61963811941}{N71963811941}
\ncline{-}{N71963811941}{N81963811941}
\ncline{-}{N81963811941}{N91963811941}
\ncline{-}{N91963811941}{N101963811941}
\ncline{-}{N101963811941}{N111963811941}
\ncline{-}{N81963811941}{N21963811941}
\ncline[doubleline=true,doublesep=0.2,arrowsize=0.6,arrowlength=0.25,arrowinset=0.6]{->}{N31963811941}{N11963811941}
\end{pspicture} ~ \\ [20pt]

        6b & $SO(5,5+n)$ & $G T^{n+4}$
            & \begin{pspicture}(0,0)(7,1)
\Cnode[fillstyle=solid,fillcolor=lightgray](0,1){N11531843743}
\Cnode(4,1){N21531843743}
\Cnode[fillstyle=solid,fillcolor=lightgray](0,0){N31531843743}
\Cnode[fillstyle=solid,fillcolor=lightgray](1,0){N41531843743}
\Cnode[fillstyle=solid,fillcolor=lightgray](2,0){N51531843743}
\Cnode[fillstyle=solid,fillcolor=black](3,0){N61531843743}
\Cnode(4,0){N71531843743}
\Cnode(5,0){N81531843743}
\Cnode(6,0){N91531843743}
\Cnode(7,0){N101531843743}
\ncline{-}{N31531843743}{N41531843743}
\ncline{-}{N41531843743}{N51531843743}
\ncline{-}{N51531843743}{N61531843743}
\ncline{-}{N61531843743}{N71531843743}
\ncline{-}{N71531843743}{N81531843743}
\ncline{-}{N81531843743}{N91531843743}
\ncline{-}{N91531843743}{N101531843743}
\ncline[doubleline=true,doublesep=0.2,arrowsize=0.6,arrowlength=0.25,arrowinset=0.6]{->}{N31531843743}{N11531843743}
\ncline{-}{N71531843743}{N21531843743}
\end{pspicture} ~
            & \begin{pspicture}(0,0)(8,1)
\Cnode[fillstyle=solid,fillcolor=lightgray](1,1){N11980953788}
\Cnode(5,1){N21980953788}
\Cnode[fillstyle=solid,fillcolor=lightgray](0,0){N31980953788}
\Cnode[fillstyle=solid,fillcolor=lightgray](1,0){N41980953788}
\Cnode[fillstyle=solid,fillcolor=lightgray](2,0){N51980953788}
\Cnode[fillstyle=solid,fillcolor=lightgray](3,0){N61980953788}
\Cnode[fillstyle=solid,fillcolor=black](4,0){N71980953788}
\Cnode(5,0){N81980953788}
\Cnode(6,0){N91980953788}
\Cnode(7,0){N101980953788}
\Cnode(8,0){N111980953788}
\ncline{-}{N31980953788}{N41980953788}
\ncline{-}{N41980953788}{N51980953788}
\ncline{-}{N51980953788}{N61980953788}
\ncline{-}{N61980953788}{N71980953788}
\ncline{-}{N71980953788}{N81980953788}
\ncline{-}{N81980953788}{N91980953788}
\ncline{-}{N91980953788}{N101980953788}
\ncline{-}{N101980953788}{N111980953788}
\ncline{-}{N81980953788}{N21980953788}
\ncline{-}{N41980953788}{N11980953788}
\end{pspicture} ~
            & \begin{pspicture}(0,0)(8,1)
\Cnode[fillstyle=solid,fillcolor=lightgray](0,1){N11963811941}
\Cnode(5,1){N21963811941}
\Cnode[fillstyle=solid,fillcolor=lightgray](0,0){N31963811941}
\Cnode[fillstyle=solid,fillcolor=lightgray](1,0){N41963811941}
\Cnode[fillstyle=solid,fillcolor=lightgray](2,0){N51963811941}
\Cnode[fillstyle=solid,fillcolor=lightgray](3,0){N61963811941}
\Cnode[fillstyle=solid,fillcolor=black](4,0){N71963811941}
\Cnode(5,0){N81963811941}
\Cnode(6,0){N91963811941}
\Cnode(7,0){N101963811941}
\Cnode(8,0){N111963811941}
\ncline{-}{N31963811941}{N41963811941}
\ncline{-}{N41963811941}{N51963811941}
\ncline{-}{N51963811941}{N61963811941}
\ncline{-}{N61963811941}{N71963811941}
\ncline{-}{N71963811941}{N81963811941}
\ncline{-}{N81963811941}{N91963811941}
\ncline{-}{N91963811941}{N101963811941}
\ncline{-}{N101963811941}{N111963811941}
\ncline{-}{N81963811941}{N21963811941}
\ncline[doubleline=true,doublesep=0.2,arrowsize=0.6,arrowlength=0.25,arrowinset=0.6]{->}{N31963811941}{N11963811941}
\end{pspicture} ~ \\ [20pt]

         5 & $\R^+ \times SO(5,5+n)$ & $G V^{n+5}$
            & \begin{pspicture}(0,0)(7,1)
\Cnode[fillstyle=solid,fillcolor=lightgray](0,1){N11726241267}
\Cnode[fillstyle=solid,fillcolor=black](4,1){N21726241267}
\Cnode[fillstyle=solid,fillcolor=lightgray](0,0){N31726241267}
\Cnode[fillstyle=solid,fillcolor=lightgray](1,0){N41726241267}
\Cnode[fillstyle=solid,fillcolor=lightgray](2,0){N51726241267}
\Cnode[fillstyle=solid,fillcolor=black](3,0){N61726241267}
\Cnode(4,0){N71726241267}
\Cnode(5,0){N81726241267}
\Cnode(6,0){N91726241267}
\Cnode(7,0){N101726241267}
\ncline{-}{N31726241267}{N41726241267}
\ncline{-}{N41726241267}{N51726241267}
\ncline{-}{N51726241267}{N61726241267}
\ncline{-}{N61726241267}{N71726241267}
\ncline{-}{N71726241267}{N81726241267}
\ncline{-}{N81726241267}{N91726241267}
\ncline{-}{N91726241267}{N101726241267}
\ncline[doubleline=true,doublesep=0.2,arrowsize=0.6,arrowlength=0.25,arrowinset=0.6]{->}{N31726241267}{N11726241267}
\ncline{-}{N71726241267}{N21726241267}
\end{pspicture} ~
            & \begin{pspicture}(0,0)(8,1)
\Cnode[fillstyle=solid,fillcolor=lightgray](1,1){N1566184472}
\Cnode[fillstyle=solid,fillcolor=black](5,1){N2566184472}
\Cnode[fillstyle=solid,fillcolor=lightgray](0,0){N3566184472}
\Cnode[fillstyle=solid,fillcolor=lightgray](1,0){N4566184472}
\Cnode[fillstyle=solid,fillcolor=lightgray](2,0){N5566184472}
\Cnode[fillstyle=solid,fillcolor=lightgray](3,0){N6566184472}
\Cnode[fillstyle=solid,fillcolor=black](4,0){N7566184472}
\Cnode(5,0){N8566184472}
\Cnode(6,0){N9566184472}
\Cnode(7,0){N10566184472}
\Cnode(8,0){N11566184472}
\ncline{-}{N3566184472}{N4566184472}
\ncline{-}{N4566184472}{N5566184472}
\ncline{-}{N5566184472}{N6566184472}
\ncline{-}{N6566184472}{N7566184472}
\ncline{-}{N7566184472}{N8566184472}
\ncline{-}{N8566184472}{N9566184472}
\ncline{-}{N9566184472}{N10566184472}
\ncline{-}{N10566184472}{N11566184472}
\ncline{-}{N8566184472}{N2566184472}
\ncline{-}{N4566184472}{N1566184472}
\end{pspicture} ~
            & \begin{pspicture}(0,0)(8,1)
\Cnode[fillstyle=solid,fillcolor=lightgray](0,1){N11497894161}
\Cnode[fillstyle=solid,fillcolor=black](5,1){N21497894161}
\Cnode[fillstyle=solid,fillcolor=lightgray](0,0){N31497894161}
\Cnode[fillstyle=solid,fillcolor=lightgray](1,0){N41497894161}
\Cnode[fillstyle=solid,fillcolor=lightgray](2,0){N51497894161}
\Cnode[fillstyle=solid,fillcolor=lightgray](3,0){N61497894161}
\Cnode[fillstyle=solid,fillcolor=black](4,0){N71497894161}
\Cnode(5,0){N81497894161}
\Cnode(6,0){N91497894161}
\Cnode(7,0){N101497894161}
\Cnode(8,0){N111497894161}
\ncline{-}{N31497894161}{N41497894161}
\ncline{-}{N41497894161}{N51497894161}
\ncline{-}{N51497894161}{N61497894161}
\ncline{-}{N61497894161}{N71497894161}
\ncline{-}{N71497894161}{N81497894161}
\ncline{-}{N81497894161}{N91497894161}
\ncline{-}{N91497894161}{N101497894161}
\ncline{-}{N101497894161}{N111497894161}
\ncline{-}{N81497894161}{N21497894161}
\ncline[doubleline=true,doublesep=0.2,arrowsize=0.6,arrowlength=0.25,arrowinset=0.6]{->}{N31497894161}{N11497894161}
\end{pspicture} ~ \\ [20pt]

         4 & $\begin{array}{c} SO(6,6+n) \times \\ \times SL(2,\R) \end{array}$ & $G V^{n+6}$
            & \begin{pspicture}(0,0)(7,1)
\Cnode[fillstyle=solid,fillcolor=lightgray](0,1){N1557754221}
\Cnode[fillstyle=solid,fillcolor=lightgray](4,1){N2557754221}
\Cnode[fillstyle=solid,fillcolor=lightgray](0,0){N3557754221}
\Cnode[fillstyle=solid,fillcolor=lightgray](1,0){N4557754221}
\Cnode[fillstyle=solid,fillcolor=lightgray](2,0){N5557754221}
\Cnode[fillstyle=solid,fillcolor=lightgray](3,0){N6557754221}
\Cnode[fillstyle=solid,fillcolor=black](4,0){N7557754221}
\Cnode(5,0){N8557754221}
\Cnode(6,0){N9557754221}
\Cnode(7,0){N10557754221}
\ncline{-}{N3557754221}{N4557754221}
\ncline{-}{N4557754221}{N5557754221}
\ncline{-}{N5557754221}{N6557754221}
\ncline{-}{N6557754221}{N7557754221}
\ncline{-}{N7557754221}{N8557754221}
\ncline{-}{N8557754221}{N9557754221}
\ncline{-}{N9557754221}{N10557754221}
\ncline[doubleline=true,doublesep=0.2,arrowsize=0.6,arrowlength=0.25,arrowinset=0.6]{->}{N3557754221}{N1557754221}
\ncline{-}{N7557754221}{N2557754221}
\end{pspicture} ~
            & \begin{pspicture}(0,0)(8,1)
\dualityNode{1,1}{N11872301448}
\dualityNode{5,1}{N21872301448}
\dualityNode{0,0}{N31872301448}
\dualityNode{1,0}{N41872301448}
\dualityNode{2,0}{N51872301448}
\dualityNode{3,0}{N61872301448}
\dualityNode{4,0}{N71872301448}
\disabledNode{5,0}{N81872301448}
\normalNode{6,0}{N91872301448}
\normalNode{7,0}{N101872301448}
\normalNode{8,0}{N111872301448}
\singleConnection{N31872301448}{N41872301448}
\singleConnection{N41872301448}{N51872301448}
\singleConnection{N51872301448}{N61872301448}
\singleConnection{N61872301448}{N71872301448}
\singleConnection{N71872301448}{N81872301448}
\singleConnection{N81872301448}{N91872301448}
\singleConnection{N91872301448}{N101872301448}
\singleConnection{N101872301448}{N111872301448}
\singleConnection{N81872301448}{N21872301448}
\singleConnection{N41872301448}{N11872301448}
\end{pspicture} ~
            & \begin{pspicture}(0,0)(8,1)
\Cnode[fillstyle=solid,fillcolor=lightgray](0,1){N12031233393}
\Cnode[fillstyle=solid,fillcolor=lightgray](5,1){N22031233393}
\Cnode[fillstyle=solid,fillcolor=lightgray](0,0){N32031233393}
\Cnode[fillstyle=solid,fillcolor=lightgray](1,0){N42031233393}
\Cnode[fillstyle=solid,fillcolor=lightgray](2,0){N52031233393}
\Cnode[fillstyle=solid,fillcolor=lightgray](3,0){N62031233393}
\Cnode[fillstyle=solid,fillcolor=lightgray](4,0){N72031233393}
\Cnode[fillstyle=solid,fillcolor=black](5,0){N82031233393}
\Cnode(6,0){N92031233393}
\Cnode(7,0){N102031233393}
\Cnode(8,0){N112031233393}
\ncline{-}{N32031233393}{N42031233393}
\ncline{-}{N42031233393}{N52031233393}
\ncline{-}{N52031233393}{N62031233393}
\ncline{-}{N62031233393}{N72031233393}
\ncline{-}{N72031233393}{N82031233393}
\ncline{-}{N82031233393}{N92031233393}
\ncline{-}{N92031233393}{N102031233393}
\ncline{-}{N102031233393}{N112031233393}
\ncline{-}{N82031233393}{N22031233393}
\ncline[doubleline=true,doublesep=0.2,arrowsize=0.6,arrowlength=0.25,arrowinset=0.6]{->}{N32031233393}{N12031233393}
\end{pspicture} ~ \\ [20pt]

         3 & $ SO(8,8+n)$ & $G V^{n+7}$
            & \begin{pspicture}(0,0)(7,1)
\Cnode[fillstyle=solid,fillcolor=lightgray](0,1){N1389328345}
\Cnode[fillstyle=solid,fillcolor=lightgray](4,1){N2389328345}
\Cnode[fillstyle=solid,fillcolor=lightgray](0,0){N3389328345}
\Cnode[fillstyle=solid,fillcolor=lightgray](1,0){N4389328345}
\Cnode[fillstyle=solid,fillcolor=lightgray](2,0){N5389328345}
\Cnode[fillstyle=solid,fillcolor=lightgray](3,0){N6389328345}
\Cnode[fillstyle=solid,fillcolor=lightgray](4,0){N7389328345}
\Cnode[fillstyle=solid,fillcolor=black](5,0){N8389328345}
\Cnode(6,0){N9389328345}
\Cnode(7,0){N10389328345}
\ncline{-}{N3389328345}{N4389328345}
\ncline{-}{N4389328345}{N5389328345}
\ncline{-}{N5389328345}{N6389328345}
\ncline{-}{N6389328345}{N7389328345}
\ncline{-}{N7389328345}{N8389328345}
\ncline{-}{N8389328345}{N9389328345}
\ncline{-}{N9389328345}{N10389328345}
\ncline[doubleline=true,doublesep=0.2,arrowsize=0.6,arrowlength=0.25,arrowinset=0.6]{->}{N3389328345}{N1389328345}
\ncline{-}{N7389328345}{N2389328345}
\end{pspicture} ~
            & \begin{pspicture}(0,0)(8,1)
\Cnode[fillstyle=solid,fillcolor=lightgray](1,1){N11528605172}
\Cnode[fillstyle=solid,fillcolor=lightgray](5,1){N21528605172}
\Cnode[fillstyle=solid,fillcolor=lightgray](0,0){N31528605172}
\Cnode[fillstyle=solid,fillcolor=lightgray](1,0){N41528605172}
\Cnode[fillstyle=solid,fillcolor=lightgray](2,0){N51528605172}
\Cnode[fillstyle=solid,fillcolor=lightgray](3,0){N61528605172}
\Cnode[fillstyle=solid,fillcolor=lightgray](4,0){N71528605172}
\Cnode[fillstyle=solid,fillcolor=lightgray](5,0){N81528605172}
\Cnode[fillstyle=solid,fillcolor=black](6,0){N91528605172}
\Cnode(7,0){N101528605172}
\Cnode(8,0){N111528605172}
\ncline{-}{N31528605172}{N41528605172}
\ncline{-}{N41528605172}{N51528605172}
\ncline{-}{N51528605172}{N61528605172}
\ncline{-}{N61528605172}{N71528605172}
\ncline{-}{N71528605172}{N81528605172}
\ncline{-}{N81528605172}{N91528605172}
\ncline{-}{N91528605172}{N101528605172}
\ncline{-}{N101528605172}{N111528605172}
\ncline{-}{N81528605172}{N21528605172}
\ncline{-}{N41528605172}{N11528605172}
\end{pspicture} ~
            & \begin{pspicture}(0,0)(8,1)
\Cnode[fillstyle=solid,fillcolor=lightgray](0,1){N11984645155}
\Cnode[fillstyle=solid,fillcolor=lightgray](5,1){N21984645155}
\Cnode[fillstyle=solid,fillcolor=lightgray](0,0){N31984645155}
\Cnode[fillstyle=solid,fillcolor=lightgray](1,0){N41984645155}
\Cnode[fillstyle=solid,fillcolor=lightgray](2,0){N51984645155}
\Cnode[fillstyle=solid,fillcolor=lightgray](3,0){N61984645155}
\Cnode[fillstyle=solid,fillcolor=lightgray](4,0){N71984645155}
\Cnode[fillstyle=solid,fillcolor=lightgray](5,0){N81984645155}
\Cnode[fillstyle=solid,fillcolor=black](6,0){N91984645155}
\Cnode(7,0){N101984645155}
\Cnode(8,0){N111984645155}
\ncline{-}{N31984645155}{N41984645155}
\ncline{-}{N41984645155}{N51984645155}
\ncline{-}{N51984645155}{N61984645155}
\ncline{-}{N61984645155}{N71984645155}
\ncline{-}{N71984645155}{N81984645155}
\ncline{-}{N81984645155}{N91984645155}
\ncline{-}{N91984645155}{N101984645155}
\ncline{-}{N101984645155}{N111984645155}
\ncline{-}{N81984645155}{N21984645155}
\ncline[doubleline=true,doublesep=0.2,arrowsize=0.6,arrowlength=0.25,arrowinset=0.6]{->}{N31984645155}{N11984645155}
\end{pspicture} ~ \\ [10pt]

        \hline
    \end{tabular}
    \caption{
        The decompositions of $B_7^{+++}$, $D_8^{+++}$ and $B_8^{+++}$ with
        respect to the possible gravity lines. The duality groups $G_D$ and the
        multiplet structures (where $G$ is the graviton, $V$ the vector and $T$ the self-dual tensor multiplet) are also given.
    }\label{B7dec}
}
\end{landscape}

\noindent
of the
fundamental ones.
Suppressing the duality indices we obtain that
the generators $\pf{2}$, $\pf{(D-3)}$, $\pf{(D-2)}$, $\pf{(D-1)}$ and $\pf{D}$
 occur at the levels $(0,2)$, $(1,1)$, $(1,2)$, $(1,3)$ and   $(1,4)$,
respectively. These results are in agreement with the tables in appendix
C\,\footnote{Note that we refer to the generic situation. There are
special cases. For instance, pure $D=10$ half--maximal supergravity has
no 1--form generators and the fundamental generators are a 2--form and a
6--form. Another exception is pure $D=9$ half--maximal supergravity
which has an $\R^+\times SO(1,1)$ duality group. We now need three
level numbers $(\ell_1, \ell_2, \ell_3)$ in order to distinguish between
the different Poincare dualities under $SO(1,1)$.}.

We will now turn to the potentials associated to the
different generators. We will first discuss the potentials
corresponding to the physical degrees of freedom  of
half--maximal supergravity. A summary of these can be found in appendix \ref{app:half--max}
and in particular table~\ref{tableii}. Afterwards we will discuss the
non--propagating deformation and top-form potentials.

The $\pf 2$ corresponds to the two-form potential, which is indeed
present in any half--maximal supergravity. Note that the
first commutator in \eqref{gen-alg} tells us that the two-form potential
transforms in a Chern-Simons way under the vector gauge transformations:
\be
\delta A_{(2)} = \partial \Lambda_{(1)} + \partial
\Lambda^M_{(0)}\,A_{(1)}^N\, \eta_{MN}\,.
\ee
Hence, the Kac--Moody approach automatically leads to
the Chern--Simons gauge
transformations that are crucial for anomaly cancellations in
string theory \cite{Green:1984sg}.
The $\pf{(D-3)}^M$, $\pf{(D-2)}$ and
$\pf{(D-2)}^{[MN]}$ correspond to the duals of the vectors, the dilaton
and the scalar coset, respectively. Note that the number of
$(D-2)^{[MN]}$--forms exceeds that of the scalars, since the latter
take values in the scalar coset $G/K$ and hence are modded out by
the compact subgroup $K$ of $G$. Therefore, we expect that there will
be a number of linear relations between the field strengths of the
$(D-2)$--forms, similar to what has been found for the 8--forms
of IIB supergravity \cite{Dall'Agata:1998va}.

Extreme cases occur when the symmetry group is compact, i.e.~$m=0$
or $m+n=0$. These correspond to ten dimensions or pure supergravity,
without vector multiplets, respectively. These theories do not have
any other scalars than the dilaton and hence one expects
supersymmetry to require all of the field strengths of the
$(D-2)^{[MN]}$-forms to vanish. Although we are not aware of a
discussion of this phenomenon in the context of half--maximal
supergravity, it has recently been encountered in pure $D=5$, $N=2$
supergravity \cite{Gomis:2007gb}.

We should also mention two exceptions that differ from the above
pattern. The $D=10$ theory without vector multiplets and the $D=6b$
theories do not contain any vectors. Rather, the simple roots
correspond to a $\pf 2$ and a $\pf 6$ in the $D=10$ case and to a $\pf 2^M$ in
the $D=6b$ case. These generate the following gauge transformations:
\begin{equation}
  D=10, \; n=0: \quad \left\{ \quad
  \begin{aligned}
   \, [\pf 2,\pf 2] &= 0 \,,\\
   [\pf 2,\pf 6] &= \pf 8 \,, \\
   [\pf 2,\pf 8] &= \pf{10} \,,
  \end{aligned} \right.
\end{equation}
for the ten-dimensional theory and
\begin{equation}
  D=6b:  \quad \left\{ \quad
  \begin{aligned}
  \, [\pf 2^M, \pf 2^N] & = \pf 4^{[MN]} \,,\\
   [ \pf 2^M, \pf 4^{[NP]} ] & =  \eta^{M [N} \pf 6^{P]} + \pf 6^{(M[N)P]} \,,
   \end{aligned} \right.
\end{equation}
for the six--dimensional case.

In yet lower dimensions a similar pattern occurs. The main difference,
however, is that for half--maximal supergravity in $D \leq 5$
dimensions, all fundamental $p$--forms are vectors.
This is in contrast to higher dimensions where there
are also fundamental $p$--forms of higher rank.
The explicit formulae are a bit messier
in the lower dimensions, as can also be
seen from table \ref{tableii}, and hence we will refrain from giving
them. It should be stressed that they follow exactly the same
pattern as above. The same applies to the deformation and
top-form potentials discussed in the next two subsections where
there is a plethora of representations in the lower dimensions, see
table \ref{halfmax}.

This finishes the discussion of the physical degrees of freedom and
their duals \cite{Kleinschmidt:2003mf, Schnakenburg:2004vd}.
The generators corresponding to these potentials
are already present in the affine
Kac--Moody extension \cite{Keurentjes:2002xc}. Potentials of yet higher rank
do not correspond to propagating degrees of freedom and only occur
in the over-- and very--extended Kac--Moody algebras. We now discuss
the deformation and top--form potentials of half--maximal supergravity.

\subsection{Deformation potentials} \label{sec:de-forms}

Turning first to the $(D-1)$-forms, it follows from \eqref{gen-alg} that
in generic dimensions these occur in  a fundamental and
an anti-symmetric three-form representation.
In dimensions $8 \leq D \leq 10$ this is the complete story as well.

In lower dimensions, however, there are more possibilities. For
example, in $D=7$ one can generate an additional deformation
potential $6$ from a multiple commutator of the dual two-form $\pf{D-4}$,
which is a $\pf 3$ in this case. This corresponds to the singlet in
table \ref{halfmax}. In addition there is another top-form
representation $7^M$. The additional commutators are:
\begin{equation}
  D=7: \quad
  \left\{ \quad \begin{aligned}
   \, [ \pf 3, \pf 3] & = \pf 6 \,, \\
   [\pf 3, \pf 4^M] & = \pf 7^M \,. \\
   \end{aligned} \right.
  \label{D=7}
\end{equation}
In fact, also the commutator $[\pf 1^M, \pf 6]$ is non-vanishing and leads
to a $\pf{7'}^M$. However, this commutator is related to the one above
by the Jacobi identity and hence $\pf 7^M$ and $\pf{7'}^M$ are linearly
dependent.

\TABLE[t]{
    \begin{tabular}{|c||c|c|c||c|c|}
    \hline
    \multirow{2}{*}{$D$}
        & \multicolumn{3}{c||}{$(D-1)$--forms}
        & \multicolumn{2}{c|}{$D$--forms} \\
        & \multicolumn{1}{c}{$p=1$}         & \multicolumn{1}{c}{$p=2$} & $p=3$
        & \multicolumn{1}{c}{constraints on $p=1$}   & other \\
    \hline
    \hline
    &&&&& \\ [-10pt]
    10,\,9,\,8  & $\fund \;\;\threeform$                    &           &
        & $1 \; \twoform \;\; \fourform$                &\\ [10pt]
    \hline &&&&& \\ [-10pt]
    7           & $\fund \;\;\threeform$                    &           & 1
        & $1 \;  \twoform \;\; \fourform$               & $\fund$ \\ [10pt]
    \hline &&&&& \\ [-10pt]
    6a          & $\fund \;\; \threeform$                   & $\fund$   &
        & $1 \; \twoform \;\; \fourform$                & 1  1  \symm \;\; \twoform \\ [10pt]
    \hline &&&&& \\ [-10pt]
    6b          &                                           &           &
        && $\fund \;\; \twohook$ \\ [6pt]
    \hline &&&&& \\ [-10pt]
    5           & $\fund \;\; \twoform \;\; \threeform$     &           &
        & $1 \; \fund \;\; \fund \;\; \twoform \;\; \twohook \;\; \threeform \;\; \fourform$&\\ [10pt]
    \hline &&&&& \\ [-10pt]
    4           & $\left(\bf 2,\fund \right) \; \left(\bf 2, \threeform \, \right)$ &           &
        & $({\bf 3},1) \; \left(\bf 3, \twoform\,\right) \; \left(\bf 3, \fourform\,\right) \; \twoform \;\; \twoform \;\; \threehook$ &\\ [10pt]
    \hline &&&&& \\ [-10pt]
    3           & $1 \; \symm \;\; \fourform$           &           &
        & $\symm \;\; \twoform \;\; \fourform \;\; \threehook \;\; \fivehook$ &\\ [15pt]
    \hline
    \end{tabular}
    \caption{The representations of deformation-- and top--forms in all
    half--maximal supergravities.
    The representations refer to the duality group $G_D$ given in table
    \ref{B7dec}. We also indicate which type $p$ of deformations they
    correspond to, and to which top--forms one can associate a quadratic
    constraint on type $1$ deformation parameters.}
    \label{halfmax}
}

In $D=6$ one has $D-4=2$, i.e.~the dual of the two-form is itself
again a two-form. To avoid confusion, we will denote the fundamental 2--form
by $\pf 2$ and the one coming from the commutator of the vectors by
$\pf{2'}$. In this theory there are again a number of extra commutators
that contribute to the deformation and top-form potentials:
\begin{equation}
  D=6a: \quad \left\{ \quad
  \begin{aligned}
  \, [\pf 2, \pf 3^M]         & = \pf{5'}^M \,, \\
  [\pf 2, \pf 4^{[MN]} ]   & = \pf{6'}^{[MN]} \,, \\
  [\pf 2, \pf 4]           & = \pf{6'} \,, \\
  [\pf 1^M, \pf{5'}^N]     & = \eta^{MN} \pf{6'} + \pf 6^{(MN)} \,.
  \end{aligned} \right.
 \end{equation}
There are also other non-vanishing commutators but these are
related by the Jacobi identity.

We now turn to the question to which types of deformations the
deformation potentials correspond.
Given that there are only a few of these, we will start with the
massive deformations. As can be seen from the previous discussion,
type 3 deformations of half--maximal supergravity are only possible
in $D=7$, for the simple reason that only here there is a fundamental
3-form. The deformation is
a singlet of the symmetry group $\mathbb{R}^+ \times SO(3,3+n)$ and
has been explicitly constructed for $n=-3$ \cite{Townsend:1983kk}.
Similarly, type 2 deformations of half--maximal supergravity are only
possible in $D=6a$ and occur in the fundamental representation of
the symmetry group $\mathbb{R}^+ \times SO(4,4+n)$. The deformed
theory has been explicitly constructed for the special cases of
$n=-4$ \cite{Romans:1985tw} and $n=16$ \cite{Haack:2001iz}.

All remaining deformation potentials correspond to type 1
deformations, i.e.~to gaugings. Note that in every dimension $D \geq
4$ there is a fundamental and three-form representation of such
deformation potentials. To be able to do more general gaugings one needs more
space-time vectors than only the fundamental representation, which
is present in all these dimensions. For example, in $D=5$ an
additional vector is provided by the dual of the two-form, giving
rise to an extra two-form representation of possible gaugings. In
$D=4$ the extra vectors are the Hodge duals of the original ones,
leading to an $SL(2,\R)$ doublet of possible gaugings. Finally, in
$D=3$ scalars are dual to vectors. This is the underlying reason for
the symmetry enhancement in three dimensions, and also gives rise to
the more general possibilities of gaugings in this case.

Many of these gaugings have been obtained in the literature.
Explicit calculations of the possible gaugings using the embedding
tensor formalism in $D=3,4,5$ have uncovered exactly the same
representations \cite{Nicolai:2001sv,Schon:2006kz}.
In addition, one can obtain
components of the three-form representation of gauging in any
dimension $D$ by a Scherk-Schwarz reduction from $D+1$ dimensions
using the $SO(m,m+n)$ symmetry, see e.g.~\cite{Kaloper:1999yr,
Behrndt:2001ab}.

A prediction that follows from the above analysis is that in the
dimensions where the possible gaugings have not yet been fully
analyzed, i.e.~in $D \geq 6$, it will be possible to introduce a
fundamental and a three-form representation of gaugings. In terms of
the embedding tensor, which describes the embedding of the gauge
group in the duality group $G$ \cite{deWit:2002vt, deWit:2003hr}, this would read
\begin{equation}
\begin{split}
  \Theta_M{}^{NP} &= f_M{}^{NP} + \delta_M^{[N} \xi^{P]} \,, \\
\Theta_N{}^0 & = \xi_N  \,,
  \label{5D-lin}
\end{split}
\end{equation}
where $\xi_M  = \eta_{MN} \xi^N$ and $f_{MNP}$ are the fundamental and three-form
representations of gaugings, respectively. The notation here is as
follows: the subscript index $M = 1, \ldots, 2m + n$ refers to the
generators of the gauge group and the superscript indices $\{ 0, MN
\}$ label the generators of the duality group $\R^+ \times SO(m,m+n)$.
The embedding tensor thus encodes which subgroup
is gauged by the vectors $\pf 1^M$. Note that the $\R^+$ factor is crucial
for the introduction of $\xi_N$, as can be seen from the $
\Theta_N{}^0$ component. The different components of $\Theta_M{}^{NP}$
and $\Theta_N{}^0$ specify which linear combinations of the gauge fields
are used to gauge $\R^+$ and a subgroup $H \subset SO(m,m+n)$, respectively:
\begin{eqnarray}
\Theta_M{}^{NP} {\pf 1}^M&:& \hskip 2truecm H\subset SO(m,m+n)\,,\cr
\Theta_N{}^0  {\pf 1}^N&:& \hskip 2truecm  \R^+\,.
\end{eqnarray}

\subsection{Top--form potentials} \label{sec:top-forms}

Subsequently, we consider the top--form potentials and their relation to the
quadratic constraints.

In generic dimensions these top--forms
occur in  a singlet and anti-symmetric two- and four-form representations.
Using the embedding tensor approach, an analysis of the quadratic
constraints on the possible deformations has been explicitly carried
out in $D=3,4,5$ \cite{Nicolai:2001sv, Schon:2006kz}.
It turns out that the representations of the quadratic
constraints exactly coincide with the representations of the possible
top-forms in these dimensions.

For $D\geq 6$ the embedding approach has not yet been applied and the
Kac--Moody approach leads to a prediction. The generic top-forms occur
in the singlet, two- and four-form representations. In addition, from
the lower-dimensional analysis \cite{Nicolai:2001sv, Schon:2006kz}
one would expect them to correspond to a quadratic constraint.
In terms of the embedding tensor $\Theta$ these would take the following form:
\begin{align}
 \label{quadcon}
 f_{MNP}\xi^P &=0 \,, \nonumber \\
 3 f_{R[MN}f_{PQ]}{}^R & = 2f_{[MNP}\xi_{Q]}\,,  \\
 \eta^{MN}\xi_M\xi_N & = 0\, . \nonumber
\end{align}

The prediction is that the most general gauging of half--maximal
supergravity in $D \geq 6$ is described by the embedding tensor
\eqref{5D-lin} subject to the quadratic constraints \eqref{quadcon}. It would
be interesting to explicitly construct these gauged theories. Note
that the fundamental representation $\xi_M$ cannot be non-zero in
$D=10$, since the quadratic constraint requiring it to be a null
vector can not be satisfied for an $SO(n)$ representations.
The gauging with deformation parameters $f_{MNP}$ can be viewed as a
gauging of a subgroup $H \subset SO(n)$ with structure constants
$f_{MNP}$. In this sense $D=10$ half-maximal matter--coupled supergravity with
gauge groups $SO(32)$ or $E_8\times E_8$, i.e.~the low--energy limit of
type I or heterotic string theory, can be viewed as the gauged
deformation of $D=10$ half-maximal supergravity coupled to 496
Maxwell multiplet\,\footnote{We thank Axel Kleinschmidt for a discussion on
this point.}.

In addition, it would be interesting to investigate the possibilities of
including the type 2 and 3 massive deformations in six and seven dimensions,
respectively, in the gauged theories; in other words, to see
which types of deformations can be turned on simultaneously.

\section{Deformations} \label{sec:def}

Up to this point we have only considered the role of the deformation
potentials in the Kac-Moody or $p$--form algebra but not the
deformation parameters themselves. These parameters  can be seen as
the duals of (the field strengths of) the deformation
potentials, see eqs.~\eqref{dualm} and \eqref{emb-ten}. This is in
contradistinction with the lower-rank potentials in which case the
$p$--form algebra gives rise to both the potentials and their duals.
In this section we will briefly consider how the inclusion of the
deformation parameters in supergravity effects the $p$--form
algebra. In particular, we will discuss how the bosonic gauge
transformations could be truncated to a Lie algebra in the deformed
case, first for massive IIA supergravity and in the next subsection
for gauged half--maximal supergravity.

\subsection{Massive IIA supergravity}

For massless IIA supergravity the fundamental generators are a 1-form
generator $\pf 1$ and a 2--form generator $\pf 2$, see figure \ref{E11-IIA}.
The other generators are the R-R
generators $\pf 3,\pf 5,\pf 7,\pf 9$ and the NS-NS generators
$\pf 6,\pf 8,\pf{10},\pf{10'}$. The basic
commutators are given by
\begin{align}
\label{cIIA}
 [\pf 2,\pf 1]&=\pf 3\,, & [\pf 1,\pf 7]&=\pf 8 \,,  \nonumber \\
 [\pf 2,\pf 3]&=\pf 5\,, & [\pf 2,\pf 7]&=\pf 9\,, \nonumber\\
 [\pf 1,\pf 5]&=\pf 6\,, & [\pf 2,\pf 8]&=\pf{10} \,, \nonumber\\
 [\pf 2,\pf 5]&=\pf 7\,, & [\pf 1,\pf 9]&=\pf{10'} \,.
\end{align}
We have not included the $0$--form generator corresponding to the duality
group $\R^+$. In fact, also the commutator $[\pf 2,\pf 6]$ is non-vanishing and leads
to an $\pf{8'}$. However, this commutator is related to $[\pf 1,\pf 7]$
by the Jacobi identity and hence $\pf 8$ and $\pf{8'}$ are linearly dependent.
As can be seen from \eqref{cIIA}, the IIA theory has a type 2 deformation
potential and two top--form potentials. There is no quadratic
constraint associated to either of the top--forms.

Let us now turn to the realisation of this symmetry on the IIA potentials.
In the following we will only consider the truncation to the low--level
potentials corresponding to $\pf 1, \pf 2, \pf 3$ as this will be
sufficient for our purpose. The gauge transformations of massless IIA
supergravity are given by
 \begin{align}
  \delta A_{(1)}  & = \Lambda_{(1)} \,, \nonumber \\
  \delta A_{(2)}  & = \Lambda_{(2)} \,, \nonumber \\
  \delta A_{(3)}  & = \Lambda_{(3)} + 3 \Lambda_{(1)} A_{(2)} \,,
 \label{IIA-gauge-transf}
 \end{align}
where the gauge parameters are all closed and hence
$\Lambda_{(p)} = \partial \lambda_{(p-1)}$. The restriction to the
$p$-form algebra corresponds to constant $\Lambda_{(p)}$'s, or
equivalently $\lambda_{(p-1)}$'s with linear coordinate dependence,
as discussed around \eqref{Lietruncation}. It can easily be verified
that this truncation to a Lie algebra satisfies the first commutator of
\eqref{cIIA}. Similarly, it is possible to include all potentials of
IIA and truncate to the $p$--form Lie algebra, that satisfies the full
\eqref{cIIA} \cite{Bergshoeff:2006qw}.

There are several formulations of the massive IIA theory. The
original formulation by Romans \cite{Romans:1985tz} contains a
constant mass parameter $m$ and no deformation potential. Later, it
was shown that there is an alternative description with a scalar
mass function $m(x)$ and a 9--form deformation potential
\cite{Bergshoeff:1996ui}. There is
even a third formulation \cite{Bergshoeff:1996ui} with only a
deformation potential and no parameter but the bosonic gauge
transformations of this formulation are highly non-linear and have
not been explicitly worked out yet. We will consider the original
Romans formulation here.

It turns out that the IIA bosonic gauge transformations can be written as in the
massless case:
 \begin{align}
  \delta A_{(1)}  & = \tilde \Lambda_{(1)} \,, \nonumber \\
  \delta A_{(2)}  & = \Lambda_{(2)} \,, \nonumber \\
  \delta A_{(3)}  & = \Lambda_{(3)} + 3 \tilde \Lambda_{(1)} A_{(2)} \,,
 \label{mIIA-gauge-transf}
 \end{align}
but with a different parameter $\tilde \Lambda_{(1)}$
that is not closed:
$\partial \tilde \Lambda_{(1)} = - m  \Lambda_{(2)}$.
Note that, up to this level, the massive modification of the gauge
transformations only occurs via $\tilde \Lambda_{(1)}$.
The
$\Lambda, \tilde \Lambda$
parameters can be expressed in terms of the
local gauge parameters $\lambda_{(p-1)}$ as follows:
 \begin{align}
  \tilde \Lambda_{(1)} & = \partial \lambda_{(0)} - m \lambda_{(1)} \,,
\nonumber \\
  \Lambda_{(2)}  & = \partial \lambda_{(1)} \,, \nonumber \\
  \Lambda_{(3)}  & = \partial \lambda_{(2)} \,. \label{gauge-para}
 \end{align}

 If we perform
the same truncation \eqref{Lietruncation} as in the massless case
to $\lambda_{(p)}$'s with linear coordinate dependence\footnote{Note that
in the massive case
the constant part of $\lambda_{(1)}$ does not drop out, but it can be
absorbed by a redefinition of $\Lambda_{(1)}$. Hence we will not consider this
constant part.},
the massive transformation parameters reduce to
  \begin{align}
  \tilde \Lambda_{(1)} & = \Lambda_{(1)} - m x \cdot \Lambda_{(2)} \,,
 \label{KM-para}
 \end{align}
and $\Lambda_{(2)}$ and $\Lambda_{(3)}$, with constant $\Lambda_{(i)}$'s.

A new feature is the appearance of an explicit coordinate dependence
in the massive case. This has been interpreted from the point of
view of the $p$--form algebra in the following way. The coordinate
$x^{\mu}$ can be seen as a new potential with its associated
symmetry being the translations \cite{Schnakenburg:2002xx}. Using
the terminology of \cite{Lavrinenko:1999xi}\footnote{Actually,
reference \cite{Lavrinenko:1999xi} proposes a different way of
introducing the deformation parameters which will be discussed
later.} we will call $x^\mu$ a ``(\dash 1)--form potential''.
Following  \cite{Schnakenburg:2002xx} the massive deformation
parameter $m$ can be introduced to the $p$--form algebra by
including an additional generator, which will be denoted by
$\pf{-1}$. Subsequently one must define the commutators between the
translation generator and the fundamental generators of the
$p$--form algebra. In the case of massive IIA, the non-zero
commutators are \cite{Schnakenburg:2002xx}
 \begin{equation}\label{defcomnew}
  [\pf 2_{\mu \nu}, \pf{-1}^\rho] = m\ \pf
 1_{[\mu} \delta_{\nu]}^\rho.
 \end{equation}
This commutator is realized by the truncated massive IIA gauge
transformations \eqref{KM-para} due to the term with explicit
coordinate dependence.

The commutator \eqref{defcomnew}, or equivalently the gauge
transformation \eqref{gauge-para}, tells us that the 1--form is
transforming with a shift, proportional to $m$, under the gauge
transformations of the 2--form. Therefore, the 1--form is ``eaten
up'' by the 2--form and the two potentials $(2,1)$ together form a
so--called St\"uckelberg pair describing a \emph{massive} 2--form.
The commutator \eqref{defcomnew} defines a deformation of the direct
sum of the $p$--form algebra and the translation generator
\cite{Schnakenburg:2002xx}.

It is not guaranteed that the truncation \eqref{Lietruncation}
is consistent in the massive case, since  $\lambda_{(1)}$ also appears
without an accompanying derivative. Therefore, closing the
algebra might force us to introduce more symmetries.
Indeed, we find that the following commutator does not
close:
 \begin{align}
  [ \delta_2 , \delta_{2'} ] A_{\mu \nu \rho} &
    = 3 m (x^\sigma \Lambda_{\sigma [\mu} \Lambda_{\nu \rho]}'
    - x^\sigma \Lambda_{\sigma [\mu}' \Lambda_{\nu \rho]}) \,.
 \end{align}
Although the three-form potential transforms with a shift by a closed
three-form, this is not covered by the present Ansatz for the
gauge parameter $\lambda_{(2)}$, as it leads to a
constant, $x$--independent, shift only.
To obtain closure one must introduce an additional
term of the form
 \begin{align}
  \lambda_{\mu \nu} = x^\sigma \Lambda_{\sigma \mu \nu} + x^{\sigma} x^{\tau} \Lambda_{\sigma, \tau \mu \nu} \,, \quad \Rightarrow \quad
  \tilde \Lambda_{\mu \nu \rho} = \Lambda_{\mu \nu \rho} + \tfrac43 x^{\sigma} \Lambda_{\sigma, \mu \nu \rho} \,.
  \label{quadratic-truncation}
 \end{align}
The algebra then closes provided
 \begin{align}
  \Lambda_{\sigma, \mu \nu \rho} = \tfrac 9 4 m (\Lambda_{\sigma [\mu} \Lambda_{\nu \rho]}' - \Lambda_{\sigma [\mu}' \Lambda_{\nu \rho]})\,.
 \end{align}
The additional parameter is anti-symmetric in the last three-indices
and satisfies $\Lambda_{[\sigma, \tau \mu \nu]} = 0$. In terms of
Lorentz representations, this corresponds to a $\pf{(3,1)}$
representation with mixed symmetry and its trace, which is a $\pf 2$.
Since the trace properties play no role here, we will denote both
together by $\Lambda_{(3,1)}$.

One can see the need to include such a symmetry also from the
$p$--form algebra point of view. Given the commutator
\eqref{defcomnew} between the translation generator and the
fundamental $\pf 2$, the Jacobi identity between the $\{\pf 2, \pf
2, - \pf 1\}$ generators implies
 \begin{align}
  [[\pf 2_{\mu \nu}, \pf 2_{\rho \sigma}], -\pf 1^\tau]
    = m \pf 3_{\mu \nu [\rho} \delta_{\sigma]}^\tau
    + m \pf 3_{\rho \sigma [\mu} \delta_{\nu]}^\tau \,.
  \label{Jacobi}
 \end{align}
Hence, in the massive case, $[\pf 2_{\mu \nu},\pf 2_{\rho \sigma}]$ must
be non-vanishing. It is anti-symmetric in a pair of two anti-symmetric indices
and hence has $\tfrac18 (D+1)D(D-1)(D-2)$ components in $D$ dimensions.
This is equal to the number of components of a (traceful) $\pf{(3,1)}$
representation. Therefore we write
 \begin{align}
  [\pf 2_{\mu \nu}, \pf 2_{\rho \sigma}]
    = m \pf{(3,1)}_{[\mu,\nu] \rho \sigma}
    - m \pf{(3,1)}_{[\rho,\sigma] \mu \nu} \,.
  \label{2,2-commutator}
 \end{align}
The above Jacobi identity is then satisfied provided
 \begin{align}
  [\pf{(3,1)}_{\mu,\nu \rho \sigma} , -\pf 1^\tau ] = \delta_\mu^\tau \pf 3_{\nu \rho \sigma} - \delta_{[\mu}^\tau \pf 3_{\nu \rho \sigma]} \,.
   \label{3,1-commutator}
 \end{align}

We have checked that the first commutator of \eqref{cIIA} together
with \eqref{defcomnew}, \eqref{2,2-commutator} and
\eqref{3,1-commutator} lead to a closed Lie algebra. Schematically
we have
 \begin{align}
  [ \pf 2, \pf 1 ] & = \pf 3 \,, &
  [ \pf 2, - \pf 1] & = m \pf 1 \,, \nonumber \\
  [ \pf 2, \pf 2 ] & = m \pf{(3,1)} \,, &
  [ \pf{(3,1)}, - \pf 1] & = \pf 3 \,.
 \end{align}
Note that $- \pf 1$ does not appear on the right-hand side of any commutator,
i.e.~the complementary generators form an ideal, and the former can
therefore be quotiented out. However, the same cannot be said for the
$\pf{(3,1)}$ due to the commutator \eqref{2,2-commutator}.

We conclude that a truncation of the massive IIA gauge transformations
forces us to consider extensions of the $p$--form algebra with additional
mixed symmetry generators. It is expected that more such generators are
needed when also the higher rank potentials are included. It remains to be
seen whether a consistent truncation exists
when all $p$--form generators are included.

It is interesting to compare the present result with the approach of
\cite{Lavrinenko:1999xi} which takes the same massive IIA gauge
transformation rules as their starting point. Before doing any
truncation one first rewrites the massive transformation rules such
that every parameter occurs with a derivative, like in the massless
case. This makes it possible to perform the same truncation as in
the massless case. For this to work it is crucial that one first
formulates the transformation rules in terms of forms and next
formally write the 0--form $m$ as the exterior derivative of a
``(-1)--form potential'' $A_{(-1)}$\,: \be m = d
A_{(-1)}\,.\label{nP} \ee Once every parameter occurs under a
derivative one can write the transformation rules as the non-linear
realization of an algebra that includes a formal ``(-1)--form
generator''. We understand that in this procedure one should
not convert to component notation in the presence of the (-1)--form
potential. Only after all (-1)--form potentials have been converted
into deformation parameters a transition to component notation can
be made. In particular, one should not consider a component
formulation of \eqref{nP} since this would lead us back to our
earlier discussion with the need to introduce extra mixed symmetry
generators. It would be interesting to see whether the
``(-1)--forms'' needed in this procedure can be given a rigorous
mathematical basis.

Sofar, we have discussed two ways to proceed in the massive case.
Either one starts extending the direct sum of the
$p$-form algebra and the translation generators with new mixed symmetry
generators or one extends the $p$--form algebra
with the formal concept of a new ``(-1)--form generator''. There is
even a third way to proceed in the massive case
which uses $E_{10}$ instead of $E_{11}$ \cite{Kleinschmidt:2004dy}.
 The spectrum of $E_{10}$
leads to precisely the same representations as $E_{11}$ except for
the top--forms which only follow from $E_{11}$. By using $E_{10}$ one
is able to not only consider kinematics but also dynamics consistent with
$E_{10}$. By using
the dynamics the authors of \cite{Kleinschmidt:2004dy} seem to be able
to derive the equations of motion of massive IIA supergravity
without the need to introduce new symmetry generators. It would be
interesting to more carefully compare the different approaches
and to obtain a better understanding of what the role of the
dynamics is.

\subsection{Half-maximal supergravity}

We now discuss the case of half--maximal supergravity.
For simplicity we will
consider only the three--form representation $f_{MNP}$ and not the
most general deformation.
 This will be sufficient for the present purpose.

The starting point will be the original ungauged $p$--form algebra of
half--maximal supergravity, which we truncate to the vectors $\pf 1^M$.
In addition we include the scalars $\pf 0^{MN}$, which are
the generators of the special orthogonal part of the duality group $G_D = \R^+ \times SO(m,m+n)$.
These generators satisfy
 \begin{align}
 [\pf 1^M, \pf 0^{NP}] = \pf 1^{[N} \eta^{P]M}\,,
  \end{align}
while other commutators vanish (including $[\pf 1^M, \pf 1^N ]$ in this
truncation).

Subsequently we introduce the three-form deformation $f_{MNP}$, which
is defined by the following non--zero commutators between the translation
generator $\pf{-1}$ and the fundamental generators, see also \cite{Riccioni:2007au}:
\begin{equation}
    [\pf 1^M, \pf{-1} ] = f^M{}_{NP}\ \pf 0^{NP} \,.
   \label{c1}
\end{equation}
Based on our experience with the massive IIA case we do not
expect the above deformation to lead to a closed algebra.
Indeed, from
the $\{\pf 1, \pf 1, \pf{-1}\}$ Jacobi identity it follows that one is
led to extend the algebra with a new generator that transforms in the
symmetric  $(1,1)$ representation.
The additional commutators take the form
 \begin{align}
  [ \pf 1^M_\mu , \pf 1^N_\nu ]
    & = 2 f^{MN}{}_P \pf{(1,1)}^P_{\mu \nu} \,, \notag \\
  [ \pf{(1,1)}^M_{\mu \nu}, - \pf 1^\tau ]
    & = \delta_{(\mu}^\tau \pf 1_{\nu)}^M \,.
    \label{extra-comm}
 \end{align}
The above Jacobi identity then vanishes.

Unlike in the massive IIA case, there are additional
non-trivial Jacobi identities, for example of the
form $\{\pf 1, \pf{-1}, \pf{(1,1)}\}$.
To satisfy these one needs to introduce additional symmetric three-index
tensor generators with commutator relations similar to
\eqref{extra-comm}. Subsequently one finds that there are Jacobi
identities involving the symmetric three-index tensors,
that require the introduction of symmetric four-index tensors.
This iterative procedure does not terminate.
In terms of the local gauge parameters $\lambda^M$ of the vector
transformations, the new symmetries can be understood as the expansion
 \begin{align}
   \lambda^M = \Lambda^M_\mu x^\mu + \Lambda^M_{\mu, \nu} x^\mu x^\nu +
\ldots \,,
 \end{align}
where $\Lambda^M_\mu$ and $\Lambda_{\mu, \nu}$ are the parameters
corresponding to the $\pf 1^M$ and $\pf{(1,1)}^M$ generators,
respectively. Hence it appears that the gauge transformations of gauged half-maximal supergravities
can only be truncated to an infinite number of generators.

It would be interesting to see if there exist an interpretation (or modification) of the approaches
\cite{Schnakenburg:2002xx, Riccioni:2007au, Lavrinenko:1999xi, Kleinschmidt:2004dy} that
can reproduce all the results that follow from the embedding tensor method \cite{Nicolai:2001sv,
deWit:2002vt,deWit:2003hr}.


\section{Conclusions}\label{sec:conclusions}

In the first part of this paper we have refined the correspondence between
the Kac--Moody spectrum of deformation and top--form potentials and the
gaugings and massive deformations of the associated supergravity.
It was shown that there is a truncation of the Kac--Moody algebra
to a Lie algebra of $p$--forms, which encodes all the relevant information
for the physical states (apart from gravity and scalars) plus the
non-propagating deformation and top--form potentials.
A special role is played by the fundamental $p$--forms, from which all
other potentials can be constructed via commutators.
In particular, one has commutators of the form \eqref{deform-commutator}
giving rise to the $(D-1)$--forms, from which the corresponding type of
supergravity deformation can be deduced. In addition, the $p$--form
algebra contains commutators leading to $D$--forms, and these may be
associated to quadratic constraints on the deformation
parameters. We should stress that the properties derived from
\eqref{deform-commutator} and the relation to the quadratic constraints
are empirical observations. It would be interesting to understand
how these follow from the bosonic gauge transformations of supergravity.

In the second part we have established that the correspondence also holds
for half--maximal supergravity. In particular, in table \ref{halfmax}
the spectrum of deformation and top--form potentials of the associated
Kac--Moody algebras is summarized. These possibilities agree perfectly
with the known gaugings and massive deformations of half--maximal
supergravity and the ensuing quadratic constraints, respectively.
In addition it gives a prediction for the most general gaugings in
$6 \leq D \leq 10$: these are encoded in a fundamental and three-form
representation of the duality groups subject to the quadratic
constraints \eqref{quadcon}.

Note that we have only realized a finite-dimensional part of the
Kac-Moody algebra as a symmetry. However, in different dimensions,
this $p$--form algebra constitutes a different truncation of the
Kac-Moody algebra. The latter contains all symmetry groups of
half-maximal supergravity in $D$ dimensions.
This shows how the very extended Kac--Moody algebra  $SO(8,8+n)^{+++}$
plays a unifying role in describing the symmetries of half--maximal
supergravity coupled to $10-D+n$ vector multiplets.

Finally, we considered the effect of the deformation itself on the
$p$--form algebra. It was found that in the deformed case, the bosonic
gauge algebra can not be truncated to a $p$--form algebra. Instead, to
obtain a closed algebra, one needs to include additional generators
with mixed symmetries whose role from the Kac-Moody point of view
remains to be clarified.

In addition to the open issues mentioned above, we see a number of
interesting venues to extend the present results. First of all, a
relevant question is whether the above correspondence,
which holds for maximal and half--maximal supergravity, can also
be extended to theories with less supersymmetry. A number of such
supergravities are given by a scalar coset $G / K(G)$ after reduction
to three dimensions. Restricting to groups $G$ with a
real split form these cosets have been classified \cite{Cremmer:1999du}.
It is natural to investigate whether the over and very extensions of $G$
contain deformation and top--form potentials corresponding to all
deformations of the associated supergravities as well.
Furthermore, these coset models $G / K(G)$ are special points in a
landscape of more general geometries. It would be
interesting to learn more about the deformation and top--form potentials
associated to the general non--coset geometries.

Recently, an example where the correspondence between supergravity and very extended algebras does not
 hold straightforwardly was found in the theory that reduces to the coset model $G_2 / SO(4)$ in three
dimensions. While minimal $D=5$ simple
supergravity allows for a triplet of deformation potentials, related to the gauging of a $U(1)$ subgroup
of the $SU(2)$ R-symmetry, there are no such potentials in the associated
Kac--Moody algebra $G_2^{+++}$ \cite{Gomis:2007gb}. A possible explanation for this phenomenon
may be that in this case the R--symmetry does not act on the original bosonic fields of the theory
\cite{Riccioni:2007hm}. Another possibility may be that there is an extension of $G_2^{+++}$ that does take
the gauging into account. It would be worthwhile to find more examples of this phenomenon and to
understand it in more detail.

It would also be interesting to study  the brane interpretation of the
deformation  and top--form potentials. They naturally couple
to domain walls and space--filling branes, respectively. It is known
that in IIA supergravity the deformation potential $\pf 9$ couples
to the half-supersymmetric D8--brane and that the top--form
potential $\pf {10}^\prime$ couples to a
half-supersymmetric space-filling brane whose string interpretation
has yet to be clarified \cite{Bergshoeff:2006qw}. The other top-form
potential $\pf {10}$ couples to a non-supersymmetric space-filling brane.
Similarly, the quadruplet $\pf 4$ of top--form potentials of
IIB supergravity couples to
a half--supersymmetric nonlinear doublet of 9--branes, including the
D9--brane \cite{Bergshoeff:2006ic}.
The doublet $\pf 2$ of top--form potentials couples to
half--supersymmetric space--filling branes whose string interpretation is yet
unclear. It would be interesting to perform a similar analysis for the
other dimensions as well and see how all these branes fit into string theory.

Furthermore, while in this paper the possibilities of adding
deformation and top--form potentials to matter coupled
supergravity theories have been discussed, one may ask whether matter
multiplets not coupled to supergravity can be extended with such potentials as well. It turns
out that this is indeed the case. In fact, it has been suggested that
a domain wall structure on a D--brane, interpolating between different values of
the brane tension, should be described by a worldvolume deformation
potential \cite{Bergshoeff:1998ys}, similar to the way strings ending on such a brane are
described by a worldvolume vector. In the case of the D9--brane
this means that the $D=10$ Maxwell multiplet can be extended with a
nine--form potential, which is indeed possible \cite{Bergshoeff:1998ys}.
This could correspond to the fundamental
representation of deformation forms in table
\ref{halfmax}.
This fundamental representation does not correspond to
a deformation of supergravity due to the
third quadratic constraint in \eqref{quadcon}.
We expect that all fundamental representations in table \ref{halfmax}
correspond to possible extensions of the $D < 10$ vector multiplets with
deformation and top--form potentials as well.
It might be worthwhile to consider the brane interpretation of
these possibilities in further detail.


\acknowledgments

We acknowledge useful discussions with Paul Cook, Mees de Roo, Olaf Hohm,
Axel Kleinschmidt, Fabio Riccioni, Henning Samtleben, Ergin Sezgin,
Kelly Stelle, Mario Trigiante, and Peter West.
We thank the Galileo Galilei Institute for Theoretical
Physics in Firenze for its hospitality and the INFN for partial support.
E.B.~and T.N.~are supported by the European Commission FP6 program
MRTN-CT-2004-005104 in which E.B.~ and T.N.~ are associated to Utrecht
University. The work of E.B.~is partially supported by the Spanish
grant BFM2003-01090 and  by a Breedte Strategie grant of the University of Groningen.
The work of T.N. is supported by the FOM programme Nr. 57
\emph{String Theory and Quantum Gravity}.
J.G.~and D.R.~have been supported by the European EC-RTN project
MRTN-CT-2004-005104, MCYT FPA 2004-04582-C02-01 and CIRIT GC
2005SGR-00564.



\appendix

\section{Terminology and notation} \label{app:defs}

Below we shortly summarize the terminology we have introduced in this paper.

\begin{description}
    \item[Deformation potential]
        A $(D-1)$--form potential in $D$ dimensions.
    \item[Top--form potential]
        A $D$--form potential in $D$ dimensions.
    \item[$p$--form algebra]
        Truncation of the Kac--Moody algebra in a particular $G_D \otimes SL(D,\R)$ decomposition
        by restricting to
        only the generators at positive levels in a purely antisymmetric $SL(D,\R)$
        tensor representation of rank $1 \leq p \leq D$.
        Also arises by considering the bosonic gauge algebra with constant
        gauge parameters of the associated supergravity.
    \item[Fundamental $p$--form]
        A $p$-form corresponding to a positive simple root of one of the disabled nodes
        in the decomposed Dynkin diagram of the Kac-Moody algebra.
        Generates the $p$--form algebra.
    \item[Type p deformation]
        A deformation of the $p$--form algebra in which a fundamental
        $p$--form becomes massive.
\end{description}
Furthermore we indicate components of $p$--forms by a boldface
italic number equal to
their rank, e.g.~$\pf 5$ stands for $\pf 5_{\mu_1\cdots \mu_5}$.
This is not to be confused with
group representations, which are represented with a boldface number equal to
their dimension. Our convention for the normalisation of products of $p$--forms
is the same as in \cite{Bergshoeff:2006qw}; in particular, we (anti-)symmetrize
 with weight one.

\section{Physical states of half--maximal supergravity} \label{app:half--max}

We consider half--maximal supergravity in any
dimension and coupled to an arbitrary number of vector multiplets.
Starting with the graviton multiplet of $D$-dimensional half--maximal
supergravity, its bosonic part consists of a metric, $m$ vector
gauge fields with $m = 10 -D$, a two-form gauge field and a single
scalar which is the dilaton. It has a global $SO(m)$ symmetry, under
which the vectors transform in the fundamental representation. The
only exceptions are $D=4$ and $D=3$ where there are non-trivial
hidden symmetries. In $D=4$ there is one extra scalar due to the
duality of the two-form potential to an axionic scalar. Together
with the dilaton this leads to an enhanced $SL(2,\mathbb{R}) \times
SO(6)$ hidden symmetry. Similarly, in $D=3$ there are 7 extra
scalars due to the duality in $D=3$ dimensions between the vectors
and scalars. In this case all physical degrees of freedom are
carried by a scalar coset $SO(8,1)/SO(8)$.

The other possible multiplet in generic dimensions is the vector multiplet,
which contains a vector and $m$ scalars. The effect of
adding $m+n$ vector multiplets is to enlarge the symmetry group from
$SO(m)$ to $SO(m,m+n)$. The scalars parameterize the corresponding
scalar coset while the vectors transform in the fundamental
representation. In four dimensions the symmetry becomes
$SL(2,\mathbb{R}) \times SO(6,6+n)$ while in three dimensions it is
given by $SO(8,8+n)$. In the latter case there again is symmetry
enhancement due to the equivalence between scalars and vectors. The
entire theory can be described in terms of the corresponding scalar
coset (coupled to gravity).

\TABLE[ht]{
    \begin{tabular}{|c|c|*{4}{r@{\ $\times$ }l|}}
        \hline
        $D$ & Cont
            & \multicolumn{2}{c|}{$g_{\mu\nu}$} & \multicolumn{2}{c|}{$p = 0$}
            & \multicolumn{2}{c|}{$p = 1$}      & \multicolumn{2}{c|}{$p = 2$} \\

        \hline
        \hline

        10  & $G V^{n}$
            & \bf 35    & \bf 1                 & \bf 1     & 1
            & \bf 8     & \bi n                 & \bf 28    & \bf 1 \\

        9   & $G V^{n+1}$
            & \bf 27    & \bf 1                 & \bf 1     & $\Bigl( 1 + (1n + 1) \Bigr)$
            & \bf 7     & (\bi n + \textbf 2)   & \bf 21    & \bf 1 \\

        8   & $G V^{n+2}$
            & \bf 20    & \bf 1                 & \bf 1     & $\Bigl( 1 + (2n + 4) \Bigr)$
            & \bf 6     & (\bi n + \textbf 4)   & \bf 15    & \bf 1 \\

        7   & $G V^{n+3}$
            & \bf 14    & \bf 1                 & \bf 1     & $\Bigl( 1 + (3n + 9) \Bigr)$
            & \bf 5     & (\bi n + \textbf 6)   & \bf 10    & \bf 1 \\

        6a  & $G V^{n+4}$
            & \bf 9     & \bf 1                 & \bf 1     & $\Bigl( 1 + (4n + 16) \Bigr)$
            & \bf 4     & (\bi n + \textbf 8)   & \bf 6     & \bf 1 \\

        6b  & $G T^{n+4}$
            & \bf 9     & \bf 1                 & \bf 1     & $(5n + 25)$
            & \multicolumn{2}{c|}{}             & \bf 6     & $\frac 1 2$ (\textbf{10} + \textbf n) \\

        5   & $G V^{n+5}$
            & \bf 5     & \bf 1                 & \bf 1     & $\Bigl( 1 + (5n + 25) \Bigr)$
            & \bf 3     & (\bi n + \textbf{11}) & \multicolumn{2}{c|}{} \\

        4   & $G V^{n+6}$
            & \bf 2     & \bf 1                 & \bf 1     & $\Bigl( (1,2) + (6n + 36,1) \Bigr)$
            & \bf 2     & (\bi n + \textbf{12}) & \multicolumn{2}{c|}{} \\

        3   & $G V^{n+7}$
            & --        & \bf 1                 & \bf 1         & $(8n + 64)$
            & \multicolumn{2}{c|}{}             & \multicolumn{2}{c|}{} \\

        \hline
    \end{tabular}
    \caption{
        The physical states of all $D=10-m$ half--maximal  supergravities
        coupled to $m+n$ vector multiplets. The multiplet structures
        (where $G$ is the graviton, $V$ the vector and $T$ the self-dual
        tensor multiplet) are also given.
    }
    \label{tableii}
}

The above multiplets belong to non-chiral half--maximal supergravity
and are the correct and complete story in generic dimensions. In six
dimensions, however, the half--maximal theory can be chiral or
non-chiral, similar to the maximal theory in ten dimensions. The
non-chiral theory is denoted by $D=6a$ and follows the above
pattern. The chiral theory, $D=6b$, instead has different
multiplets. In particular, the graviton multiplet contains gravity,
five scalars and five self-dual plus one anti-self-dual two-form
gauge fields. The global symmetry is given by $SO(5,1)$. The other
possible multiplet is that of the tensor, which contains an
anti-self-dual two-form and five scalars. Adding $4+n$ of such
tensor multiplets to the graviton multiplet enhances the symmetry to
$SO(5,5+n)$.

Upon dimensional reduction over a circle, the graviton multiplet
splits up into a graviton multiplet plus a vector multiplet. A
vector (or tensor) multiplet reduces to a vector multiplet in the
lower dimensions. This was the reason for adding $m+n$ instead of
$n$ vector or tensor multiplets in any dimension; it can easily be
seen that $n$ remains invariant under dimensional reduction. That
is, a theory with a certain value of $n$ reduces to a theory with
the same value of $n$ in lower dimensions.

For the reader's convenience we have given the physical states corresponding to
$D=10-m$ half--maximal supergravity coupled to $m + n$ vector (or
tensor) multiplets in table \ref{tableii}, see, e.g.,
\cite{Tanii:1998px}.

\section{Group theory} \label{app:group_theory}

In this appendix we will generalize the analysis of \cite{Bergshoeff:2007qi} to
allow for non--simply laced Dynkin diagrams. The key difference between a simply
laced and a non--simply laced diagram is that for the latter the associated Cartan
matrix is not symmetric, and no longer fulfills the role of a metric on
the root space. Moreover, the metric on the weight space is no longer given
by the inverse of the Cartan matrix.

The root space metric is important in constructing the root system -- one needs
it to compute inner products between roots. The weight space metric plays a
similar role for the highest weight representations, which are a necessary
ingredient for the level decomposition. We will show how both metrics can be
obtained from appropriate symmetrizations of the (inverse) Cartan matrix.

We start out from the defining equation for the Cartan matrix, which reads

\be
    A_{ij} = 2 \frac{\norm{\alpha_i}{\alpha_j}}{\norm{\alpha_j}{\alpha_j}} \,.
\ee
Here $\alpha_i$ are the simple roots which span the whole root system $\Delta$,
and $\norm{\cdot}{\cdot}$ is the norm inferred from the Killing norm.
The indices run over the rank of the associated Lie algebra.

Any root $\alpha$ of $\Delta$ can be expressed as a linear combination of simple
roots,

\be
    \alpha = m^i \alpha_i \,,
\ee
where contracted indices are being summed over. The values of $m^i$ are also
known as the root labels.

Because the Killing norm is symmetric and bilinear, an inner product between two
roots $\alpha = m^i \alpha_i$ and $\beta = n^i \alpha_i$ can be written as
\be
\label{root_product}
    \norm{\alpha}{\beta} = B_{ij} m^i n^j \,,
\ee
where the metric $B$ on the root space is defined as

\be
\label{root_metric}
    B_{ij}
    \equiv \norm{\alpha_i}{\alpha_j}
    = A_{ij} \frac{\norm{\alpha_j}{\alpha_j}}{2} \,,
\ee
which is symmetric by construction.
Note that in this case the repeated index is not summed over because it is not
contracted.

From \eqref{root_product} and \eqref{root_metric} it is apparent that we must
first determine the norms of the simple roots before inner products on $\Delta$
can be computed. To that end we reshuffle the defining equation for the
Cartan matrix to obtain

\be
    \norm{\alpha_i}{\alpha_i} = \frac{A_{ij}}{A_{ji}} \norm{\alpha_j}{\alpha_j} \,.
\ee
So once a normalization for one of the simple roots has been chosen, all others
are also fixed. A common normalization is to choose $\alpha^2 = 2$ for the
longest simple root (i.e. the simple root which has the highest norm).
Instead, we will adhere to $\alpha^2 = 2$ for the \emph{shortest} simple root
(the simple root with the lowest norm). The latter normalization is particularly
convenient for computer-based calculations, because then the root metric $B$
has only integer values.

We now turn to the metric on the weight space. The weight space itself is
spanned by the fundamental weights $\lambda^i$, which are defined via

\be
 \lambda^i \bar \alpha_j = \delta^i_j \,,
\ee
where the simple coroots $\bar \alpha_i$ are given by
$\bar \alpha_i = \frac{2 \alpha_i}{\norm{\alpha_i}{\alpha_i}}$.
The basis specified by the fundamental weights is also known as the Dynkin
basis. Every weight $\lambda$ can be expanded on this basis as

\be
\lambda = p_i \lambda^i \,.
\ee
The values of the $p_i$ are also known as the Dynkin labels of the weight.
The relation between the Dynkin labels and the components of the root
is given by

\be
    \label{dynkin_labels} p_i = A_{ji} m^j \,.
\ee
As the Dynkin basis is the dual basis of the simple coroots, the metric
$G$ on the weight space is the inverse of the simple coroot metric.
The latter is given by

\be
    {\norm{\bar\alpha_i}{\bar\alpha_j}}
        = \frac{2 A_{ij}}{{\norm{\alpha_i}{\alpha_i}}} \,.
\ee
Therefore $G$ is given by

\be
    G^{ij}
    \equiv {\norm{\lambda_i}{\lambda_j}}
    =\frac{\norm{\alpha_j}{\alpha_j}}{2} \left( A^{-1} \right)^{ij} .
\ee
By construction $G$ is symmetric, just like the root metric $B$.

As explained in \cite{Nicolai:2003fw,Kleinschmidt:2003mf}, the level decomposition of
infinite-dimensional Lie algebra entails scanning for subalgebra representations at given levels.
The subalgebra representations are defined by their Dynkin labels, and have to
satisfy three conditions:

\begin{enumerate}[label=(\roman*)]
    \item The Dynkin labels all have to be integer and non-negative.
    \item \label{A:cond2} The associated root labels have to be integers.
    \item \label{A:cond3} The length squared of the root must not exceed the maximum value.
\end{enumerate}

The subalgebra is obtained by `disabling' nodes from the Dynkin diagram.
We can then split up the index of the full algebra into $i = (a,s)$, where $a$
runs over the disabled nodes and $s$ over the subalgebra.
To see whether condition \ref{A:cond2} is satisfied for particular values of
$p_s$, we can invert equation \eqref{dynkin_labels} in order to obtain

\be
\label{sub_root_labels}
m^s = \left( A^{-1}_{\text{sub}} \right)^{ts} \left( p_t - l^a A_{at} \right) ,
\ee
where $m^s$ are the root labels associated to the Dynkin labels $p_s$,
$A_{\text{sub}}$ is the Cartan matrix of the subalgebra, and $l^a$ are the levels.
Condition \ref{A:cond3} may be verified by decomposing \eqref{root_product}
into its contributions from the deleted nodes and the subalgebra:

\be
\label{root_squared}
\alpha^2
    = G_{\text{sub}}^{st} \left( p_s p_t - A_{as} A_{bt} l^a l^b \right)
    + B_{ab} l^a l^b \leq \alpha^2_{\text{max}} \,.
\ee
Here $G_{\text{sub}}$ is the weight metric of the subalgebra, and $\alpha^2_{\text{max}}$
is given by the norm of the longest simple root.
Note that for this formula to be valid, we have to make sure that a long
(or short) root in the full algebra is also a long (short) root in the
subalgebra, which in general is not automatically the case. Luckily we are
always free to choose a normalization such that root lengths match.

When using \eqref{root_squared} to scan for representations, it is important for
$G_{\text{sub}}$ to only have non-negative entries. If this is not the case,
then the root norm $\alpha^2$ is not a monotonically increasing function of the
Dynkin labels $p_s$ at fixed levels $l^a$, and one might miss representations
using a simple scanning algorithm. However, as we always shall be decomposing
with respect to (direct products of) finite dimensional subalgebras,
$G_{\text{sub}}$ will never contain negative entries.

\section{Low level $D_8^{+++}$ decompositions} \label{app:results}

Here we list the output of SimpLie \cite{SimpLie} at low levels, using
the various decompositions of $D_8^{+++}$ as indicated by the Dynkin
diagram accompanying the tables.
The regular subalgebra splits into a part belonging to the gravity line
$A_n$ (the white nodes) and a part belonging to the internal duality
group $G_D$ (the grey nodes).

In the following tables we respectively list the levels, the Dynkin
labels of $A_n$ and $G_D$, the root labels, the root length, the
dimension of the representations of $A_n$ and $G$, the multiplicity of
the root, the outer multiplicity, and the interpretation as a physical
field.
The deformation-- and top--form potentials are indicated by `de' and `top',
respectively. When the internal group does not exist, we do not list
the corresponding columns. In all cases the Dynkin labels of the lowest
weights of the representations are given. All tables are truncated at
the point when the number of indices of the gravity subalgebra
representations exceed the dimension. The order of the levels, Dynkin
labels, and root labels as they appear in the tables are determined by
the order of the node labels on the Dynkin diagram. This ordering is
always first from left to right, then from top to bottom.

The interpretation of the representations at level zero as the graviton is,
unlike the $p$-forms at higher levels, not quite straightforward. The graviton
emerges when one combines the adjoint representation of $A_n$ with a scalar
coming from one of the disabled nodes, see \cite{Damour:2002cu,Kleinschmidt:2003mf}.
We have indicated these parts of the graviton by $\bar g_{\mu\nu}$ and
$\hat g_{\mu\nu}$, respectively.

\vspace{2cm}

\setlongtables
\footnotesize
\psset{unit=0.8cm,radius=.2}
\begin{figure}[H]
\begin{center}
\begin{pspicture}(0,0)(8,1)
\disabledNode{1,1}{N11784499420}
\nodeLabel{N11784499420}{1}
\disabledNode{5,1}{N21784499420}
\nodeLabel{N21784499420}{2}
\normalNode{0,0}{N31784499420}
\nodeLabel{N31784499420}{3}
\normalNode{1,0}{N41784499420}
\nodeLabel{N41784499420}{4}
\normalNode{2,0}{N51784499420}
\nodeLabel{N51784499420}{5}
\normalNode{3,0}{N61784499420}
\nodeLabel{N61784499420}{6}
\normalNode{4,0}{N71784499420}
\nodeLabel{N71784499420}{7}
\normalNode{5,0}{N81784499420}
\nodeLabel{N81784499420}{8}
\normalNode{6,0}{N91784499420}
\nodeLabel{N91784499420}{9}
\normalNode{7,0}{N101784499420}
\nodeLabel{N101784499420}{10}
\normalNode{8,0}{N111784499420}
\nodeLabel{N111784499420}{11}
\singleConnection{N31784499420}{N41784499420}
\singleConnection{N41784499420}{N51784499420}
\singleConnection{N51784499420}{N61784499420}
\singleConnection{N61784499420}{N71784499420}
\singleConnection{N71784499420}{N81784499420}
\singleConnection{N81784499420}{N91784499420}
\singleConnection{N91784499420}{N101784499420}
\singleConnection{N101784499420}{N111784499420}
\singleConnection{N81784499420}{N21784499420}
\singleConnection{N41784499420}{N11784499420}
\end{pspicture}
\end{center}
\caption{$D_{8}^{+++}$ decomposed as $A_{9}^{}$}
\end{figure}
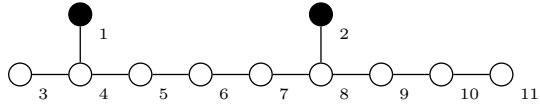

\begin{longtable}{|r@{\ }r|r@{\ }r@{\ }r@{\ }r@{\ }r@{\ }r@{\ }r@{\ }r@{\ }r|r@{\ }r@{\ }r@{\ }r@{\ }r@{\ }r@{\ }r@{\ }r@{\ }r@{\ }r@{\ }r|r|r|r|r|c|}
\caption{$A_{9}^{}$ representations in $D_{8}^{+++}$} \\
\hline
\multicolumn{2}{|c|}{$l$} &
\multicolumn{9}{|c|}{$p_{\rm{grav}}$} &
\multicolumn{11}{|c|}{$m$} &
\multicolumn{1}{|c|}{$\alpha^2$} &
\multicolumn{1}{|c|}{$d_{\rm{grav}}$} &
\multicolumn{1}{|c|}{$\textrm{mult}(\alpha)$} &
\multicolumn{1}{|c|}{$\mu$} &
\multicolumn{1}{|c|}{fields}\\
\hline
\hline
0 & 0 & 1 & 0 & 0 & 0 & 0 & 0 & 0 & 0 & 1 & 0 & 0 & -1 & -1 & -1 & -1 & -1 & -1 & -1 & -1 & -1 & 2 & 99 & 1 & 1 & $\bar g_{\mu\nu}$\\
0 & 0 & 0 & 0 & 0 & 0 & 0 & 0 & 0 & 0 & 0 & 0 & 0 & 0 & 0 & 0 & 0 & 0 & 0 & 0 & 0 & 0 & 0 & 1 & 11 & 2 & $\pf 0, \hat g_{\mu\nu}$\\
\hline
1 & 0 & 0 & 1 & 0 & 0 & 0 & 0 & 0 & 0 & 0 & 1 & 0 & 0 & 0 & 0 & 0 & 0 & 0 & 0 & 0 & 0 & 2 & 45 & 1 & 1 & $\pf 2$\\
\hline
0 & 1 & 0 & 0 & 0 & 0 & 0 & 1 & 0 & 0 & 0 & 0 & 1 & 0 & 0 & 0 & 0 & 0 & 0 & 0 & 0 & 0 & 2 & 210 & 1 & 1 & ${}^\star\, \pf 2$\\
\hline
1 & 1 & 1 & 0 & 0 & 0 & 0 & 0 & 1 & 0 & 0 & 1 & 1 & 0 & 1 & 1 & 1 & 1 & 1 & 0 & 0 & 0 & 2 & 1155 & 1 & 1 & ${}^\star\,  g_{\mu\nu}$\\
1 & 1 & 0 & 0 & 0 & 0 & 0 & 0 & 0 & 1 & 0 & 1 & 1 & 1 & 2 & 2 & 2 & 2 & 2 & 1 & 0 & 0 & 0 & 45 & 8 & 1 & ${}^\star\, \pf 0$\\
\hline
2 & 1 & 0 & 1 & 0 & 0 & 0 & 0 & 0 & 1 & 0 & 2 & 1 & 1 & 2 & 2 & 2 & 2 & 2 & 1 & 0 & 0 & 2 & 1925 & 1 & 1 & \\
2 & 1 & 1 & 0 & 0 & 0 & 0 & 0 & 0 & 0 & 1 & 2 & 1 & 1 & 3 & 3 & 3 & 3 & 3 & 2 & 1 & 0 & 0 & 99 & 8 & 1 & \\
2 & 1 & 0 & 0 & 0 & 0 & 0 & 0 & 0 & 0 & 0 & 2 & 1 & 2 & 4 & 4 & 4 & 4 & 4 & 3 & 2 & 1 & -2 & 1 & 45 & 1 & top\\
\hline
\end{longtable}

\begin{figure}[H]
\begin{center}
\begin{pspicture}(0,0)(8,1)
\disabledNode{1,1}{N12116776400}
\nodeLabel{N12116776400}{1}
\disabledNode{5,1}{N22116776400}
\nodeLabel{N22116776400}{2}
\disabledNode{0,0}{N32116776400}
\nodeLabel{N32116776400}{3}
\normalNode{1,0}{N42116776400}
\nodeLabel{N42116776400}{4}
\normalNode{2,0}{N52116776400}
\nodeLabel{N52116776400}{5}
\normalNode{3,0}{N62116776400}
\nodeLabel{N62116776400}{6}
\normalNode{4,0}{N72116776400}
\nodeLabel{N72116776400}{7}
\normalNode{5,0}{N82116776400}
\nodeLabel{N82116776400}{8}
\normalNode{6,0}{N92116776400}
\nodeLabel{N92116776400}{9}
\normalNode{7,0}{N102116776400}
\nodeLabel{N102116776400}{10}
\normalNode{8,0}{N112116776400}
\nodeLabel{N112116776400}{11}
\singleConnection{N32116776400}{N42116776400}
\singleConnection{N42116776400}{N52116776400}
\singleConnection{N52116776400}{N62116776400}
\singleConnection{N62116776400}{N72116776400}
\singleConnection{N72116776400}{N82116776400}
\singleConnection{N82116776400}{N92116776400}
\singleConnection{N92116776400}{N102116776400}
\singleConnection{N102116776400}{N112116776400}
\singleConnection{N82116776400}{N22116776400}
\singleConnection{N42116776400}{N12116776400}
\end{pspicture}
\end{center}
\caption{$D_{8}^{+++}$ decomposed as $A_{8}^{}$}
\end{figure}

\begin{longtable}{|r@{\ }r@{\ }r|r@{\ }r@{\ }r@{\ }r@{\ }r@{\ }r@{\ }r@{\ }r|r@{\ }r@{\ }r@{\ }r@{\ }r@{\ }r@{\ }r@{\ }r@{\ }r@{\ }r@{\ }r|r|r|r|r|c|}
\caption{$A_{8}^{}$ representations in $D_{8}^{+++}$} \\
\hline
\multicolumn{3}{|c|}{$l$} &
\multicolumn{8}{|c|}{$p_{\rm{grav}}$} &
\multicolumn{11}{|c|}{$m$} &
\multicolumn{1}{|c|}{$\alpha^2$} &
\multicolumn{1}{|c|}{$d_{\rm{grav}}$} &
\multicolumn{1}{|c|}{$\textrm{mult}(\alpha)$} &
\multicolumn{1}{|c|}{$\mu$} &
\multicolumn{1}{|c|}{fields}\\
\hline
\hline
0 & 0 & 0 & 1 & 0 & 0 & 0 & 0 & 0 & 0 & 1 & 0 & 0 & 0 & -1 & -1 & -1 & -1 & -1 & -1 & -1 & -1 & 2 & 80 & 1 & 1 & $\bar g_{\mu\nu}$\\
0 & 0 & 0 & 0 & 0 & 0 & 0 & 0 & 0 & 0 & 0 & 0 & 0 & 0 & 0 & 0 & 0 & 0 & 0 & 0 & 0 & 0 & 0 & 1 & 11 & 3 & $\pf 0, \hat g_{\mu\nu}$\\
\hline
1 & 0 & 0 & 1 & 0 & 0 & 0 & 0 & 0 & 0 & 0 & 1 & 0 & 0 & 0 & 0 & 0 & 0 & 0 & 0 & 0 & 0 & 2 & 9 & 1 & 1 & $\pf 1$\\
\hline
0 & 1 & 0 & 0 & 0 & 0 & 0 & 1 & 0 & 0 & 0 & 0 & 1 & 0 & 0 & 0 & 0 & 0 & 0 & 0 & 0 & 0 & 2 & 126 & 1 & 1 & ${}^\star\, \pf 2$\\
\hline
0 & 0 & 1 & 1 & 0 & 0 & 0 & 0 & 0 & 0 & 0 & 0 & 0 & 1 & 0 & 0 & 0 & 0 & 0 & 0 & 0 & 0 & 2 & 9 & 1 & 1 & $\pf 1$\\
\hline
1 & 1 & 0 & 0 & 0 & 0 & 0 & 0 & 1 & 0 & 0 & 1 & 1 & 0 & 1 & 1 & 1 & 1 & 1 & 0 & 0 & 0 & 2 & 84 & 1 & 1 & ${}^\star\, \pf 1$\\
\hline
1 & 0 & 1 & 0 & 1 & 0 & 0 & 0 & 0 & 0 & 0 & 1 & 0 & 1 & 1 & 0 & 0 & 0 & 0 & 0 & 0 & 0 & 2 & 36 & 1 & 1 & $\pf 2$\\
\hline
0 & 1 & 1 & 0 & 0 & 0 & 0 & 0 & 1 & 0 & 0 & 0 & 1 & 1 & 1 & 1 & 1 & 1 & 1 & 0 & 0 & 0 & 2 & 84 & 1 & 1 & ${}^\star\, \pf 1$\\
\hline
1 & 1 & 1 & 1 & 0 & 0 & 0 & 0 & 1 & 0 & 0 & 1 & 1 & 1 & 1 & 1 & 1 & 1 & 1 & 0 & 0 & 0 & 2 & 720 & 1 & 1 & ${}^\star\,  g_{\mu\nu}$\\
1 & 1 & 1 & 0 & 0 & 0 & 0 & 0 & 0 & 1 & 0 & 1 & 1 & 1 & 2 & 2 & 2 & 2 & 2 & 1 & 0 & 0 & 0 & 36 & 8 & 2 & ${}^\star\, \pf 0$\\
\hline
2 & 1 & 1 & 1 & 0 & 0 & 0 & 0 & 0 & 1 & 0 & 2 & 1 & 1 & 2 & 2 & 2 & 2 & 2 & 1 & 0 & 0 & 2 & 315 & 1 & 1 & \\
2 & 1 & 1 & 0 & 0 & 0 & 0 & 0 & 0 & 0 & 1 & 2 & 1 & 1 & 3 & 3 & 3 & 3 & 3 & 2 & 1 & 0 & 0 & 9 & 8 & 1 & de\\
\hline
1 & 1 & 2 & 1 & 0 & 0 & 0 & 0 & 0 & 1 & 0 & 1 & 1 & 2 & 2 & 2 & 2 & 2 & 2 & 1 & 0 & 0 & 2 & 315 & 1 & 1 & \\
1 & 1 & 2 & 0 & 0 & 0 & 0 & 0 & 0 & 0 & 1 & 1 & 1 & 2 & 3 & 3 & 3 & 3 & 3 & 2 & 1 & 0 & 0 & 9 & 8 & 1 & de\\
\hline
2 & 1 & 2 & 0 & 1 & 0 & 0 & 0 & 0 & 1 & 0 & 2 & 1 & 2 & 3 & 2 & 2 & 2 & 2 & 1 & 0 & 0 & 2 & 1215 & 1 & 1 & \\
2 & 1 & 2 & 1 & 0 & 0 & 0 & 0 & 0 & 0 & 1 & 2 & 1 & 2 & 3 & 3 & 3 & 3 & 3 & 2 & 1 & 0 & 0 & 80 & 8 & 2 & \\
2 & 1 & 2 & 0 & 0 & 0 & 0 & 0 & 0 & 0 & 0 & 2 & 1 & 2 & 4 & 4 & 4 & 4 & 4 & 3 & 2 & 1 & -2 & 1 & 45 & 2 & top\\
\hline
\end{longtable}

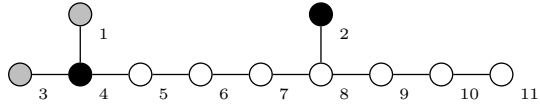
\begin{figure}[H]
\begin{center}
\begin{pspicture}(0,0)(8,1)
\dualityNode{1,1}{N1817052144}
\nodeLabel{N1817052144}{1}
\disabledNode{5,1}{N2817052144}
\nodeLabel{N2817052144}{2}
\dualityNode{0,0}{N3817052144}
\nodeLabel{N3817052144}{3}
\disabledNode{1,0}{N4817052144}
\nodeLabel{N4817052144}{4}
\normalNode{2,0}{N5817052144}
\nodeLabel{N5817052144}{5}
\normalNode{3,0}{N6817052144}
\nodeLabel{N6817052144}{6}
\normalNode{4,0}{N7817052144}
\nodeLabel{N7817052144}{7}
\normalNode{5,0}{N8817052144}
\nodeLabel{N8817052144}{8}
\normalNode{6,0}{N9817052144}
\nodeLabel{N9817052144}{9}
\normalNode{7,0}{N10817052144}
\nodeLabel{N10817052144}{10}
\normalNode{8,0}{N11817052144}
\nodeLabel{N11817052144}{11}
\singleConnection{N3817052144}{N4817052144}
\singleConnection{N4817052144}{N5817052144}
\singleConnection{N5817052144}{N6817052144}
\singleConnection{N6817052144}{N7817052144}
\singleConnection{N7817052144}{N8817052144}
\singleConnection{N8817052144}{N9817052144}
\singleConnection{N9817052144}{N10817052144}
\singleConnection{N10817052144}{N11817052144}
\singleConnection{N8817052144}{N2817052144}
\singleConnection{N4817052144}{N1817052144}
\end{pspicture}
\end{center}
\caption{$D_{8}^{+++}$ decomposed as $A_{1}^{} \otimes A_{1}^{} \otimes A_{7}^{}$}
\end{figure}

\begin{longtable}{|r@{\ }r|r@{\ }r@{\ }r@{\ }r@{\ }r@{\ }r@{\ }r|r@{\ }r|r@{\ }r@{\ }r@{\ }r@{\ }r@{\ }r@{\ }r@{\ }r@{\ }r@{\ }r@{\ }r|r|r|r|r|r|c|}
\caption{$A_{1}^{} \otimes A_{1}^{} \otimes A_{7}^{}$ representations in $D_{8}^{+++}$} \\
\hline
\multicolumn{2}{|c|}{$l$} &
\multicolumn{7}{|c|}{$p_{\rm{grav}}$} &
\multicolumn{2}{|c|}{$p_G$} &
\multicolumn{11}{|c|}{$m$} &
\multicolumn{1}{|c|}{$\alpha^2$} &
\multicolumn{1}{|c|}{$d_{\rm{grav}}$} &
\multicolumn{1}{|c|}{$d_G$} &
\multicolumn{1}{|c|}{$\textrm{mult}(\alpha)$} &
\multicolumn{1}{|c|}{$\mu$} &
\multicolumn{1}{|c|}{fields}\\
\hline
\hline
0 & 0 & 1 & 0 & 0 & 0 & 0 & 0 & 1 & 0 & 0 & 0 & 0 & 0 & 0 & -1 & -1 & -1 & -1 & -1 & -1 & -1 & 2 & 63 & 1 & 1 & 1 & $\bar g_{\mu\nu}$\\
0 & 0 & 0 & 0 & 0 & 0 & 0 & 0 & 0 & 0 & 2 & 0 & 0 & -1 & 0 & 0 & 0 & 0 & 0 & 0 & 0 & 0 & 2 & 1 & 3 & 1 & 1 & $\pf 0$\\
0 & 0 & 0 & 0 & 0 & 0 & 0 & 0 & 0 & 2 & 0 & -1 & 0 & 0 & 0 & 0 & 0 & 0 & 0 & 0 & 0 & 0 & 2 & 1 & 3 & 1 & 1 & $\pf 0$\\
0 & 0 & 0 & 0 & 0 & 0 & 0 & 0 & 0 & 0 & 0 & 0 & 0 & 0 & 0 & 0 & 0 & 0 & 0 & 0 & 0 & 0 & 0 & 1 & 1 & 11 & 2 & $\pf 0, \hat g_{\mu\nu}$\\
\hline
1 & 0 & 0 & 0 & 0 & 1 & 0 & 0 & 0 & 0 & 0 & 0 & 1 & 0 & 0 & 0 & 0 & 0 & 0 & 0 & 0 & 0 & 2 & 70 & 1 & 1 & 1 & ${}^\star\, \pf 2$\\
\hline
0 & 1 & 1 & 0 & 0 & 0 & 0 & 0 & 0 & 1 & 1 & 0 & 0 & 0 & 1 & 0 & 0 & 0 & 0 & 0 & 0 & 0 & 2 & 8 & 4 & 1 & 1 & $\pf 1$\\
\hline
1 & 1 & 0 & 0 & 0 & 0 & 1 & 0 & 0 & 1 & 1 & 0 & 1 & 0 & 1 & 1 & 1 & 1 & 1 & 0 & 0 & 0 & 2 & 56 & 4 & 1 & 1 & ${}^\star\, \pf 1$\\
\hline
2 & 0 & 1 & 0 & 0 & 0 & 0 & 0 & 1 & 0 & 0 & 0 & 2 & 0 & 0 & 0 & 1 & 2 & 3 & 2 & 1 & 0 & 2 & 63 & 1 & 1 & 1 & \\
\hline
0 & 2 & 0 & 1 & 0 & 0 & 0 & 0 & 0 & 0 & 0 & 1 & 0 & 1 & 2 & 1 & 0 & 0 & 0 & 0 & 0 & 0 & 2 & 28 & 1 & 1 & 1 & $\pf 2$\\
\hline
1 & 2 & 1 & 0 & 0 & 0 & 1 & 0 & 0 & 0 & 0 & 1 & 1 & 1 & 2 & 1 & 1 & 1 & 1 & 0 & 0 & 0 & 2 & 420 & 1 & 1 & 1 & ${}^\star\,  g_{\mu\nu}$\\
1 & 2 & 0 & 0 & 0 & 0 & 0 & 1 & 0 & 0 & 2 & 1 & 1 & 0 & 2 & 2 & 2 & 2 & 2 & 1 & 0 & 0 & 2 & 28 & 3 & 1 & 1 & ${}^\star\, \pf 0$\\
1 & 2 & 0 & 0 & 0 & 0 & 0 & 1 & 0 & 2 & 0 & 0 & 1 & 1 & 2 & 2 & 2 & 2 & 2 & 1 & 0 & 0 & 2 & 28 & 3 & 1 & 1 & ${}^\star\, \pf 0$\\
1 & 2 & 0 & 0 & 0 & 0 & 0 & 1 & 0 & 0 & 0 & 1 & 1 & 1 & 2 & 2 & 2 & 2 & 2 & 1 & 0 & 0 & 0 & 28 & 1 & 8 & 1 & ${}^\star\, \pf 0$\\
\hline
1 & 3 & 1 & 0 & 0 & 0 & 0 & 1 & 0 & 1 & 1 & 1 & 1 & 1 & 3 & 2 & 2 & 2 & 2 & 1 & 0 & 0 & 2 & 216 & 4 & 1 & 1 & \\
1 & 3 & 0 & 0 & 0 & 0 & 0 & 0 & 1 & 1 & 1 & 1 & 1 & 1 & 3 & 3 & 3 & 3 & 3 & 2 & 1 & 0 & 0 & 8 & 4 & 8 & 2 & de\\
\hline
1 & 4 & 0 & 1 & 0 & 0 & 0 & 1 & 0 & 0 & 0 & 2 & 1 & 2 & 4 & 3 & 2 & 2 & 2 & 1 & 0 & 0 & 2 & 720 & 1 & 1 & 1 & \\
1 & 4 & 1 & 0 & 0 & 0 & 0 & 0 & 1 & 0 & 2 & 2 & 1 & 1 & 4 & 3 & 3 & 3 & 3 & 2 & 1 & 0 & 2 & 63 & 3 & 1 & 1 & \\
1 & 4 & 1 & 0 & 0 & 0 & 0 & 0 & 1 & 2 & 0 & 1 & 1 & 2 & 4 & 3 & 3 & 3 & 3 & 2 & 1 & 0 & 2 & 63 & 3 & 1 & 1 & \\
1 & 4 & 1 & 0 & 0 & 0 & 0 & 0 & 1 & 0 & 0 & 2 & 1 & 2 & 4 & 3 & 3 & 3 & 3 & 2 & 1 & 0 & 0 & 63 & 1 & 8 & 1 & \\
1 & 4 & 0 & 0 & 0 & 0 & 0 & 0 & 0 & 0 & 2 & 2 & 1 & 1 & 4 & 4 & 4 & 4 & 4 & 3 & 2 & 1 & 0 & 1 & 3 & 8 & 1 & top\\
1 & 4 & 0 & 0 & 0 & 0 & 0 & 0 & 0 & 2 & 0 & 1 & 1 & 2 & 4 & 4 & 4 & 4 & 4 & 3 & 2 & 1 & 0 & 1 & 3 & 8 & 1 & top\\
1 & 4 & 0 & 0 & 0 & 0 & 0 & 0 & 0 & 0 & 0 & 2 & 1 & 2 & 4 & 4 & 4 & 4 & 4 & 3 & 2 & 1 & -2 & 1 & 1 & 45 & 2 & top\\
\hline
\end{longtable}

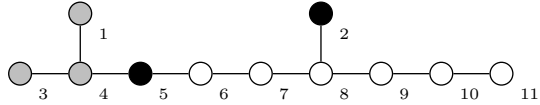
\begin{figure}[H]
\begin{center}
\begin{pspicture}(0,0)(8,1)
\dualityNode{1,1}{N1909476900}
\nodeLabel{N1909476900}{1}
\disabledNode{5,1}{N2909476900}
\nodeLabel{N2909476900}{2}
\dualityNode{0,0}{N3909476900}
\nodeLabel{N3909476900}{3}
\dualityNode{1,0}{N4909476900}
\nodeLabel{N4909476900}{4}
\disabledNode{2,0}{N5909476900}
\nodeLabel{N5909476900}{5}
\normalNode{3,0}{N6909476900}
\nodeLabel{N6909476900}{6}
\normalNode{4,0}{N7909476900}
\nodeLabel{N7909476900}{7}
\normalNode{5,0}{N8909476900}
\nodeLabel{N8909476900}{8}
\normalNode{6,0}{N9909476900}
\nodeLabel{N9909476900}{9}
\normalNode{7,0}{N10909476900}
\nodeLabel{N10909476900}{10}
\normalNode{8,0}{N11909476900}
\nodeLabel{N11909476900}{11}
\singleConnection{N3909476900}{N4909476900}
\singleConnection{N4909476900}{N5909476900}
\singleConnection{N5909476900}{N6909476900}
\singleConnection{N6909476900}{N7909476900}
\singleConnection{N7909476900}{N8909476900}
\singleConnection{N8909476900}{N9909476900}
\singleConnection{N9909476900}{N10909476900}
\singleConnection{N10909476900}{N11909476900}
\singleConnection{N8909476900}{N2909476900}
\singleConnection{N4909476900}{N1909476900}
\end{pspicture}
\end{center}
\caption{$D_{8}^{+++}$ decomposed as $A_{3}^{} \otimes A_{6}^{}$}
\end{figure}

\begin{longtable}{|r@{\ }r|r@{\ }r@{\ }r@{\ }r@{\ }r@{\ }r|r@{\ }r@{\ }r|r@{\ }r@{\ }r@{\ }r@{\ }r@{\ }r@{\ }r@{\ }r@{\ }r@{\ }r@{\ }r|r|r|r|r|r|c|}
\caption{$A_{3}^{} \otimes A_{6}^{}$ representations in $D_{8}^{+++}$} \\
\hline
\multicolumn{2}{|c|}{$l$} &
\multicolumn{6}{|c|}{$p_{\rm{grav}}$} &
\multicolumn{3}{|c|}{$p_G$} &
\multicolumn{11}{|c|}{$m$} &
\multicolumn{1}{|c|}{$\alpha^2$} &
\multicolumn{1}{|c|}{$d_{\rm{grav}}$} &
\multicolumn{1}{|c|}{$d_G$} &
\multicolumn{1}{|c|}{$\textrm{mult}(\alpha)$} &
\multicolumn{1}{|c|}{$\mu$} &
\multicolumn{1}{|c|}{fields}\\
\hline
\hline
0 & 0 & 1 & 0 & 0 & 0 & 0 & 1 & 0 & 0 & 0 & 0 & 0 & 0 & 0 & 0 & -1 & -1 & -1 & -1 & -1 & -1 & 2 & 48 & 1 & 1 & 1 & $\bar g_{\mu\nu}$\\
0 & 0 & 0 & 0 & 0 & 0 & 0 & 0 & 1 & 1 & 0 & -1 & 0 & -1 & -1 & 0 & 0 & 0 & 0 & 0 & 0 & 0 & 2 & 1 & 15 & 1 & 1 & $\pf 0$\\
0 & 0 & 0 & 0 & 0 & 0 & 0 & 0 & 0 & 0 & 0 & 0 & 0 & 0 & 0 & 0 & 0 & 0 & 0 & 0 & 0 & 0 & 0 & 1 & 1 & 11 & 2 & $\pf 0, \hat g_{\mu\nu}$\\
\hline
1 & 0 & 0 & 0 & 1 & 0 & 0 & 0 & 0 & 0 & 0 & 0 & 1 & 0 & 0 & 0 & 0 & 0 & 0 & 0 & 0 & 0 & 2 & 35 & 1 & 1 & 1 & ${}^\star\, \pf 2$\\
\hline
0 & 1 & 1 & 0 & 0 & 0 & 0 & 0 & 0 & 0 & 1 & 0 & 0 & 0 & 0 & 1 & 0 & 0 & 0 & 0 & 0 & 0 & 2 & 7 & 6 & 1 & 1 & $\pf 1$\\
\hline
1 & 1 & 0 & 0 & 0 & 1 & 0 & 0 & 0 & 0 & 1 & 0 & 1 & 0 & 0 & 1 & 1 & 1 & 1 & 0 & 0 & 0 & 2 & 35 & 6 & 1 & 1 & ${}^\star\, \pf 1$\\
\hline
2 & 0 & 0 & 0 & 0 & 0 & 0 & 1 & 0 & 0 & 0 & 0 & 2 & 0 & 0 & 0 & 1 & 2 & 3 & 2 & 1 & 0 & 2 & 7 & 1 & 1 & 1 & de\\
\hline
0 & 2 & 0 & 1 & 0 & 0 & 0 & 0 & 0 & 0 & 0 & 1 & 0 & 1 & 2 & 2 & 1 & 0 & 0 & 0 & 0 & 0 & 2 & 21 & 1 & 1 & 1 & $\pf 2$\\
\hline
2 & 1 & 1 & 0 & 0 & 0 & 0 & 1 & 0 & 0 & 1 & 0 & 2 & 0 & 0 & 1 & 1 & 2 & 3 & 2 & 1 & 0 & 2 & 48 & 6 & 1 & 1 & \\
2 & 1 & 0 & 0 & 0 & 0 & 0 & 0 & 0 & 0 & 1 & 0 & 2 & 0 & 0 & 1 & 2 & 3 & 4 & 3 & 2 & 1 & 0 & 1 & 6 & 7 & 1 & top\\
\hline
1 & 2 & 1 & 0 & 0 & 1 & 0 & 0 & 0 & 0 & 0 & 1 & 1 & 1 & 2 & 2 & 1 & 1 & 1 & 0 & 0 & 0 & 2 & 224 & 1 & 1 & 1 & ${}^\star\,  g_{\mu\nu}$\\
1 & 2 & 0 & 0 & 0 & 0 & 1 & 0 & 1 & 1 & 0 & 0 & 1 & 0 & 1 & 2 & 2 & 2 & 2 & 1 & 0 & 0 & 2 & 21 & 15 & 1 & 1 & ${}^\star\, \pf 0$\\
1 & 2 & 0 & 0 & 0 & 0 & 1 & 0 & 0 & 0 & 0 & 1 & 1 & 1 & 2 & 2 & 2 & 2 & 2 & 1 & 0 & 0 & 0 & 21 & 1 & 8 & 1 & ${}^\star\, \pf 0$\\
\hline
1 & 3 & 1 & 0 & 0 & 0 & 1 & 0 & 0 & 0 & 1 & 1 & 1 & 1 & 2 & 3 & 2 & 2 & 2 & 1 & 0 & 0 & 2 & 140 & 6 & 1 & 1 & \\
1 & 3 & 0 & 0 & 0 & 0 & 0 & 1 & 0 & 2 & 0 & 1 & 1 & 0 & 2 & 3 & 3 & 3 & 3 & 2 & 1 & 0 & 2 & 7 & 10 & 1 & 1 & de\\
1 & 3 & 0 & 0 & 0 & 0 & 0 & 1 & 2 & 0 & 0 & 0 & 1 & 1 & 2 & 3 & 3 & 3 & 3 & 2 & 1 & 0 & 2 & 7 & 10 & 1 & 1 & de\\
1 & 3 & 0 & 0 & 0 & 0 & 0 & 1 & 0 & 0 & 1 & 1 & 1 & 1 & 2 & 3 & 3 & 3 & 3 & 2 & 1 & 0 & 0 & 7 & 6 & 8 & 1 & de\\
\hline
1 & 4 & 0 & 1 & 0 & 0 & 1 & 0 & 0 & 0 & 0 & 2 & 1 & 2 & 4 & 4 & 3 & 2 & 2 & 1 & 0 & 0 & 2 & 392 & 1 & 1 & 1 & \\
1 & 4 & 1 & 0 & 0 & 0 & 0 & 1 & 1 & 1 & 0 & 1 & 1 & 1 & 3 & 4 & 3 & 3 & 3 & 2 & 1 & 0 & 2 & 48 & 15 & 1 & 1 & \\
1 & 4 & 1 & 0 & 0 & 0 & 0 & 1 & 0 & 0 & 0 & 2 & 1 & 2 & 4 & 4 & 3 & 3 & 3 & 2 & 1 & 0 & 0 & 48 & 1 & 8 & 1 & \\
1 & 4 & 0 & 0 & 0 & 0 & 0 & 0 & 1 & 1 & 0 & 1 & 1 & 1 & 3 & 4 & 4 & 4 & 4 & 3 & 2 & 1 & 0 & 1 & 15 & 8 & 2 & top\\
1 & 4 & 0 & 0 & 0 & 0 & 0 & 0 & 0 & 0 & 0 & 2 & 1 & 2 & 4 & 4 & 4 & 4 & 4 & 3 & 2 & 1 & -2 & 1 & 1 & 45 & 1 & top\\
\hline
\end{longtable}

\begin{figure}[H]
\begin{center}
\begin{pspicture}(0,0)(8,1)
\dualityNode{1,1}{N11980953788}
\nodeLabel{N11980953788}{1}
\disabledNode{5,1}{N21980953788}
\nodeLabel{N21980953788}{2}
\dualityNode{0,0}{N31980953788}
\nodeLabel{N31980953788}{3}
\dualityNode{1,0}{N41980953788}
\nodeLabel{N41980953788}{4}
\dualityNode{2,0}{N51980953788}
\nodeLabel{N51980953788}{5}
\disabledNode{3,0}{N61980953788}
\nodeLabel{N61980953788}{6}
\normalNode{4,0}{N71980953788}
\nodeLabel{N71980953788}{7}
\normalNode{5,0}{N81980953788}
\nodeLabel{N81980953788}{8}
\normalNode{6,0}{N91980953788}
\nodeLabel{N91980953788}{9}
\normalNode{7,0}{N101980953788}
\nodeLabel{N101980953788}{10}
\normalNode{8,0}{N111980953788}
\nodeLabel{N111980953788}{11}
\singleConnection{N31980953788}{N41980953788}
\singleConnection{N41980953788}{N51980953788}
\singleConnection{N51980953788}{N61980953788}
\singleConnection{N61980953788}{N71980953788}
\singleConnection{N71980953788}{N81980953788}
\singleConnection{N81980953788}{N91980953788}
\singleConnection{N91980953788}{N101980953788}
\singleConnection{N101980953788}{N111980953788}
\singleConnection{N81980953788}{N21980953788}
\singleConnection{N41980953788}{N11980953788}
\end{pspicture}
\end{center}
\caption{$D_{8}^{+++}$ decomposed as $D_{4}^{} \otimes A_{5}^{}$}
\end{figure}

\begin{longtable}{|r@{\ }r|r@{\ }r@{\ }r@{\ }r@{\ }r|r@{\ }r@{\ }r@{\ }r|r@{\ }r@{\ }r@{\ }r@{\ }r@{\ }r@{\ }r@{\ }r@{\ }r@{\ }r@{\ }r|r|r|r|r|r|c|}
\caption{$D_{4}^{} \otimes A_{5}^{}$ representations in $D_{8}^{+++}$} \\
\hline
\multicolumn{2}{|c|}{$l$} &
\multicolumn{5}{|c|}{$p_{\rm{grav}}$} &
\multicolumn{4}{|c|}{$p_G$} &
\multicolumn{11}{|c|}{$m$} &
\multicolumn{1}{|c|}{$\alpha^2$} &
\multicolumn{1}{|c|}{$d_{\rm{grav}}$} &
\multicolumn{1}{|c|}{$d_G$} &
\multicolumn{1}{|c|}{$\textrm{mult}(\alpha)$} &
\multicolumn{1}{|c|}{$\mu$} &
\multicolumn{1}{|c|}{fields}\\
\hline
\hline
0 & 0 & 1 & 0 & 0 & 0 & 1 & 0 & 0 & 0 & 0 & 0 & 0 & 0 & 0 & 0 & 0 & -1 & -1 & -1 & -1 & -1 & 2 & 35 & 1 & 1 & 1 & $\bar g_{\mu\nu}$\\
0 & 0 & 0 & 0 & 0 & 0 & 0 & 0 & 0 & 1 & 0 & -1 & 0 & -1 & -2 & -1 & 0 & 0 & 0 & 0 & 0 & 0 & 2 & 1 & 28 & 1 & 1 & $\pf 0$\\
0 & 0 & 0 & 0 & 0 & 0 & 0 & 0 & 0 & 0 & 0 & 0 & 0 & 0 & 0 & 0 & 0 & 0 & 0 & 0 & 0 & 0 & 0 & 1 & 1 & 11 & 2 & $p=0, \bar g_{\mu\nu}$\\
\hline
1 & 0 & 0 & 1 & 0 & 0 & 0 & 0 & 0 & 0 & 0 & 0 & 1 & 0 & 0 & 0 & 0 & 0 & 0 & 0 & 0 & 0 & 2 & 15 & 1 & 1 & 1 & ${}^\star\, \pf 2$\\
\hline
0 & 1 & 1 & 0 & 0 & 0 & 0 & 0 & 0 & 0 & 1 & 0 & 0 & 0 & 0 & 0 & 1 & 0 & 0 & 0 & 0 & 0 & 2 & 6 & 8 & 1 & 1 & $\pf 1$\\
\hline
1 & 1 & 0 & 0 & 1 & 0 & 0 & 0 & 0 & 0 & 1 & 0 & 1 & 0 & 0 & 0 & 1 & 1 & 1 & 0 & 0 & 0 & 2 & 20 & 8 & 1 & 1 & ${}^\star\, \pf 1$\\
\hline
0 & 2 & 0 & 1 & 0 & 0 & 0 & 0 & 0 & 0 & 0 & 1 & 0 & 1 & 2 & 2 & 2 & 1 & 0 & 0 & 0 & 0 & 2 & 15 & 1 & 1 & 1 & $\pf 2$\\
\hline
2 & 1 & 0 & 0 & 0 & 0 & 1 & 0 & 0 & 0 & 1 & 0 & 2 & 0 & 0 & 0 & 1 & 2 & 3 & 2 & 1 & 0 & 2 & 6 & 8 & 1 & 1 & de\\
\hline
1 & 2 & 0 & 0 & 0 & 1 & 0 & 0 & 0 & 1 & 0 & 0 & 1 & 0 & 0 & 1 & 2 & 2 & 2 & 1 & 0 & 0 & 2 & 15 & 28 & 1 & 1 & ${}^\star\, \pf 0$\\
1 & 2 & 1 & 0 & 1 & 0 & 0 & 0 & 0 & 0 & 0 & 1 & 1 & 1 & 2 & 2 & 2 & 1 & 1 & 0 & 0 & 0 & 2 & 105 & 1 & 1 & 1 & ${}^\star\,  g_{\mu\nu}$\\
1 & 2 & 0 & 0 & 0 & 1 & 0 & 0 & 0 & 0 & 0 & 1 & 1 & 1 & 2 & 2 & 2 & 2 & 2 & 1 & 0 & 0 & 0 & 15 & 1 & 8 & 1 & ${}^\star\, \pf 0$\\
\hline
2 & 2 & 1 & 0 & 0 & 0 & 1 & 0 & 0 & 1 & 0 & 0 & 2 & 0 & 0 & 1 & 2 & 2 & 3 & 2 & 1 & 0 & 2 & 35 & 28 & 1 & 1 & \\
2 & 2 & 0 & 1 & 0 & 1 & 0 & 0 & 0 & 0 & 0 & 1 & 2 & 1 & 2 & 2 & 2 & 2 & 2 & 1 & 0 & 0 & 2 & 189 & 1 & 1 & 1 & \\
2 & 2 & 0 & 0 & 0 & 0 & 0 & 0 & 0 & 0 & 2 & 0 & 2 & 0 & 0 & 0 & 2 & 3 & 4 & 3 & 2 & 1 & 2 & 1 & 35 & 1 & 1 & top\\
2 & 2 & 1 & 0 & 0 & 0 & 1 & 0 & 0 & 0 & 0 & 1 & 2 & 1 & 2 & 2 & 2 & 2 & 3 & 2 & 1 & 0 & 0 & 35 & 1 & 8 & 1 & \\
2 & 2 & 0 & 0 & 0 & 0 & 0 & 0 & 0 & 1 & 0 & 0 & 2 & 0 & 0 & 1 & 2 & 3 & 4 & 3 & 2 & 1 & 0 & 1 & 28 & 7 & 1 & top\\
2 & 2 & 0 & 0 & 0 & 0 & 0 & 0 & 0 & 0 & 0 & 1 & 2 & 1 & 2 & 2 & 2 & 3 & 4 & 3 & 2 & 1 & -2 & 1 & 1 & 43 & 2 & top\\
\hline
1 & 3 & 1 & 0 & 0 & 1 & 0 & 0 & 0 & 0 & 1 & 1 & 1 & 1 & 2 & 2 & 3 & 2 & 2 & 1 & 0 & 0 & 2 & 84 & 8 & 1 & 1 & \\
1 & 3 & 0 & 0 & 0 & 0 & 1 & 1 & 1 & 0 & 0 & 0 & 1 & 0 & 1 & 2 & 3 & 3 & 3 & 2 & 1 & 0 & 2 & 6 & 56 & 1 & 1 & de\\
1 & 3 & 0 & 0 & 0 & 0 & 1 & 0 & 0 & 0 & 1 & 1 & 1 & 1 & 2 & 2 & 3 & 3 & 3 & 2 & 1 & 0 & 0 & 6 & 8 & 8 & 1 & de\\
\hline
1 & 4 & 1 & 0 & 0 & 0 & 1 & 0 & 0 & 1 & 0 & 1 & 1 & 1 & 2 & 3 & 4 & 3 & 3 & 2 & 1 & 0 & 2 & 35 & 28 & 1 & 1 & \\
1 & 4 & 0 & 1 & 0 & 1 & 0 & 0 & 0 & 0 & 0 & 2 & 1 & 2 & 4 & 4 & 4 & 3 & 2 & 1 & 0 & 0 & 2 & 189 & 1 & 1 & 1 & \\
1 & 4 & 0 & 0 & 0 & 0 & 0 & 0 & 2 & 0 & 0 & 1 & 1 & 0 & 2 & 3 & 4 & 4 & 4 & 3 & 2 & 1 & 2 & 1 & 35 & 1 & 1 & top\\
1 & 4 & 0 & 0 & 0 & 0 & 0 & 2 & 0 & 0 & 0 & 0 & 1 & 1 & 2 & 3 & 4 & 4 & 4 & 3 & 2 & 1 & 2 & 1 & 35 & 1 & 1 & top\\
1 & 4 & 1 & 0 & 0 & 0 & 1 & 0 & 0 & 0 & 0 & 2 & 1 & 2 & 4 & 4 & 4 & 3 & 3 & 2 & 1 & 0 & 0 & 35 & 1 & 8 & 1 & \\
1 & 4 & 0 & 0 & 0 & 0 & 0 & 0 & 0 & 1 & 0 & 1 & 1 & 1 & 2 & 3 & 4 & 4 & 4 & 3 & 2 & 1 & 0 & 1 & 28 & 8 & 1 & top\\
1 & 4 & 0 & 0 & 0 & 0 & 0 & 0 & 0 & 0 & 0 & 2 & 1 & 2 & 4 & 4 & 4 & 4 & 4 & 3 & 2 & 1 & -2 & 1 & 1 & 45 & 1 & top\\
\hline
\end{longtable}

\begin{figure}[H]
\begin{center}
\begin{pspicture}(0,0)(8,1)
\dualityNode{1,1}{N11980953788}
\nodeLabel{N11980953788}{1}
\normalNode{5,1}{N21980953788}
\nodeLabel{N21980953788}{2}
\dualityNode{0,0}{N31980953788}
\nodeLabel{N31980953788}{3}
\dualityNode{1,0}{N41980953788}
\nodeLabel{N41980953788}{4}
\dualityNode{2,0}{N51980953788}
\nodeLabel{N51980953788}{5}
\dualityNode{3,0}{N61980953788}
\nodeLabel{N61980953788}{6}
\disabledNode{4,0}{N71980953788}
\nodeLabel{N71980953788}{7}
\normalNode{5,0}{N81980953788}
\nodeLabel{N81980953788}{8}
\normalNode{6,0}{N91980953788}
\nodeLabel{N91980953788}{9}
\normalNode{7,0}{N101980953788}
\nodeLabel{N101980953788}{10}
\normalNode{8,0}{N111980953788}
\nodeLabel{N111980953788}{11}
\singleConnection{N31980953788}{N41980953788}
\singleConnection{N41980953788}{N51980953788}
\singleConnection{N51980953788}{N61980953788}
\singleConnection{N61980953788}{N71980953788}
\singleConnection{N71980953788}{N81980953788}
\singleConnection{N81980953788}{N91980953788}
\singleConnection{N91980953788}{N101980953788}
\singleConnection{N101980953788}{N111980953788}
\singleConnection{N81980953788}{N21980953788}
\singleConnection{N41980953788}{N11980953788}
\end{pspicture}
\end{center}
\caption{$D_{8}^{+++}$ decomposed as $D_{5}^{} \otimes A_{5}^{}$}
\end{figure}

\begin{longtable}{|r|r@{\ }r@{\ }r@{\ }r@{\ }r|r@{\ }r@{\ }r@{\ }r@{\ }r|r@{\ }r@{\ }r@{\ }r@{\ }r@{\ }r@{\ }r@{\ }r@{\ }r@{\ }r@{\ }r|r|r|r|r|r|c|}
\caption{$D_{5}^{} \otimes A_{5}^{}$ representations in $D_{8}^{+++}$} \\
\hline
\multicolumn{1}{|c|}{$l$} &
\multicolumn{5}{|c|}{$p_{\rm{grav}}$} &
\multicolumn{5}{|c|}{$p_G$} &
\multicolumn{11}{|c|}{$m$} &
\multicolumn{1}{|c|}{$\alpha^2$} &
\multicolumn{1}{|c|}{$d_{\rm{grav}}$} &
\multicolumn{1}{|c|}{$d_G$} &
\multicolumn{1}{|c|}{$\textrm{mult}(\alpha)$} &
\multicolumn{1}{|c|}{$\mu$} &
\multicolumn{1}{|c|}{fields}\\
\hline
\hline
0 & 0 & 0 & 0 & 0 & 0 & 0 & 0 & 0 & 1 & 0 & -1 & 0 & -1 & -2 & -2 & -1 & 0 & 0 & 0 & 0 & 0 & 2 & 1 & 45 & 1 & 1 & $\pf 0$\\
0 & 1 & 0 & 0 & 0 & 1 & 0 & 0 & 0 & 0 & 0 & 0 & -1 & 0 & 0 & 0 & 0 & 0 & -1 & -1 & -1 & -1 & 2 & 35 & 1 & 1 & 1 & $\bar g_{\mu\nu}$\\
0 & 0 & 0 & 0 & 0 & 0 & 0 & 0 & 0 & 0 & 0 & 0 & 0 & 0 & 0 & 0 & 0 & 0 & 0 & 0 & 0 & 0 & 0 & 1 & 1 & 11 & 1 & $\hat g_{\mu\nu}$\\
\hline
1 & 0 & 1 & 0 & 0 & 0 & 0 & 0 & 0 & 0 & 1 & 0 & 0 & 0 & 0 & 0 & 0 & 1 & 0 & 0 & 0 & 0 & 2 & 15 & 10 & 1 & 1 & $\pf 2, {}^\star\, \pf 2$\\
\hline
2 & 0 & 0 & 0 & 1 & 0 & 0 & 0 & 0 & 1 & 0 & 0 & 1 & 0 & 0 & 0 & 1 & 2 & 2 & 1 & 0 & 0 & 2 & 15 & 45 & 1 & 1 & ${}^\star\, \pf 0$\\
2 & 1 & 0 & 1 & 0 & 0 & 0 & 0 & 0 & 0 & 0 & 1 & 0 & 1 & 2 & 2 & 2 & 2 & 1 & 0 & 0 & 0 & 2 & 105 & 1 & 1 & 1 & ${}^\star\,  g_{\mu\nu}$\\
\hline
3 & 1 & 0 & 0 & 0 & 1 & 0 & 0 & 1 & 0 & 0 & 0 & 1 & 0 & 0 & 1 & 2 & 3 & 3 & 2 & 1 & 0 & 2 & 35 & 120 & 1 & 1 & \\
3 & 0 & 1 & 0 & 1 & 0 & 0 & 0 & 0 & 0 & 1 & 1 & 1 & 1 & 2 & 2 & 2 & 3 & 2 & 1 & 0 & 0 & 2 & 189 & 10 & 1 & 1 & \\
3 & 0 & 0 & 0 & 0 & 0 & 0 & 0 & 0 & 1 & 1 & 0 & 2 & 0 & 0 & 0 & 1 & 3 & 4 & 3 & 2 & 1 & 2 & 1 & 320 & 1 & 1 & top\\
3 & 1 & 0 & 0 & 0 & 1 & 0 & 0 & 0 & 0 & 1 & 1 & 1 & 1 & 2 & 2 & 2 & 3 & 3 & 2 & 1 & 0 & 0 & 35 & 10 & 8 & 1 & \\
3 & 0 & 0 & 0 & 0 & 0 & 0 & 0 & 0 & 0 & 1 & 1 & 2 & 1 & 2 & 2 & 2 & 3 & 4 & 3 & 2 & 1 & -2 & 1 & 10 & 43 & 1 & top\\
\hline
\end{longtable}

\begin{figure}[H]
\begin{center}
\begin{pspicture}(0,0)(8,1)
\dualityNode{1,1}{N1566184472}
\nodeLabel{N1566184472}{1}
\disabledNode{5,1}{N2566184472}
\nodeLabel{N2566184472}{2}
\dualityNode{0,0}{N3566184472}
\nodeLabel{N3566184472}{3}
\dualityNode{1,0}{N4566184472}
\nodeLabel{N4566184472}{4}
\dualityNode{2,0}{N5566184472}
\nodeLabel{N5566184472}{5}
\dualityNode{3,0}{N6566184472}
\nodeLabel{N6566184472}{6}
\disabledNode{4,0}{N7566184472}
\nodeLabel{N7566184472}{7}
\normalNode{5,0}{N8566184472}
\nodeLabel{N8566184472}{8}
\normalNode{6,0}{N9566184472}
\nodeLabel{N9566184472}{9}
\normalNode{7,0}{N10566184472}
\nodeLabel{N10566184472}{10}
\normalNode{8,0}{N11566184472}
\nodeLabel{N11566184472}{11}
\singleConnection{N3566184472}{N4566184472}
\singleConnection{N4566184472}{N5566184472}
\singleConnection{N5566184472}{N6566184472}
\singleConnection{N6566184472}{N7566184472}
\singleConnection{N7566184472}{N8566184472}
\singleConnection{N8566184472}{N9566184472}
\singleConnection{N9566184472}{N10566184472}
\singleConnection{N10566184472}{N11566184472}
\singleConnection{N8566184472}{N2566184472}
\singleConnection{N4566184472}{N1566184472}
\end{pspicture}
\end{center}
\caption{$D_{8}^{+++}$ decomposed as $D_{5}^{} \otimes A_{4}^{}$}
\end{figure}

\begin{longtable}{|r@{\ }r|r@{\ }r@{\ }r@{\ }r|r@{\ }r@{\ }r@{\ }r@{\ }r|r@{\ }r@{\ }r@{\ }r@{\ }r@{\ }r@{\ }r@{\ }r@{\ }r@{\ }r@{\ }r|r|r|r|r|r|c|}
\caption{$D_{5}^{} \otimes A_{4}^{}$ representations in $D_{8}^{+++}$} \\
\hline
\multicolumn{2}{|c|}{$l$} &
\multicolumn{4}{|c|}{$p_{\rm{grav}}$} &
\multicolumn{5}{|c|}{$p_G$} &
\multicolumn{11}{|c|}{$m$} &
\multicolumn{1}{|c|}{$\alpha^2$} &
\multicolumn{1}{|c|}{$d_{\rm{grav}}$} &
\multicolumn{1}{|c|}{$d_G$} &
\multicolumn{1}{|c|}{$\textrm{mult}(\alpha)$} &
\multicolumn{1}{|c|}{$\mu$} &
\multicolumn{1}{|c|}{fields}\\
\hline
\hline
0 & 0 & 0 & 0 & 0 & 0 & 0 & 0 & 0 & 1 & 0 & -1 & 0 & -1 & -2 & -2 & -1 & 0 & 0 & 0 & 0 & 0 & 2 & 1 & 45 & 1 & 1 & $\pf 0$\\
0 & 0 & 1 & 0 & 0 & 1 & 0 & 0 & 0 & 0 & 0 & 0 & 0 & 0 & 0 & 0 & 0 & 0 & -1 & -1 & -1 & -1 & 2 & 24 & 1 & 1 & 1 & $\bar g_{\mu\nu}$\\
0 & 0 & 0 & 0 & 0 & 0 & 0 & 0 & 0 & 0 & 0 & 0 & 0 & 0 & 0 & 0 & 0 & 0 & 0 & 0 & 0 & 0 & 0 & 1 & 1 & 11 & 2 & $p=0,\hat g_{\mu\nu}$\\
\hline
1 & 0 & 1 & 0 & 0 & 0 & 0 & 0 & 0 & 0 & 0 & 0 & 1 & 0 & 0 & 0 & 0 & 0 & 0 & 0 & 0 & 0 & 2 & 5 & 1 & 1 & 1 & $\pf 1$\\
\hline
0 & 1 & 1 & 0 & 0 & 0 & 0 & 0 & 0 & 0 & 1 & 0 & 0 & 0 & 0 & 0 & 0 & 1 & 0 & 0 & 0 & 0 & 2 & 5 & 10 & 1 & 1 & $\pf 1$\\
\hline
1 & 1 & 0 & 1 & 0 & 0 & 0 & 0 & 0 & 0 & 1 & 0 & 1 & 0 & 0 & 0 & 0 & 1 & 1 & 0 & 0 & 0 & 2 & 10 & 10 & 1 & 1 & ${}^\star\, \pf 1$\\
\hline
0 & 2 & 0 & 1 & 0 & 0 & 0 & 0 & 0 & 0 & 0 & 1 & 0 & 1 & 2 & 2 & 2 & 2 & 1 & 0 & 0 & 0 & 2 & 10 & 1 & 1 & 1 & ${}^\star\, \pf 1$\\
\hline
1 & 2 & 0 & 0 & 1 & 0 & 0 & 0 & 0 & 1 & 0 & 0 & 1 & 0 & 0 & 0 & 1 & 2 & 2 & 1 & 0 & 0 & 2 & 10 & 45 & 1 & 1 & ${}^\star\, \pf 0$\\
1 & 2 & 1 & 1 & 0 & 0 & 0 & 0 & 0 & 0 & 0 & 1 & 1 & 1 & 2 & 2 & 2 & 2 & 1 & 0 & 0 & 0 & 2 & 40 & 1 & 1 & 1 & ${}^\star\,  g_{\mu\nu}$\\
1 & 2 & 0 & 0 & 1 & 0 & 0 & 0 & 0 & 0 & 0 & 1 & 1 & 1 & 2 & 2 & 2 & 2 & 2 & 1 & 0 & 0 & 0 & 10 & 1 & 8 & 1 & ${}^\star\, \pf 0$\\
\hline
2 & 2 & 0 & 0 & 0 & 1 & 0 & 0 & 0 & 1 & 0 & 0 & 2 & 0 & 0 & 0 & 1 & 2 & 3 & 2 & 1 & 0 & 2 & 5 & 45 & 1 & 1 & de\\
2 & 2 & 1 & 0 & 1 & 0 & 0 & 0 & 0 & 0 & 0 & 1 & 2 & 1 & 2 & 2 & 2 & 2 & 2 & 1 & 0 & 0 & 2 & 45 & 1 & 1 & 1 & \\
\hline
1 & 3 & 0 & 0 & 0 & 1 & 0 & 0 & 1 & 0 & 0 & 0 & 1 & 0 & 0 & 1 & 2 & 3 & 3 & 2 & 1 & 0 & 2 & 5 & 120 & 1 & 1 & de\\
1 & 3 & 1 & 0 & 1 & 0 & 0 & 0 & 0 & 0 & 1 & 1 & 1 & 1 & 2 & 2 & 2 & 3 & 2 & 1 & 0 & 0 & 2 & 45 & 10 & 1 & 1 & \\
1 & 3 & 0 & 0 & 0 & 1 & 0 & 0 & 0 & 0 & 1 & 1 & 1 & 1 & 2 & 2 & 2 & 3 & 3 & 2 & 1 & 0 & 0 & 5 & 10 & 8 & 1 & de\\
\hline
2 & 3 & 1 & 0 & 0 & 1 & 0 & 0 & 1 & 0 & 0 & 0 & 2 & 0 & 0 & 1 & 2 & 3 & 3 & 2 & 1 & 0 & 2 & 24 & 120 & 1 & 1 & \\
2 & 3 & 0 & 0 & 0 & 0 & 0 & 0 & 0 & 1 & 1 & 0 & 2 & 0 & 0 & 0 & 1 & 3 & 4 & 3 & 2 & 1 & 2 & 1 & 320 & 1 & 1 & top\\
2 & 3 & 0 & 1 & 1 & 0 & 0 & 0 & 0 & 0 & 1 & 1 & 2 & 1 & 2 & 2 & 2 & 3 & 3 & 1 & 0 & 0 & 2 & 75 & 10 & 1 & 1 & \\
2 & 3 & 0 & 0 & 0 & 0 & 0 & 0 & 1 & 0 & 0 & 0 & 2 & 0 & 0 & 1 & 2 & 3 & 4 & 3 & 2 & 1 & 0 & 1 & 120 & 7 & 1 & top\\
2 & 3 & 1 & 0 & 0 & 1 & 0 & 0 & 0 & 0 & 1 & 1 & 2 & 1 & 2 & 2 & 2 & 3 & 3 & 2 & 1 & 0 & 0 & 24 & 10 & 8 & 2 & \\
2 & 3 & 0 & 0 & 0 & 0 & 0 & 0 & 0 & 0 & 1 & 1 & 2 & 1 & 2 & 2 & 2 & 3 & 4 & 3 & 2 & 1 & -2 & 1 & 10 & 43 & 2 & top\\
\hline
1 & 4 & 1 & 0 & 0 & 1 & 0 & 0 & 0 & 1 & 0 & 1 & 1 & 1 & 2 & 2 & 3 & 4 & 3 & 2 & 1 & 0 & 2 & 24 & 45 & 1 & 1 & \\
1 & 4 & 0 & 0 & 0 & 0 & 1 & 1 & 0 & 0 & 0 & 0 & 1 & 0 & 1 & 2 & 3 & 4 & 4 & 3 & 2 & 1 & 2 & 1 & 210 & 1 & 1 & top\\
1 & 4 & 0 & 0 & 0 & 0 & 0 & 0 & 0 & 1 & 0 & 1 & 1 & 1 & 2 & 2 & 3 & 4 & 4 & 3 & 2 & 1 & 0 & 1 & 45 & 8 & 1 & top\\
1 & 4 & 0 & 1 & 1 & 0 & 0 & 0 & 0 & 0 & 0 & 2 & 1 & 2 & 4 & 4 & 4 & 4 & 3 & 1 & 0 & 0 & 2 & 75 & 1 & 1 & 1 & \\
1 & 4 & 1 & 0 & 0 & 1 & 0 & 0 & 0 & 0 & 0 & 2 & 1 & 2 & 4 & 4 & 4 & 4 & 3 & 2 & 1 & 0 & 0 & 24 & 1 & 8 & 1 & \\
1 & 4 & 0 & 0 & 0 & 0 & 0 & 0 & 0 & 0 & 0 & 2 & 1 & 2 & 4 & 4 & 4 & 4 & 4 & 3 & 2 & 1 & -2 & 1 & 1 & 45 & 1 & top\\
\hline
\end{longtable}

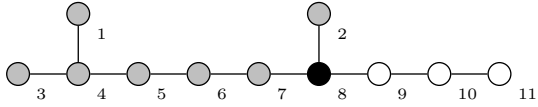
\begin{figure}[H]
\begin{center}
\begin{pspicture}(0,0)(8,1)
\dualityNode{1,1}{N11872301448}
\nodeLabel{N11872301448}{1}
\dualityNode{5,1}{N21872301448}
\nodeLabel{N21872301448}{2}
\dualityNode{0,0}{N31872301448}
\nodeLabel{N31872301448}{3}
\dualityNode{1,0}{N41872301448}
\nodeLabel{N41872301448}{4}
\dualityNode{2,0}{N51872301448}
\nodeLabel{N51872301448}{5}
\dualityNode{3,0}{N61872301448}
\nodeLabel{N61872301448}{6}
\dualityNode{4,0}{N71872301448}
\nodeLabel{N71872301448}{7}
\disabledNode{5,0}{N81872301448}
\nodeLabel{N81872301448}{8}
\normalNode{6,0}{N91872301448}
\nodeLabel{N91872301448}{9}
\normalNode{7,0}{N101872301448}
\nodeLabel{N101872301448}{10}
\normalNode{8,0}{N111872301448}
\nodeLabel{N111872301448}{11}
\singleConnection{N31872301448}{N41872301448}
\singleConnection{N41872301448}{N51872301448}
\singleConnection{N51872301448}{N61872301448}
\singleConnection{N61872301448}{N71872301448}
\singleConnection{N71872301448}{N81872301448}
\singleConnection{N81872301448}{N91872301448}
\singleConnection{N91872301448}{N101872301448}
\singleConnection{N101872301448}{N111872301448}
\singleConnection{N81872301448}{N21872301448}
\singleConnection{N41872301448}{N11872301448}
\end{pspicture}
\end{center}
\caption{$D_{8}^{+++}$ decomposed as $D_{6}^{} \otimes A_{1}^{} \otimes A_{3}^{}$}
\end{figure}

\begin{longtable}{|r|r@{\ }r@{\ }r|r@{\ }r@{\ }r@{\ }r@{\ }r@{\ }r@{\ }r|r@{\ }r@{\ }r@{\ }r@{\ }r@{\ }r@{\ }r@{\ }r@{\ }r@{\ }r@{\ }r|r|r|r|r|r|c|}
\caption{$D_{6}^{} \otimes A_{1}^{} \otimes A_{3}^{}$ representations in $D_{8}^{+++}$} \\
\hline
\multicolumn{1}{|c|}{$l$} &
\multicolumn{3}{|c|}{$p_{\rm{grav}}$} &
\multicolumn{7}{|c|}{$p_G$} &
\multicolumn{11}{|c|}{$m$} &
\multicolumn{1}{|c|}{$\alpha^2$} &
\multicolumn{1}{|c|}{$d_{\rm{grav}}$} &
\multicolumn{1}{|c|}{$d_G$} &
\multicolumn{1}{|c|}{$\textrm{mult}(\alpha)$} &
\multicolumn{1}{|c|}{$\mu$} &
\multicolumn{1}{|c|}{fields}\\
\hline
\hline
0 & 0 & 0 & 0 & 0 & 0 & 0 & 0 & 0 & 1 & 0 & -1 & 0 & -1 & -2 & -2 & -2 & -1 & 0 & 0 & 0 & 0 & 2 & 1 & 66 & 1 & 1 & $\pf 0$\\
0 & 1 & 0 & 1 & 0 & 0 & 0 & 0 & 0 & 0 & 0 & 0 & 0 & 0 & 0 & 0 & 0 & 0 & 0 & -1 & -1 & -1 & 2 & 15 & 1 & 1 & 1 & $\bar g_{\mu\nu}$\\
0 & 0 & 0 & 0 & 0 & 2 & 0 & 0 & 0 & 0 & 0 & 0 & -1 & 0 & 0 & 0 & 0 & 0 & 0 & 0 & 0 & 0 & 2 & 1 & 3 & 1 & 1 & $\pf 0$\\
0 & 0 & 0 & 0 & 0 & 0 & 0 & 0 & 0 & 0 & 0 & 0 & 0 & 0 & 0 & 0 & 0 & 0 & 0 & 0 & 0 & 0 & 0 & 1 & 1 & 11 & 1 & $\hat g_{\mu\nu}$\\
\hline
1 & 1 & 0 & 0 & 0 & 1 & 0 & 0 & 0 & 0 & 1 & 0 & 0 & 0 & 0 & 0 & 0 & 0 & 1 & 0 & 0 & 0 & 2 & 4 & 24 & 1 & 1 & $\pf 1, {}^\star\, \pf 1$\\
\hline
2 & 0 & 1 & 0 & 0 & 0 & 0 & 0 & 0 & 1 & 0 & 0 & 1 & 0 & 0 & 0 & 0 & 1 & 2 & 1 & 0 & 0 & 2 & 6 & 66 & 1 & 1 & ${}^\star\, \pf 0$\\
2 & 2 & 0 & 0 & 0 & 0 & 0 & 0 & 0 & 0 & 0 & 1 & 1 & 1 & 2 & 2 & 2 & 2 & 2 & 0 & 0 & 0 & 2 & 10 & 1 & 1 & 1 & ${}^\star\,  g_{\mu\nu}$\\
2 & 0 & 1 & 0 & 0 & 2 & 0 & 0 & 0 & 0 & 0 & 1 & 0 & 1 & 2 & 2 & 2 & 2 & 2 & 1 & 0 & 0 & 2 & 6 & 3 & 1 & 1 & ${}^\star\, \pf 0$\\
\hline
3 & 0 & 0 & 1 & 0 & 1 & 0 & 0 & 1 & 0 & 0 & 0 & 1 & 0 & 0 & 0 & 1 & 2 & 3 & 2 & 1 & 0 & 2 & 4 & 440 & 1 & 1 & de\\
3 & 1 & 1 & 0 & 0 & 1 & 0 & 0 & 0 & 0 & 1 & 1 & 1 & 1 & 2 & 2 & 2 & 2 & 3 & 1 & 0 & 0 & 2 & 20 & 24 & 1 & 1 & \\
3 & 0 & 0 & 1 & 0 & 1 & 0 & 0 & 0 & 0 & 1 & 1 & 1 & 1 & 2 & 2 & 2 & 2 & 3 & 2 & 1 & 0 & 0 & 4 & 24 & 8 & 1 & de\\
\hline
4 & 0 & 0 & 0 & 0 & 0 & 0 & 0 & 1 & 0 & 1 & 0 & 2 & 0 & 0 & 0 & 1 & 2 & 4 & 3 & 2 & 1 & 2 & 1 & 2079 & 1 & 1 & top\\
4 & 1 & 0 & 1 & 0 & 0 & 0 & 1 & 0 & 0 & 0 & 0 & 2 & 0 & 0 & 1 & 2 & 3 & 4 & 2 & 1 & 0 & 2 & 15 & 495 & 1 & 1 & \\
4 & 0 & 0 & 0 & 0 & 2 & 0 & 1 & 0 & 0 & 0 & 0 & 1 & 0 & 0 & 1 & 2 & 3 & 4 & 3 & 2 & 1 & 2 & 1 & 1485 & 1 & 1 & top\\
4 & 1 & 0 & 1 & 0 & 0 & 0 & 0 & 0 & 0 & 2 & 1 & 2 & 1 & 2 & 2 & 2 & 2 & 4 & 2 & 1 & 0 & 2 & 15 & 77 & 1 & 1 & \\
4 & 0 & 2 & 0 & 0 & 0 & 0 & 0 & 0 & 1 & 0 & 1 & 2 & 1 & 2 & 2 & 2 & 3 & 4 & 2 & 0 & 0 & 2 & 20 & 66 & 1 & 1 & \\
4 & 1 & 0 & 1 & 0 & 2 & 0 & 0 & 0 & 1 & 0 & 1 & 1 & 1 & 2 & 2 & 2 & 3 & 4 & 2 & 1 & 0 & 2 & 15 & 198 & 1 & 1 & \\
4 & 1 & 0 & 1 & 0 & 0 & 0 & 0 & 0 & 1 & 0 & 1 & 2 & 1 & 2 & 2 & 2 & 3 & 4 & 2 & 1 & 0 & 0 & 15 & 66 & 8 & 1 & \\
4 & 0 & 0 & 0 & 0 & 2 & 0 & 0 & 0 & 1 & 0 & 1 & 1 & 1 & 2 & 2 & 2 & 3 & 4 & 3 & 2 & 1 & 0 & 1 & 198 & 8 & 1 & top\\
4 & 0 & 0 & 0 & 0 & 0 & 0 & 0 & 0 & 1 & 0 & 1 & 2 & 1 & 2 & 2 & 2 & 3 & 4 & 3 & 2 & 1 & -2 & 1 & 66 & 43 & 2 & top\\
4 & 2 & 1 & 0 & 0 & 0 & 0 & 0 & 0 & 0 & 0 & 2 & 2 & 2 & 4 & 4 & 4 & 4 & 4 & 1 & 0 & 0 & 2 & 45 & 1 & 1 & 1 & \\
4 & 0 & 2 & 0 & 0 & 2 & 0 & 0 & 0 & 0 & 0 & 2 & 1 & 2 & 4 & 4 & 4 & 4 & 4 & 2 & 0 & 0 & 2 & 20 & 3 & 1 & 1 & \\
4 & 1 & 0 & 1 & 0 & 2 & 0 & 0 & 0 & 0 & 0 & 2 & 1 & 2 & 4 & 4 & 4 & 4 & 4 & 2 & 1 & 0 & 0 & 15 & 3 & 8 & 1 & \\
4 & 1 & 0 & 1 & 0 & 0 & 0 & 0 & 0 & 0 & 0 & 2 & 2 & 2 & 4 & 4 & 4 & 4 & 4 & 2 & 1 & 0 & -2 & 15 & 1 & 44 & 2 & \\
4 & 0 & 0 & 0 & 0 & 2 & 0 & 0 & 0 & 0 & 0 & 2 & 1 & 2 & 4 & 4 & 4 & 4 & 4 & 3 & 2 & 1 & -2 & 1 & 3 & 45 & 1 & top\\
\hline
\end{longtable}

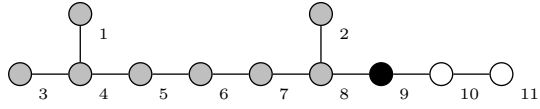
\begin{figure}[H]
\begin{center}
\begin{pspicture}(0,0)(8,1)
\dualityNode{1,1}{N11528605172}
\nodeLabel{N11528605172}{1}
\dualityNode{5,1}{N21528605172}
\nodeLabel{N21528605172}{2}
\dualityNode{0,0}{N31528605172}
\nodeLabel{N31528605172}{3}
\dualityNode{1,0}{N41528605172}
\nodeLabel{N41528605172}{4}
\dualityNode{2,0}{N51528605172}
\nodeLabel{N51528605172}{5}
\dualityNode{3,0}{N61528605172}
\nodeLabel{N61528605172}{6}
\dualityNode{4,0}{N71528605172}
\nodeLabel{N71528605172}{7}
\dualityNode{5,0}{N81528605172}
\nodeLabel{N81528605172}{8}
\disabledNode{6,0}{N91528605172}
\nodeLabel{N91528605172}{9}
\normalNode{7,0}{N101528605172}
\nodeLabel{N101528605172}{10}
\normalNode{8,0}{N111528605172}
\nodeLabel{N111528605172}{11}
\singleConnection{N31528605172}{N41528605172}
\singleConnection{N41528605172}{N51528605172}
\singleConnection{N51528605172}{N61528605172}
\singleConnection{N61528605172}{N71528605172}
\singleConnection{N71528605172}{N81528605172}
\singleConnection{N81528605172}{N91528605172}
\singleConnection{N91528605172}{N101528605172}
\singleConnection{N101528605172}{N111528605172}
\singleConnection{N81528605172}{N21528605172}
\singleConnection{N41528605172}{N11528605172}
\end{pspicture}
\end{center}
\caption{$D_{8}^{+++}$ decomposed as $D_{8}^{} \otimes A_{2}^{}$}
\end{figure}

\begin{longtable}{|r|r@{\ }r|r@{\ }r@{\ }r@{\ }r@{\ }r@{\ }r@{\ }r@{\ }r|r@{\ }r@{\ }r@{\ }r@{\ }r@{\ }r@{\ }r@{\ }r@{\ }r@{\ }r@{\ }r|r|r|r|r|r|c|}
\caption{$D_{8}^{} \otimes A_{2}^{}$ representations in $D_{8}^{+++}$} \\
\hline
\multicolumn{1}{|c|}{$l$} &
\multicolumn{2}{|c|}{$p_{\rm{grav}}$} &
\multicolumn{8}{|c|}{$p_G$} &
\multicolumn{11}{|c|}{$m$} &
\multicolumn{1}{|c|}{$\alpha^2$} &
\multicolumn{1}{|c|}{$d_{\rm{grav}}$} &
\multicolumn{1}{|c|}{$d_G$} &
\multicolumn{1}{|c|}{$\textrm{mult}(\alpha)$} &
\multicolumn{1}{|c|}{$\mu$} &
\multicolumn{1}{|c|}{fields}\\
\hline
\hline
0 & 0 & 0 & 0 & 0 & 0 & 0 & 0 & 0 & 0 & 1 & -1 & -1 & -1 & -2 & -2 & -2 & -2 & -2 & 0 & 0 & 0 & 2 & 1 & 120 & 1 & 1 & $\pf 0$\\
0 & 1 & 1 & 0 & 0 & 0 & 0 & 0 & 0 & 0 & 0 & 0 & 0 & 0 & 0 & 0 & 0 & 0 & 0 & 0 & -1 & -1 & 2 & 8 & 1 & 1 & 1 & $\bar g_{\mu\nu}$\\
0 & 0 & 0 & 0 & 0 & 0 & 0 & 0 & 0 & 0 & 0 & 0 & 0 & 0 & 0 & 0 & 0 & 0 & 0 & 0 & 0 & 0 & 0 & 1 & 1 & 11 & 1 & $\hat g_{\mu\nu}$\\
\hline
1 & 1 & 0 & 0 & 0 & 0 & 0 & 0 & 0 & 0 & 1 & 0 & 0 & 0 & 0 & 0 & 0 & 0 & 0 & 1 & 0 & 0 & 2 & 3 & 120 & 1 & 1 & ${}^\star\, \pf 0$\\
\hline
2 & 0 & 1 & 0 & 0 & 0 & 0 & 0 & 1 & 0 & 0 & 0 & 1 & 0 & 0 & 0 & 0 & 1 & 2 & 2 & 1 & 0 & 2 & 3 & 1820 & 1 & 1 & de\\
2 & 2 & 0 & 0 & 0 & 0 & 0 & 0 & 0 & 0 & 1 & 1 & 1 & 1 & 2 & 2 & 2 & 2 & 2 & 2 & 0 & 0 & 2 & 6 & 120 & 1 & 1 & \\
2 & 0 & 1 & 0 & 2 & 0 & 0 & 0 & 0 & 0 & 0 & 1 & 0 & 1 & 2 & 2 & 2 & 2 & 2 & 2 & 1 & 0 & 2 & 3 & 135 & 1 & 1 & de\\
2 & 0 & 1 & 0 & 0 & 0 & 0 & 0 & 0 & 0 & 0 & 2 & 2 & 2 & 4 & 4 & 4 & 4 & 4 & 2 & 1 & 0 & -2 & 3 & 1 & 44 & 1 & de\\
\hline
3 & 0 & 0 & 0 & 1 & 0 & 0 & 1 & 0 & 0 & 0 & 0 & 1 & 0 & 0 & 0 & 1 & 2 & 3 & 3 & 2 & 1 & 2 & 1 & 60060 & 1 & 1 & top\\
3 & 1 & 1 & 0 & 0 & 0 & 1 & 0 & 0 & 0 & 0 & 0 & 2 & 0 & 0 & 1 & 2 & 3 & 4 & 3 & 1 & 0 & 2 & 8 & 8008 & 1 & 1 & \\
3 & 1 & 1 & 0 & 1 & 0 & 0 & 0 & 0 & 1 & 0 & 1 & 1 & 1 & 2 & 2 & 2 & 2 & 3 & 3 & 1 & 0 & 2 & 8 & 7020 & 1 & 1 & \\
3 & 0 & 0 & 0 & 1 & 0 & 0 & 0 & 0 & 1 & 0 & 1 & 1 & 1 & 2 & 2 & 2 & 2 & 3 & 3 & 2 & 1 & 0 & 1 & 7020 & 8 & 1 & top\\
3 & 1 & 1 & 0 & 0 & 0 & 0 & 0 & 1 & 0 & 0 & 1 & 2 & 1 & 2 & 2 & 2 & 3 & 4 & 3 & 1 & 0 & 0 & 8 & 1820 & 8 & 1 & \\
3 & 0 & 0 & 0 & 0 & 0 & 0 & 0 & 1 & 0 & 0 & 1 & 2 & 1 & 2 & 2 & 2 & 3 & 4 & 3 & 2 & 1 & -2 & 1 & 1820 & 43 & 1 & top\\
3 & 3 & 0 & 0 & 0 & 0 & 0 & 0 & 0 & 0 & 1 & 2 & 2 & 2 & 4 & 4 & 4 & 4 & 4 & 3 & 0 & 0 & 2 & 10 & 120 & 1 & 1 & \\
3 & 1 & 1 & 0 & 2 & 0 & 0 & 0 & 0 & 0 & 0 & 2 & 1 & 2 & 4 & 4 & 4 & 4 & 4 & 3 & 1 & 0 & 0 & 8 & 135 & 8 & 1 & \\
3 & 1 & 1 & 0 & 0 & 0 & 0 & 0 & 0 & 0 & 1 & 2 & 2 & 2 & 4 & 4 & 4 & 4 & 4 & 3 & 1 & 0 & -2 & 8 & 120 & 44 & 2 & \\
3 & 0 & 0 & 0 & 2 & 0 & 0 & 0 & 0 & 0 & 0 & 2 & 1 & 2 & 4 & 4 & 4 & 4 & 4 & 3 & 2 & 1 & -2 & 1 & 135 & 45 & 1 & top\\
3 & 0 & 0 & 0 & 0 & 0 & 0 & 0 & 0 & 0 & 1 & 2 & 2 & 2 & 4 & 4 & 4 & 4 & 4 & 3 & 2 & 1 & -4 & 1 & 120 & 195 & 1 & top\\
3 & 1 & 1 & 0 & 0 & 0 & 0 & 0 & 0 & 0 & 0 & 3 & 3 & 3 & 6 & 6 & 6 & 6 & 6 & 3 & 1 & 0 & -4 & 8 & 1 & 192 & 1 & \\
\hline
\end{longtable}

\normalsize


\end{document}